\documentstyle[eqsecnum,feynmf,ifthen,aps]{revtex}

\newcommand{\be}{\begin{eqnarray}}
\newcommand{\ee}{\end{eqnarray}}
\newcommand{\bc}{\begin{center}}
\newcommand{\ec}{\end{center}}
\newcommand{\s}{\footnotesize{s}}
\newcommand{\h}{\footnotesize{h}}
\newcommand{\slashed}[1]%              gives the Feynman-slash
{{%                                    using e.g.:
\setbox0=\hbox{$#1$}%                    \slashed{p}
\makebox[0pt][l]{$#1$}%
\makebox[\wd0][c]{/}%
}}
\newcommand{\G}{\footnotesize{G}}
\newcommand{\nx}{\noexpand}

\newcounter{test}
\newcounter{rueck}
\newcounter{diagh}[rueck]

\newcommand{\refmy}[1]{\ifthenelse{\equal{\ref{#1}}{1}\or\equal{\ref{#1}}{2}
\or\equal{\ref{#1}}{3}\or\equal{\ref{#1}}{4}\or\equal{\ref{#1}}{5}\or\equal{\ref{#1}}{6}
\or\equal{\ref{#1}}{7}\or\equal{\ref{#1}}{8}\or\equal{\ref{#1}}{9}\or\equal{\ref{#1}}{10}
\or\equal{\ref{#1}}{11}\or\equal{\ref{#1}}{12}\or\equal{\ref{#1}}{13}\or\equal{\ref{#1}}{14}
\or\equal{\ref{#1}}{15}\or\equal{\ref{#1}}{16}\or\equal{\ref{#1}}{17}\or\equal{\ref{#1}}{18}
\or\equal{\ref{#1}}{19}\or\equal{\ref{#1}}{20}\or\equal{\ref{#1}}{21}\or\equal{\ref{#1}}{22}
\or\equal{\ref{#1}}{23}\or\equal{\ref{#1}}{24}\or\equal{\ref{#1}}{25}\or\equal{\ref{#1}}{26}}{\setcounter{test}{\ref{#1}}\Alph{test}}{{\bf ??}}}

\newcommand{\arrowliner}[6]{
  \fmfforce{(#1,#2)}{v20}
  \fmfforce{(#3,#4)}{v21}
  \fmf{plains,#5}{v20,v21}
  \fmf{#6,#5}{v20,v21}
}

\newcommand{\arrowlinel}[6]{
  \fmfforce{#1,#2}{v20}
  \fmfforce{#3,#4}{v21}
  \fmf{plains,#5}{v21,v20}
  \fmf{#6,#5}{v21,v20}
}

\newcommand{\stargl}{
  \fmfsurround{v1,v2,v3}
  \fmfforce{(0.5w,h)}{v1}
  \fmfforce{(.933w,.25h)}{v3}
  \fmfforce{(.067w,.25h)}{v2}
  \fmfforce{(.5w,.5h)}{v4}
  \fmfsurround{v5,v6,v7}
  \fmfforce{(0.5w,h)}{v5}
  \fmfforce{(.933w,.25h)}{v7}
  \fmfforce{(.067w,.25h)}{v6}
  \fmfforce{(.5w,.5h)}{v8}
}

\newcommand{\starplace}[8]{ 
  \ifthenelse{\equal{#1}{}}{}{\fmfv{label=#1, l.a=#2, l.d=.1w}{v1}}
  \ifthenelse{\equal{#3}{}}{}{\fmfv{label=#3, l.a=#4, l.d=.1w}{v2}}
  \ifthenelse{\equal{#5}{}}{}{\fmfv{label=#5, l.a=#6, l.d=.1w}{v3}}
  \ifthenelse{\equal{#7}{}}{}{\fmfv{label=#7, l.a=#8, l.d=.1w}{v4}}
}

\newcommand{\starplacedummy}[8]{ 
  \ifthenelse{\equal{#1}{}}{}{\fmfv{label=#1, l.a=#2, l.d=.1w}{v5}}
  \ifthenelse{\equal{#3}{}}{}{\fmfv{label=#3, l.a=#4, l.d=.1w}{v6}}
  \ifthenelse{\equal{#5}{}}{}{\fmfv{label=#5, l.a=#6, l.d=.1w}{v7}}
  \ifthenelse{\equal{#7}{}}{}{\fmfv{label=#7, l.a=#8, l.d=.1w}{v8}}
}

\newcommand{\startypes}[6]{
\ifthenelse{\equal{#2}{arrowr} \or \equal{#2}{arrowl}}{
\ifthenelse{\equal{#2}{arrowr}}{
\fmf{#1,left=0.578}{v2,v1}\arrowliner{.104w}{.613h}{.204w}{.786h}{left=.1}{#2}}{\fmf{#1,left=0.578}{v2,v1}\arrowlinel{.104w}{.613h}{.204w}{.786h}{right=.1}{#2}}}{\fmf{#1,left=0.578,label=#2,l.s=right,l.d=0.05w}{v2,v1}}
\ifthenelse{\equal{#4}{arrowr} \or \equal{#4}{arrowl}}{
\ifthenelse{\equal{#4}{arrowr}}{
\fmf{#3,left=0.578}{v1,v3}\arrowliner{.896w}{.613h}{.796w}{.786h}{right=.1}{#4}}{\fmf{#3,left=0.578}{v1,v3}\arrowlinel{.896w}{.613h}{.796w}{.786h}{left=.1}{#4}}}{\fmf{#3,left=0.578,label=#4,l.s=right,l.d=0.05w}{v1,v3}}
\ifthenelse{\equal{#6}{arrowr} \or \equal{#6}{arrowl}}{
\ifthenelse{\equal{#6}{arrowr}}{
\fmf{#5,left=0.578}{v3,v2}\arrowliner{.4w}{.1h}{.6w}{.1h}{right=.1}{#6}}{\fmf{#5,left=0.578}{v3,v2}\arrowlinel{.4w}{.1h}{.6w}{.1h}{left=.1}{#6}}}{\fmf{#5,left=0.578,label=#6,l.s=right,l.d=0.05w}{v3,v2}}
}

\newcommand{\startypec}[6]{
\ifthenelse{\equal{#2}{arrowr} \or \equal{#2}{arrowl}}{
\ifthenelse{\equal{#2}{arrowr}}{
\fmf{#1}{v1,v4}\arrowliner{.575w}{.65h}{.575w}{.85h}{left=0}{#2}}{\fmf{#1}{v1,v4}\arrowlinel{.575w}{.65h}{.575w}{.85h}{right=0}{#2}}}{\fmf{#1,label=#2,l.s=left,l.d=0.05w}{v1,v4}}
\ifthenelse{\equal{#4}{arrowr} \or \equal{#4}{arrowl}}{
\ifthenelse{\equal{#4}{arrowr}}{
\fmf{#3}{v4,v2}\arrowliner{.163w}{.425h}{.337w}{.525h}{right=0}{#4}}{\fmf{#3}{v4,v2}\arrowlinel{.163w}{.425h}{.337w}{.525h}{left=0}{#4}}}{\fmf{#3,label=#4,l.s=left,l.d=0.05w}{v4,v2}}
\ifthenelse{\equal{#6}{arrowr} \or \equal{#6}{arrowl}}{
\ifthenelse{\equal{#6}{arrowr}}{
\fmf{#5}{v3,v4}\arrowliner{.837w}{.425h}{.663w}{.525h}{right=0}{#6}}{\fmf{#5}{v3,v4}\arrowlinel{.837w}{.425h}{.663w}{.525h}{left=0}{#6}}}{\fmf{#5,label=#6,l.s=left,l.d=0.05w}{v3,v4}}
}

\newcommand{\starproj}[6]{
  \ifthenelse{\equal{#1}{proj_v}}{\fmf{projs,right=0.578}{v1,v2}}{ }
  \ifthenelse{\equal{#1}{proj_h}}{\fmf{projs,left=0.578}{v2,v1}}{ }
  \ifthenelse{\equal{#2}{proj_v}}{\fmf{projs,right=0.578}{v3,v1}}{ }
  \ifthenelse{\equal{#2}{proj_h}}{\fmf{projs,left=0.578}{v1,v3}}{ }
  \ifthenelse{\equal{#3}{proj_v}}{\fmf{projs,right=0.578}{v2,v3}}{ }
  \ifthenelse{\equal{#3}{proj_h}}{\fmf{projs,left=0.578}{v3,v2}}{ }
  \ifthenelse{\equal{#4}{proj_v}}{\fmf{projc}{v1,v4}}{ }
  \ifthenelse{\equal{#4}{proj_h}}{\fmf{projc}{v4,v1}}{ }
  \ifthenelse{\equal{#5}{proj_v}}{\fmf{projc}{v2,v4}}{ }
  \ifthenelse{\equal{#5}{proj_h}}{\fmf{projc}{v4,v2}}{ }
  \ifthenelse{\equal{#6}{proj_v}}{\fmf{projc}{v3,v4}}{ }
  \ifthenelse{\equal{#6}{proj_h}}{\fmf{projc}{v4,v3}}{ }
}

\newcommand{\selfgl}{
 \fmfsurround{v1,v2,v3,v4}
 \fmfsurround{v5,v6,v7,v8}
 \fmfforce{(.146w,.854h)}{v2}
 \fmfforce{(.854w,.854h)}{v1}
 \fmfforce{(.146w,.146h)}{v3}
 \fmfforce{(.854w,.146h)}{v4}
 \fmfforce{(.146w,.854h)}{v6}
 \fmfforce{(.854w,.854h)}{v5}
 \fmfforce{(.146w,.146h)}{v7}
 \fmfforce{(.854w,.146h)}{v8}
} 

\newcommand{\selfplace}[8]{
 \ifthenelse{\equal{#1}{}}{}{\fmfv{label=#1, l.a=#2, l.d=.1w}{v1}}
 \ifthenelse{\equal{#3}{}}{}{\fmfv{label=#3, l.a=#4, l.d=.1w}{v2}}
 \ifthenelse{\equal{#5}{}}{}{\fmfv{label=#5, l.a=#6, l.d=.1w}{v3}}
 \ifthenelse{\equal{#7}{}}{}{\fmfv{label=#7, l.a=#8, l.d=.1w}{v4}}
}

\newcommand{\selfplacedummy}[8]{
 \ifthenelse{\equal{#1}{}}{}{\fmfv{label=#1, l.a=#2, l.d=.1w}{v5}}
 \ifthenelse{\equal{#3}{}}{}{\fmfv{label=#3, l.a=#4, l.d=.1w}{v6}}
 \ifthenelse{\equal{#5}{}}{}{\fmfv{label=#5, l.a=#6, l.d=.1w}{v7}}
 \ifthenelse{\equal{#7}{}}{}{\fmfv{label=#7, l.a=#8, l.d=.1w}{v8}}
}

\newcommand{\selftypeb}[8]{
\ifthenelse{\equal{#2}{arrowr} \or \equal{#2}{arrowl}}{
\ifthenelse{\equal{#2}{arrowr}}{
\fmf{#1,right=0.414}{v1,v2}\arrowliner{.4w}{.9h}{.6w}{.9w}{left=0.1}{#2}}{
\fmf{#1,right=0.414}{v1,v2}\arrowlinel{.4w}{.9h}{.6w}{.9h}{right=0.1}{#2}}}{
\fmf{#1,right=0.414,label=#2,l.s=left,l.d=0.05w}{v1,v2}}
\ifthenelse{\equal{#4}{arrowr} \or \equal{#4}{arrowl}}{
\ifthenelse{\equal{#4}{arrowr}}{
\fmf{#3,right=0.414}{v2,v1}\arrowliner{.4w}{.808h}{.6w}{.808w}{right=0.1}{#4}}{
\fmf{#3,right=0.414}{v2,v1}\arrowlinel{.4w}{.808h}{.6w}{.808h}{left=0.1}{#4}}}{
\fmf{#3,right=0.414,label=#4,l.s=left,l.d=0.05w}{v2,v1}}
\ifthenelse{\equal{#6}{arrowr} \or \equal{#6}{arrowl}}{
\ifthenelse{\equal{#6}{arrowr}}{
\fmf{#5,right=0.414}{v3,v4}\arrowliner{.4w}{.1h}{.6w}{.1h}{right=0.1}{#6}}{
\fmf{#5,right=0.414}{v3,v4}\arrowlinel{.4w}{.1h}{.6w}{.1h}{left=0.1}{#6}}}{
\fmf{#5,right=0.414,label=#6,l.s=left,l.d=0.05w}{v3,v4}}
\ifthenelse{\equal{#8}{arrowr} \or \equal{#8}{arrowl}}{
\ifthenelse{\equal{#8}{arrowr}}{
\fmf{#7,right=0.414}{v4,v3}\arrowliner{.4w}{.192h}{.6w}{.192h}{left=0.1}{#8}}{
\fmf{#7,right=0.414}{v4,v3}\arrowlinel{.4w}{.192h}{.6w}{.192h}{right=0.1}{#8}}}{\fmf{#7,right=0.414,label=#8,l.s=left,l.d=0.05w}{v4,v3}}
}

\newcommand{\selftypes}[4]{
\ifthenelse{\equal{#2}{arrowr} \or \equal{#2}{arrowl}}{
\ifthenelse{\equal{#2}{arrowr}}{
\fmf{#1,right=0.414}{v4,v1}\arrowliner{.9w}{.4h}{.9w}{.6h}{right=0.1}{#2}}{
\fmf{#1,right=0.414}{v4,v1}\arrowlinel{.9w}{.4h}{.9w}{.6h}{left=0.1}{#2}}}{\fmf{#1,right=0.414,label=#2,l.s=left,l.d=0.05w}{v4,v1}}
\ifthenelse{\equal{#4}{arrowr} \or \equal{#4}{arrowl}}{
\ifthenelse{\equal{#4}{arrowl}}{
\fmf{#3,right=0.414}{v2,v3}\arrowliner{.1w}{.4h}{.1w}{.6h}{left=0.1}{#4}}{
\fmf{#3,right=0.414}{v2,v3}\arrowlinel{.1w}{.4h}{.1w}{.6h}{right=0.1}{#4}}}{\fmf{#3,right=0.414,label=#4,l.s=left,l.d=0.05w}{v2,v3}}
}

\newcommand{\selfproj}[6]{
  \ifthenelse{\equal{#1}{proj_v}}{\fmf{projs,right=0.414}{v1,v2}}{ }
  \ifthenelse{\equal{#1}{proj_h}}{\fmf{projs,left=0.414}{v2,v1}}{ }
  \ifthenelse{\equal{#2}{proj_v}}{\fmf{projs,right=0.414}{v2,v1}}{ }
  \ifthenelse{\equal{#2}{proj_h}}{\fmf{projs,left=0.414}{v1,v2}}{ }
  \ifthenelse{\equal{#3}{proj_v}}{\fmf{projs,right=0.414}{v4,v3}}{ }
  \ifthenelse{\equal{#3}{proj_h}}{\fmf{projs,left=0.414}{v3,v4}}{ }
  \ifthenelse{\equal{#4}{proj_v}}{\fmf{projs,right=0.414}{v3,v4}}{ }
  \ifthenelse{\equal{#4}{proj_h}}{\fmf{projs,left=0.414}{v4,v3}}{ }
  \ifthenelse{\equal{#5}{proj_v}}{\fmf{projs,right=0.414}{v4,v1}}{ }
  \ifthenelse{\equal{#5}{proj_h}}{\fmf{projs,left=0.414}{v1,v4}}{ }
  \ifthenelse{\equal{#6}{proj_v}}{\fmf{projs,right=0.414}{v2,v3}}{ }
  \ifthenelse{\equal{#6}{proj_h}}{\fmf{projs,left=0.414}{v3,v2}}{ }
}

\newcommand{\selftwogl}{
 \fmfsurround{v1,v2,v3}
 \fmfforce{(0.5w,.75h)}{v1}
 \fmfforce{(0.175w,0.1875h)}{v2}
 \fmfforce{(.825w,0.1875h)}{v3}
 \fmfforce{(.5w,h)}{v7}
 \fmfsurround{v4,v5,v6}
 \fmfforce{(0.5w,.75h)}{v4}
 \fmfforce{(0.175w,0.1875h)}{v5}
 \fmfforce{(.825w,0.1875h)}{v6}
}

\newcommand{\selftwoplace}[8]{
 \ifthenelse{\equal{#1}{}}{}{\fmfv{label=#1, l.a=#2, l.d=.1w}{v1}}
 \ifthenelse{\equal{#3}{}}{}{\fmfv{label=#3, l.a=#4, l.d=.1w}{v2}}
 \ifthenelse{\equal{#5}{}}{}{\fmfv{label=#5, l.a=#6, l.d=.1w}{v3}}
 \ifthenelse{\equal{#7}{}}{}{\fmfv{label=#7, l.a=#8, l.d=.1w}{v7}}
}

\newcommand{\selftwoplacedummy}[6]{
 \ifthenelse{\equal{#1}{}}{}{\fmfv{label=#1, l.a=#2, l.d=.1w}{v4}}
 \ifthenelse{\equal{#3}{}}{}{\fmfv{label=#3, l.a=#4, l.d=.1w}{v5}}
 \ifthenelse{\equal{#5}{}}{}{\fmfv{label=#5, l.a=#6, l.d=.1w}{v6}}
}

\newcommand{\selftwotypes}[8]{ 
\ifthenelse{\equal{#2}{arrowr} \or \equal{#2}{arrowl}}{
\ifthenelse{\equal{#2}{arrowr}}{
\fmf{#1,left=0.578}{v2,v1}\arrowliner{.19w}{.438h}{.29w}{.612h}{left=0.1}{#2}}{
\fmf{#1,left=0.578}{v2,v1}\arrowlinel{.19w}{.438h}{.29w}{.612h}{right=0.1}{#2}}}{
\fmf{#1,left=0.578,label=#2,l.s=right,l.d=0.05w}{v2,v1}}
\ifthenelse{\equal{#4}{arrowr} \or \equal{#4}{arrowl}}{
\ifthenelse{\equal{#4}{arrowr}}{
\fmf{#3,left=0.578}{v1,v3}\arrowliner{.81w}{.438h}{.71w}{.612h}{right=0.1}{#4}}{
\fmf{#3,left=0.578}{v1,v3}\arrowlinel{.81w}{.438h}{.71w}{.612h}{left=0.1}{#4}}}{
\fmf{#3,left=0.578,label=#4,l.s=right,l.d=0.05w}{v1,v3}}
\ifthenelse{\equal{#6}{arrowr} \or \equal{#6}{arrowl}}{
\ifthenelse{\equal{#6}{arrowr}}{
\fmf{#5,right=0.578}{v3,v2}\arrowliner{.4w}{.1h}{.6w}{.1h}{right=0.1}{#6}}{
\fmf{#5,right=0.578}{v3,v2}\arrowlinel{.4w}{.1h}{.6w}{.1h}{left=0.1}{#6}}}{
\fmf{#5,right=0.578,label=#6,l.s=left,l.d=0.05w}{v3,v2}}
\ifthenelse{\equal{#8}{arrowr} \or \equal{#8}{arrowl}}{
\ifthenelse{\equal{#8}{arrowr}}{
\fmf{#7,right=0.578}{v2,v3}\arrowliner{.4w}{.275h}{.6w}{.275h}{left=0.1}{#8}}{
\fmf{#7,right=0.578}{v2,v3}\arrowlinel{.4w}{.275h}{.6w}{.275h}{right=0.1}{#8}}}{\fmf{#7,right=0.578,label=#8,l.s=left,l.d=0.05w}{v2,v3}}
}

\newcommand{\selftwotypeb}[2]{
\fmf{#1,left,label=#2,l.s=right,l.d=0.05w}{v1,v7}
\fmf{#1,left}{v7,v1}
}

\newcommand{\selftwoproj}[5]{
  \ifthenelse{\equal{#1}{proj_v}}{\fmf{projs,left=0.578}{v2,v1}}{ }
  \ifthenelse{\equal{#1}{proj_h}}{\fmf{projs,right=0.578}{v1,v2}}{ }
  \ifthenelse{\equal{#2}{proj_v}}{\fmf{projs,right=0.578}{v3,v1}}{ }
  \ifthenelse{\equal{#2}{proj_h}}{\fmf{projs,left=0.578}{v1,v3}}{ }
  \ifthenelse{\equal{#3}{proj_v}}{\fmf{projs,right=0.578}{v3,v2}}{ }
  \ifthenelse{\equal{#3}{proj_h}}{\fmf{projs,left=0.578}{v2,v3}}{ }
  \ifthenelse{\equal{#4}{proj_v}}{\fmf{projs,right=0.578}{v2,v3}}{ }
  \ifthenelse{\equal{#4}{proj_h}}{\fmf{projs,left=0.578}{v3,v2}}{ }
  \ifthenelse{\equal{#5}{proj_v}}{\fmf{projs,right}{v1,v7}}{ }
  \ifthenelse{\equal{#5}{proj_h}}{\fmf{projs,left}{v1,v7}}{ }
}

\newcommand{\pacgl}{
  \fmfsurround{v1,v2,v3}
  \fmfforce{(0,.5h)}{v1}
  \fmfforce{(.75w,0.067h)}{v2}
  \fmfforce{(.75w,.933h)}{v3}
  \fmfsurround{v4,v5,v6}
  \fmfforce{(0,.5h)}{v4}
  \fmfforce{(.75w,0.067)}{v5}
  \fmfforce{(.75w,.933h)}{v6}
}

\newcommand{\pacplace}[6]{
  \ifthenelse{\equal{#1}{}}{}{\fmfv{label=#1, l.a=#2, l.d=.1w}{v1}}
 \ifthenelse{\equal{#3}{}}{}{\fmfv{label=#3, l.a=#4, l.d=.1w}{v2}}
 \ifthenelse{\equal{#5}{}}{}{\fmfv{label=#5, l.a=#6, l.d=.1w}{v3}}
}

\newcommand{\pacplacedummy}[6]{
 \ifthenelse{\equal{#1}{}}{}{\fmfv{label=#1, l.a=#2, l.d=.1w}{v4}}
 \ifthenelse{\equal{#3}{}}{}{\fmfv{label=#3, l.a=#4, l.d=.1w}{v5}}
 \ifthenelse{\equal{#5}{}}{}{\fmfv{label=#5, l.a=#6, l.d=.1w}{v6}}
}

\newcommand{\pactypes}[6]{
\ifthenelse{\equal{#2}{arrowr} \or \equal{#2}{arrowl}}{
\ifthenelse{\equal{#2}{arrowr}}{
\fmf{#1,right=0.578}{v1,v2}\arrowliner{.387w}{.104h}{.214w}{.204h}{left=0.1}{#2}}{
\fmf{#1,right=0.578}{v1,v2}\arrowlinel{.387w}{.104h}{.214w}{.204h}{right=0.1}{#2}}}{
\fmf{#1,right=0.578,label=#2,l.s=left,l.d=0.05w}{v1,v2}}
 \ifthenelse{\equal{#4}{arrowr} \or \equal{#4}{arrowl}}{
\ifthenelse{\equal{#4}{arrowr}}{
\fmf{#3,left=0.578}{v3,v2}\arrowliner{.387w}{.896h}{.214w}{.796h}{right=0.1}{#4}}{
\fmf{#3,left=0.578}{v3,v2}\arrowlinel{.387w}{.896h}{.214w}{.796h}{left=0.1}{#4}}}{
\fmf{#3,left=0.578,label=#4,l.s=right,l.d=0.05w}{v3,v2}}
\ifthenelse{\equal{#6}{arrowr} \or \equal{#6}{arrowl}}{
\ifthenelse{\equal{#6}{arrowr}}{
\fmf{#5,right=0.578}{v3,v1}\arrowliner{.9w}{.4h}{.9w}{.6h}{right=0.1}{#6}}{
\fmf{#5,right=0.578}{v3,v1}\arrowlinel{.9w}{.4h}{.9w}{.6h}{left=0.1}{#6}}}{
\fmf{#5,right=0.578,label=#6,l.s=left,l.d=0.05w}{v3,v1}}
}

\newcommand{\pactypec}[4]{
\ifthenelse{\equal{#2}{arrowr} \or \equal{#2}{arrowl}}{
\ifthenelse{\equal{#2}{arrowr}}{
\fmf{#1}{v1,v3}\arrowliner{.271w}{.757h}{.479w}{.877h}{left=0}{#2}}{
\fmf{#1}{v1,v2}\arrowlinel{.271w}{.757h}{.479w}{.877h}{left=0}{#2}}}{
\fmf{#1,label=#2,l.s=left,l.d=0.05w}{v1,v2}}
\ifthenelse{\equal{#4}{arrowr} \or \equal{#4}{arrowl}}{
\ifthenelse{\equal{#4}{arrowr}}{
\fmf{#3}{v1,v3}\arrowliner{.271w}{.444h}{.479w}{.324h}{right=0}{#4}}{
\fmf{#3}{v1,v3}\arrowlinel{.271w}{.444h}{.479w}{.324h}{left=0}{#4}}}{
\fmf{#3,label=#4,l.s=right,l.d=0.05w}{v1,v3}}
}

\newcommand{\pacproj}[5]{
  \ifthenelse{\equal{#1}{proj_v}}{\fmf{projs,right=0.578}{v1,v2}}{ }
  \ifthenelse{\equal{#1}{proj_h}}{\fmf{projs,left=0.578}{v2,v1}}{ }
  \ifthenelse{\equal{#2}{proj_v}}{\fmf{projs,right=0.578}{v2,v3}}{ }
  \ifthenelse{\equal{#2}{proj_h}}{\fmf{projs,left=0.578}{v3,v2}}{ }
  \ifthenelse{\equal{#3}{proj_v}}{\fmf{projs,right=0.578}{v3,v1}}{ }
  \ifthenelse{\equal{#3}{proj_h}}{\fmf{projs,left=0.578}{v1,v3}}{ }
  \ifthenelse{\equal{#4}{proj_v}}{\fmf{projs}{v1,v2}}{ }
  \ifthenelse{\equal{#4}{proj_h}}{\fmf{projs}{v2,v1}}{ }
  \ifthenelse{\equal{#5}{proj_v}}{\fmf{projs}{v1,v3}}{ }
  \ifthenelse{\equal{#5}{proj_h}}{\fmf{projs}{v3,v1}}{ }
}
 
\begin{document}

%%%%%%%%%%%%%%%%%%%%%%%%%%%%%%%%%%%%%%%%%%%%%%%%%%%%%%%%%%%%%%%%%%%%%%%
%                                                                     %
% Starting the 1.font-file and adding additional Metafont-Definitions %
%                                                                     %
%%%%%%%%%%%%%%%%%%%%%%%%%%%%%%%%%%%%%%%%%%%%%%%%%%%%%%%%%%%%%%%%%%%%%%%

\begin{fmffile}{paperf1}

%%%%%%%%%%%%%%%%%%%%%%%%%%%%%%%%%%%%%%%%%%%%%%%%%%%%%%%%%%%%%%%%%%%%%%%
%                                                                     %
%       Some Metafont-definitions in addition to feynmf.mf            %
%                                                                     %
%%%%%%%%%%%%%%%%%%%%%%%%%%%%%%%%%%%%%%%%%%%%%%%%%%%%%%%%%%%%%%%%%%%%%%%
%
% Variables:
%

\fmfcurved
\fmfpen{thin} 
\fmfset{curly_len}{3mm}
\fmfset{wiggly_len}{2mm}

%
% Styles:
%

\fmfcmd{ %
 vardef cross_bar (expr p, len, ang) =
  ((-len/2,0)--(len/2,0))
    rotated (ang + angle direction length(p)/2 of p)
    shifted point length(p)/2 of p
 enddef;
 style_def crossed expr p = 
   ccutdraw cross_bar (p, 5mm, 45);
   ccutdraw cross_bar (p, 5mm, -45)
 enddef;}

\fmfcmd{ %
 vardef proj_tarrow (expr p, frac) =
  save a, t, z;
  pair z;
  t1 = frac*length p;
  a = angle direction t1 of p;
  z = point t1 of p;
  (t2,whatever) = p intersectiontimes
    (halfcircle scaled 3arrow_len rotated (a-90) shifted z);
  arrow_head (p, t1, t2, arrow_ang)
enddef;
 style_def projs expr p  =
    cfill (proj_tarrow (reverse p, 0.85));
 enddef;
 style_def projc expr p =
    cfill (proj_tarrow (reverse p, 0.7));
 enddef;}

\fmfcmd{ 
 vardef projl_tarrow (expr p, frac) =
  save a, t, z;
  pair z;
  t1 = frac*length p;
  a = angle direction t1 of p;
  z = point t1 of p;
  (t2,whatever) = p intersectiontimes
    (halfcircle scaled 2arrow_len rotated (a-90) shifted z);
  arrow_head (p, t1, t2, arrow_ang)
 enddef;
 style_def arrowr expr p =
   cfill (projl_tarrow (reverse p, 0.6));
 enddef;
 style_def arrowl expr p  =
   cfill (projl_tarrow (reverse p, 0.6));
 enddef;
}

\fmfcmd{
style_def plains expr p =
 draw p;
 undraw subpath (0,.15) of p;
enddef;}

%%%%%%%%%%%%%%%%%%%%%%%%%%%%%%%%%%%%%%%%%%%%%%%%%%%%%%%%%%%%%%%%%%%%%%%
%                                                                     %
%                  Here starts the paper                              %
%                                                                     %
%%%%%%%%%%%%%%%%%%%%%%%%%%%%%%%%%%%%%%%%%%%%%%%%%%%%%%%%%%%%%%%%%%%%%%%

\draft
\preprint{TRP-96-23}

\title{
 Gauge Invariance of Resummation Schemes:\\
 The QCD Partition Function
}

\author{M.~Achhammer$^1$ , U.~Heinz$^{2}$\cite{Uli} , 
        S.~Leupold$^{1,3}$, U.~A.~Wiedemann$^1$}
\address{
   $~^1$Institut f\"ur Theoretische Physik, Universit\"at Regensburg,\\
        D-93040 Regensburg, Germany\\
   $~^{2}$Cern/TH, CH-1211 Geneva 23\\
        Switzerland\\
   $~^{3}$Institut f\"ur Theoretische Physik, Justus-Liebig-Universit\"at 
          Giessen,\\ 
        D-35392 Giessen, Germany
}
\date{\today}

\maketitle

\begin{abstract}

We pick up a method originally developed by Cheng and Tsai for vacuum 
perturbation theory which allows to test the consistency of different sets of 
Feynman rules on a purely diagrammatic level, making explicit loop 
calculations superfluous. We generalize it to perturbative  calculations in 
thermal field theory and we show that it can be adapted to check the gauge 
invariance of physical quantities calculated in improved perturbation schemes.
Specifically, we extend this diagrammatic technique to a simple resummation 
scheme in imaginary time perturbation theory. As an application, we check up 
to {$\cal O$}({\em g}$\,^4$) in general covariant gauge the gauge invariance 
of the result for the QCD partition function which was recently obtained in 
Feynman gauge.

\end{abstract} 


%%%%%%%%%%%%%%%%%%%%%%%%%%%%%%%%%%%%%%%%%%%%%%%%%%%%%%%%%%%%%%%%%%%%%%
\section{Introduction} \label{sec1}
%%%%%%%%%%%%%%%%%%%%%%%%%%%%%%%%%%%%%%%%%%%%%%%%%%%%%%%%%%%%%%%%%%%%%%

In finite temperature quantum field theory for massless degrees of freedom, 
the loop expansion does not coincide with an expansion in orders of the 
coupling constant. Naive perturbative calculations lead to gauge dependent 
and infrared divergent results for physical quantities \cite{bra90b}. 
This necessitates the reorganization of the perturbative expansion in 
consistent resummation schemes. One of the most important consistency checks 
for such schemes is the test of their gauge invariance. In practice, this is 
done mostly by explicit calculations \cite{bra90a}, though Ward identities 
can be used as well \cite{reb90}. In the vacuum sector, yet another technical 
tool is at hand. This is a purely diagrammatic method developed by Cheng and 
Tsai \cite{cheMIT} which allows to establish the gauge invariance of a set of 
Feynman diagrams without carrying out explicit loop calculations. Recently a 
similar diagrammatic method was independently developed by Feng and Lam 
\cite{fen96}. The main motivation of the present work is to undertake a first 
step in extending the Cheng/Tsai method to resummed perturbative calculations.
To this end, we review and slightly extend the approach of Cheng and Tsai 
presenting a complete set of diagrammatic rules. As an example we show how to 
check in this approach the gauge invariance of the two loop contribution to 
the vacuum partition function to {$\cal O$}({\em g}$\,^2$) in general 
covariant $\alpha$-gauge. Then, we extend these rules to a simple resummation 
scheme.

As a first application, we focus  on the gauge invariance up to 
{$\cal O$}({\em g}$\,^{4}$) of the QCD partition function. The latter has 
recently been calculated up to {$\cal O$}({\em g}$\,^5$) 
\cite{arn94,arn95,zha95} using a gluon propagator with a resummed static mode. 
Indeed, there are several reasons for choosing this example. First, the 
calculations to order {$\cal O$}({\em g}$\,^4$) and {$\cal O$}({\em g}$\,^5$) 
exist in Feynman gauge only, and the authors did not check the gauge 
invariance of their results. Since the chosen resummation scheme amounts to 
the ``free'' Lagrangian of a massive Yang-Mills field, which is not gauge 
invariant, invariance of the final result is not automatic. This is different 
from naive perturbation theory where gauge invariance of observable quantities
is automatic if one uses a complete and consistent set of Feynman rules. 
(For certain gauges this may be a non-trivial problem in itself.) In fact, 
the method of Cheng and Tsai was invented to study the consistency of 
different sets of Feynman rules in this respect. Second, in contrast to 
dynamic quantities of finite temperature QCD which in general require resummed
propagators {\em and} vertices \cite{bra90a}, the calculation of the partition
function requires only resummed propagators. So, our work extends the method 
of Cheng and Tsai to the simplest case of a resummation scheme.\\
In Sec.~\ref{sec2} we outline the method of Cheng and Tsai \cite{cheMIT} and 
give a short example. In Sec.~\ref{sec3} we discuss the resummation scheme and
some general aspects of the free energy (resp.~partition function). As an 
application we extend the example discussed in Sec.~\ref{sec2} to finite 
temperature and prove the gauge invariance of the QCD free energy up to 
{$\cal O$}({\em g}$\,^3$). In Sec.~\ref{sec4} we discuss a technical issue 
which arose in the Secs.~\ref{sec2} and \ref{sec3}, the problem of shifting 
momentum variables. In Sec.~\ref{sec5} we then prove the gauge invariance of 
the free energy in finite temperature QCD up to {$\cal O$}({\em g}$\,^4$).\\

%%%%%%%%%%%%%%%%%%%%%%%%%%%%%%%%%%%%%%%%%%%%%%%%%%%%%%%%%%%%%%%%%%%%%
\section{The diagrammatic method of Cheng and Tsai}\label{sec2}
%%%%%%%%%%%%%%%%%%%%%%%%%%%%%%%%%%%%%%%%%%%%%%%%%%%%%%%%%%%%%%%%%%%%%

In the Feynman rules of QCD, the chosen gauge manifests itself in the 
$k_\mu$-dependent part of the free gluon propagator $D_{\mu\nu}^{ab}$ 
and in the way $D_{\mu\nu}^{ab}$ couples to ghosts. We write the gluon 
propagator in the general form

\be
D_{\mu\nu}^{ab} &=& 
   -i\delta^{ab}\left[g_{\mu\nu} + a_\mu (-k)k_\nu - a_\nu (k)k_\mu \right]
    \frac{1}{k^2 + i\varepsilon} \nonumber\\
            & \equiv &  
   D_{F,\mu\nu}^{ab}(k) + \Delta_{\mu}^{ab}(-k)k_{\nu} 
   - \Delta_{\nu}^{ab}(k)k_{\mu}.
\label{prop}
\ee

\be
 \parbox{3cm}{\bc
  \begin{fmfgraph}(2.5,2)
   \fmfforce{(0,.5h)}{v1}
   \fmfforce{(w,.5h)}{v2}
   \fmf{boson}{v1,v2}
 \end{fmfgraph}
\ec}
\quad = \quad
 \parbox{2.5cm}{\bc
  \begin{fmfgraph*}(2.5,2)
   \fmfforce{(0,.5h)}{v1}
   \fmfforce{(w,.5h)}{v2}
   \fmf{boson,label=$D_F$,label.s=left}{v1,v2}
 \end{fmfgraph*}
\ec}
\quad + \quad
 \parbox{2.5cm}{\bc
  \begin{fmfgraph*}(2.5,2)
   \fmfforce{(0,.5h)}{v1}
   \fmfforce{(w,.5h)}{v2}
   \fmfforce{(.85w,.5h)}{v3}
   \fmf{boson}{v1,v3}
   \fmfv{label=$\nx\Delta$,label.a=35}{v1}
   \fmf{projc}{v2,v1}
 \end{fmfgraph*}
\ec}
\quad - \quad
 \parbox{2.5cm}{\bc
  \begin{fmfgraph*}(2.5,2)
   \fmfforce{(0,.5h)}{v1}
   \fmfforce{(w,.5h)}{v2}
   \fmfforce{(.15w,.5h)}{v3}
   \fmf{boson}{v3,v2}
   \fmfv{label=$\nx\Delta$,label.a=145}{v2}
   \fmf{projc}{v1,v2}
 \end{fmfgraph*}
\ec}\nonumber
\ee
Diagrammatically we denote the factor $k_\mu$ by an arrow-head. The 
Feynman-propagator $D_{F,\mu\nu}^{ab}$ is given by 

\be
  D_{F,\mu\nu}^{ab}(k) = -i\delta^{ab} g_{\mu\nu} \frac{1}{k^2 + i\varepsilon} 
\ee
and $\Delta_{\mu}^{ab}$ is defined as

\be
  \Delta_{\mu}^{ab}(k) = -i\delta^{ab}\frac{a_{\mu}(k)}{k^{2}+i\varepsilon}.
\ee

The gauge dependent pieces are contained in the function $a_\mu (k)$,
e.g.~$a_\mu (k)= \frac{1}{2}(1-\alpha)\frac{k_\mu}{k^2}$ for covariant 
$\alpha $-gauges. Checking the gauge invariance of a set of Feynman graphs 
amounts then to the assertion that the result is unaffected by changing 
$a_\mu$(k), i.e.~by choosing a different gauge. In the following we review 
the method of Cheng and Tsai \cite{cheMIT} to do this check diagrammatically.\\

The free gluon propagator in (\ref{prop}) has been separated diagrammatically 
into its Feynman gauge part $D_F$ and explicitly gauge dependent parts. In 
this way the method of Cheng and Tsai allows to disentangle arbitrary diagrams
into their Feynman gauge parts and remainder terms. A set of diagrammatic 
rules is used to test whether the sum of these remainder terms vanishes. If 
this is the case, the calculation is independent of the particular choice of 
$a_\mu (k)$. The original motivation of Cheng and Tsai was to check by this 
technique the consistency of the Feynman rules for different gauges
\cite{cheMIT,che86a,che86b}. We shall employ the same technique to show the 
gauge invariance for quantities calculated in resummation schemes.

\subsection{The Basic Idea}
\label{sec2.1.1}

We start with $-i\delta^{ab}\frac{a_\mu (k)k_\nu}{k^2+i\varepsilon}$, 
the gauge-dependent part of the propagator. We split this term into the two 
factors $\Delta_{\mu}^{ab}(k)$ and $k_\nu$. If this is a contribution to an 
internal line of a Feynman diagram, then $k_\nu$ is Lorentz-contracted with a 
vertex. For the case of a 3-gluon-vertex, this contraction results in a very 
simple structure,

\be
  k^\nu \Gamma_{\nu\mu\rho}^{abc}(k,q,p) =
     \mbox{\em{g}}f^{abc} \left[\left(p^2g_{\mu\rho}-p_\mu p_\rho \right)
                             -\left(q^2g_{\mu\rho}-q_\mu q_\rho \right)\right].
\label{projec}
\ee

Each one of the two terms in square brackets has the structure of a transverse
projector ${\cal P}$, 
e.g.~${\cal P}_{\mu\rho}(q) = q^2g_{\mu\rho} - q_\mu q_\rho$, acting 
on one of the 2 outgoing lines of the vertex. Note that in what follows we 
refer to ${\cal P}$ as a ``projector'' though it does not satisfy the 
normalization, ${\cal P}^{2} \not= {\cal P}$.  Diagrammatically, we express 
(\ref{projec}) in the following way:

\be
 \parbox{2cm}{\bc
  \begin{fmfgraph*}(2,2)
   \fmfoutgoing{o1}
   \fmfforce{(0,h)}{v2}
   \fmfforce{(0,0)}{v3}
   \fmfforce{(.4w,.5h)}{v1}
   \fmf{boson,label=k,label.d=.04w}{v2,v1}
   \fmfset{arrow_ang}{25}
   \fmf{projc}{v1,v2}
   \fmfset{arrow_ang}{15}
   \fmf{boson,label=q,label.side=left,label.d=.04w}{v3,v1}
   \fmf{boson,label=p,label.d=.04w}{v1,o1}
   \fmfdot{v1}
  \end{fmfgraph*}
 \ec }
\quad = \quad + \quad
 \parbox{2cm}{\bc
  \begin{fmfgraph*}(2,2)
   \fmfoutgoing{o1}
   \fmfforce{(0,h)}{v2}
   \fmfforce{(0,0)}{v3}
   \fmfforce{(.4w,.5h)}{v1}
   \fmf{boson,label=k,label.d=0.04w}{v2,v1}
   \fmf{boson,label=q,label.d=0.04w}{v3,v1}
   \fmf{boson,label=p,label.d=0.04w}{v1,o1}
   \fmfv{label=${\nx\cal P}$,l.a=30}{v1}    
  \end{fmfgraph*}
\ec}
\quad - \quad
 \parbox{2cm}{\bc
  \begin{fmfgraph*}(2,2)
   \fmfoutgoing{o1}
   \fmfforce{(0,h)}{v2}
   \fmfforce{(0,0)}{v3}
   \fmfforce{(.4w,.5h)}{v1}
   \fmfv{label=${\nx\cal P}$,l.a=-95}{v1}
   \fmf{boson,label=k,label.d=0.04w}{v2,v1}
   \fmf{boson,label=q,label.d=0.04w}{v3,v1}
   \fmf{boson,label=p,label.d=0.04w}{v1,o1}
  \end{fmfgraph*}
 \ec}.
\label{projecdia}
\ee

The blob at the vertex on the l.h.s.~indicates that this vertex carries a 
complete Lorentz structure which has disappeared on the r.h.s.~upon 
contraction. Note that the diagrams on the r.h.s.~have opposite sign. \\
${\cal P}$ acts on the next propagator in the following way:

\be
  \mbox{\em{g}} f^{abc} {\cal P}_\mu^\rho (p) D_{\rho\sigma}^{cd}(p) 
\qquad = \qquad 
  - \qquad i\mbox{\em{g}} f^{abd}g_{\mu\sigma} \qquad - 
    \qquad G_{\mu}^{abd} (p)p_\sigma \; .
\label{main}
\ee
We represent this equation graphically as

\be
 \parbox{3cm}{\bc
  \begin{fmfgraph*}(2.5,2)
   \fmfforce{(0,.5h)}{v1}
   \fmfforce{(w,.5h)}{v2}
   \fmfforce{(0,.5h)}{v3}
   \fmf{boson}{v1,v2}
   \fmfv{label=${\nx\cal P}$,l.a=35}{v3}
 \end{fmfgraph*}
\ec }
\quad = \quad \mbox{--} \quad
 \parbox{2.5cm}{\bc
  \begin{fmfgraph}(2.5,2)
   \fmfforce{(0,.5h)}{v1}
   \fmfforce{(w,.5h)}{v2}
   \fmfforce{.4w,.6h}{v20}
   \fmfforce{.6w,.6h}{v21}
   \fmf{plains}{v21,v20}
   \fmf{arrowr}{v21,v20}
   \fmf{boson}{v1,v2}
  \end{fmfgraph}
\ec}
\quad \mbox{--} \quad
 \parbox{2.5cm}{\bc
  \begin{fmfgraph*}(2.5,2)
   \fmfforce{(0,.5h)}{v1}
   \fmfforce{(w,.5h)}{v2}
   \fmfforce{(.85w,.5h)}{v3}
   \fmf{boson}{v1,v3}
   \fmfv{label=\nx\G,l.a=35}{v1}
   \fmf{projc}{v2,v1}
  \end{fmfgraph*}
 \ec}\; .
\label{maindia}
\ee

The first term on the r.h.s.~of (\ref{main}) is momentum independent;
sandwiched between two vertices attached to either end of the line it leads
to a contraction of the vertices (cf.~(\ref{rel1})). It is graphically 
represented by the so called ``contraction arrow''. The $G$ in the second 
term on the r.h.s.~is nothing but the product of a ghost-gluon vertex and a 
ghost propagator 

\be
  G_{\mu}^{abd} (k) = -i\mbox{\em{g}}f^{abd}
                      \left[\left(a\cdot k - 1\right)k_\mu - k^2 a_\mu \right]
                      \frac{1}{k^2+i\epsilon}
\label{ghost}
\ee
and is therefore called the ghost contribution. For any covariant gauge $G(k)$
reduces to 

\be
  G_{F,\mu}^{abd}(k)=i\mbox{\em{g}}f^{abd}\frac{k_{\mu}}{k^{2}+i\varepsilon}.
\ee
Putting (\ref{projec}) and (\ref{main}), respectively (\ref{projecdia}) and
(\ref{maindia}), together, we have 

\be
 \parbox{2cm}{\bc
  \begin{fmfgraph}(2,2)
   \fmfoutgoing{o1}
   \fmfforce{(0,h)}{v2}
   \fmfforce{(0,0)}{v3}
   \fmfforce{(.4w,.5h)}{v1}
   \fmf{boson}{v2,v1}
   \fmfset{arrow_ang}{25}
   \fmf{projc}{v1,v2}
   \fmfset{arrow_ang}{15}
   \fmf{boson}{v3,v1,o1}
   \fmfdot{v1}
  \end{fmfgraph}
 \ec}
\quad = \quad -
 \parbox{2cm}{\bc
  \begin{fmfgraph}(2,2)
   \fmfoutgoing{o1}
   \fmfforce{(0,h)}{v2}
   \fmfforce{(0,0)}{v3}
   \fmfforce{(.4w,.5h)}{v1}
   \fmf{boson}{v2,v1}
   \fmf{boson}{v3,v1}
   \fmfforce{.6w,.4h}{v20}
   \fmfforce{.8w,.4h}{v21}
   \fmf{plains}{v21,v20}
   \fmf{arrowr}{v21,v20}
   \fmf{boson}{v1,o1}
  \end{fmfgraph}
 \ec}
\quad -
 \parbox{2cm}{\bc
  \begin{fmfgraph*}(2,2)
   \fmfoutgoing{o1}
   \fmfforce{(0,h)}{v2}
   \fmfforce{(0,0)}{v3}
   \fmfforce{(.4w,.5h)}{v1}
   \fmfforce{(.9w,.5h)}{v4}
   \fmfv{label=\nx\G, l.a=35, l.d=0.05w}{v1}
   \fmf{boson}{v2,v1}
   \fmf{boson}{v3,v1}
   \fmf{boson}{v1,v4}
   \fmfset{arrow_ang}{30}
   \fmf{projc}{o1,v1}
   \fmfset{arrow_ang}{15}
  \end{fmfgraph*}
 \ec}
\quad +
 \parbox{2cm}{\bc
  \begin{fmfgraph}(2,2)
   \fmfoutgoing{o1}
   \fmfforce{(0,h)}{v2}
   \fmfforce{(0,0)}{v3}
   \fmfforce{(.4w,.5h)}{v1}
   \fmf{boson}{v2,v1}
   \fmf{boson}{v1,o1}
   \fmfforce{.322w,.253h}{v20}
   \fmfforce{.198w,.097h}{v21}
   \fmf{plains}{v21,v20}
   \fmf{arrowr}{v21,v20}
   \fmf{boson}{v1,v3}
  \end{fmfgraph}
 \ec}
\quad +
 \parbox{2cm}{\bc
  \begin{fmfgraph*}(2,2)
   \fmfoutgoing{o1}
   \fmfforce{(0,h)}{v2}
   \fmfforce{(0,0)}{v3}
   \fmfforce{(.4w,.5h)}{v1}
   \fmfforce{(.08w,.1h)}{v4}
   \fmfv{label=\nx\G, l.a=-85, l.d=0.1w}{v1}
   \fmf{boson}{v2,v1}
   \fmf{boson}{v4,v1}
   \fmf{boson}{v1,o1}
   \fmfset{arrow_ang}{30}
   \fmf{projc}{v3,v1}
   \fmfset{arrow_ang}{15}
  \end{fmfgraph*}
 \ec}\quad .
\label{maindiagram}
\ee

There are three more diagrammatic rules needed for the decomposition of Feynman
diagrams into explicitly gauge dependent and gauge independent parts. The 
first specifies how ${\cal P}$ acts on $\Delta$:

\be
  \mbox{\em{g}}f^{abc}{\cal P}_\rho^{\mu}(q)\Delta_\mu^{cd}(q) = 
     G_{\rho}^{abd}(q) - G_{F,\rho}^{abd}(q).
\label{ghocov}
\ee

Note that for all covariant gauges $G_F = G$ and thus the r.h.s.~is zero. 
Therefore we get for covariant gauges using (\ref{projec}), (\ref{main}) and 
(\ref{ghocov})

\be
 \parbox{2cm}{\bc
  \begin{fmfgraph*}(2,2)
   \fmfoutgoing{o1}
   \fmfforce{(0,h)}{v2}
   \fmfforce{(0,0)}{v3}
   \fmfforce{(.4w,.5h)}{v1}
   \fmf{boson}{v2,v1}
   \fmfv{label=$\nx\Delta$,label.a=-100}{v1}
   \fmfset{arrow_ang}{25}
   \fmf{projc}{v1,v2}
   \fmfset{arrow_ang}{15}
   \fmf{boson}{v3,v1,o1}
   \fmfdot{v1}
  \end{fmfgraph*}
 \ec}
\quad = \quad - \quad
 \parbox{2cm}{\bc
  \begin{fmfgraph*}(2,2)
   \fmfoutgoing{o1}
   \fmfforce{(0,h)}{v2}
   \fmfforce{(0,0)}{v3}
   \fmfforce{(.4w,.5h)}{v1}
   \fmfforce{.6w,.4h}{v20}
   \fmfforce{.8w,.4h}{v21}
   \fmf{plains}{v21,v20}
   \fmf{arrowr}{v21,v20}
   \fmf{boson}{v2,v1}
   \fmf{boson}{v3,v1}
   \fmfv{label=$\nx\Delta$,label.a=-100}{v1}
   \fmf{boson}{v1,o1}
  \end{fmfgraph*}
 \ec}
\quad - \quad
 \parbox{2cm}{\bc
  \begin{fmfgraph*}(2,2)
   \fmfoutgoing{o1}
   \fmfforce{(0,h)}{v2}
   \fmfforce{(0,0)}{v3}
   \fmfforce{(.4w,.5h)}{v1}
   \fmfforce{(.4w,.5h)}{v4}
   \fmfforce{(.9w,.5h)}{v5}
   \fmfv{label=\nx\G, l.a=35, l.d=0.05w}{v1}
   \fmf{boson}{v2,v1}
   \fmf{boson}{v3,v1}
   \fmf{boson}{v1,v5}
   \fmfv{label=$\nx\Delta$,label.a=-100}{v4}
   \fmfset{arrow_ang}{30}
   \fmf{projc}{o1,v1}
   \fmfset{arrow_ang}{15}
  \end{fmfgraph*}
 \ec}\quad .
\label{ghocovz2}
\ee

We will work in general covariant gauge in the following applications, so
rule (\ref{ghocovz2}) will be often used.\\
The second rule concerns the case that an arrow-head $k_{\mu}$ acts on a 
ghost-contribution. It reads

\be
  k^\nu G_{\nu}^{abc} (q) 
         &=& - q^{\nu}G_{\nu}^{abc} (q) - p^\nu G_{\nu}^{abc} (q) \nonumber \\
         &=& - i\mbox{\em{g}}f^{abc} - p^\nu G_{\nu}^{abc} (q).
  \label{energimp}
\ee
Graphically we have

\be
 \parbox{2cm}{
  \begin{fmfgraph*}(2,2)
   \fmfoutgoing{o1}
   \fmfforce{(0,h)}{v2}
   \fmfforce{(0,0)}{v3}
   \fmfforce{(.4w,.5h)}{v1}
   \fmfv{l=\nx\G,l.a=-85,l.d=0.1w}{v1}
   \fmf{boson,label=k,label.d=0.04w}{v2,v1}
   \fmfset{arrow_ang}{25}
   \fmf{projc}{v1,v2}
   \fmfset{arrow_ang}{15}
   \fmf{boson,label=p,label.d=0.04w}{v1,o1}
   \fmf{boson,label=q,label.d=0.04w}{v3,v1}
  \end{fmfgraph*}
 } 
\quad = \quad - \quad
 \parbox{2cm}{
  \begin{fmfgraph*}(2,2)
   \fmfoutgoing{o1}
   \fmfforce{(0,h)}{v2}
   \fmfforce{(0,0)}{v3}
   \fmfforce{(.4w,.5h)}{v1}
   \fmfv{l={\nx \bf 1},l.a=-85,l.d=0.1w}{v1}
   \fmf{boson,label=k,label.d=0.04w}{v2,v1}
   \fmf{boson,label=q,label.d=0.04w}{v3,v1}
   \fmf{boson,label=p,label.d=0.04w}{v1,o1}
  \end{fmfgraph*}
 } 
\quad - \quad
 \parbox{2cm}{
  \begin{fmfgraph*}(2,2)
   \fmfoutgoing{o1}
   \fmfforce{(0,h)}{v2}
   \fmfforce{(0,0)}{v3}
   \fmfforce{(.4w,.5h)}{v1}
   \fmfv{l=\nx\G,l.a=-85,l.d=0.1w}{v1}
   \fmf{boson,label=p,label.d=0.04w}{v1,o1}
   \fmfset{arrow_ang}{25}
   \fmf{projc}{v1,o1}
   \fmfset{arrow_ang}{15}
   \fmf{boson,label=q,label.d=0.04w}{v3,v1}
   \fmf{boson,label=k,label.d=0.04w}{v2,v1}
  \end{fmfgraph*}
 }\; ,
\label{momentumcon}
\ee
where we denote $i\mbox{\em{g}}f^{abc}$ by {\bf 1}. This relation is a direct 
consequence of energy-momentum conservation at the vertex.\\
The last rule concerns the case that $k_\mu$ is connected to a fermion-gluon 
vertex $F$. We have

\be
 S^{mi}(q)k^\mu F_{\mu,ij}^a S^{jn}(p) 
    & = & 
 i\mbox{\em{g}}\delta^{im} \delta^{jn} \frac{1}{(\slashed{q}-m_f+i\varepsilon)}
 k^{\mu}\gamma_\mu t^a_{ij}\frac{1}{(\slashed{p}-m_f+i\varepsilon)}\nonumber\\
    & = & 
 i\mbox{\em{g}} (t^a)^{mn} \left( \frac{1}{\slashed{q}-m_f+i\varepsilon} -
                                 \frac{1}{\slashed{p}-m_f+i\varepsilon}\right),
\label{fermionrule}  
\ee

which will be represented graphically as

\be
\parbox{2cm}{
 \begin{fmfgraph*}(2,2)
   \fmfoutgoing{o1}
   \fmfforce{(.5w,h)}{v2}
   \fmfforce{(0,.5h)}{v3}
   \fmfforce{(.5w,.5h)}{v1}
   \fmf{boson,label=k,label.d=0.04w}{v2,v1}
   \fmfset{arrow_ang}{25}
   \fmf{projc}{v1,v2}
   \fmfset{arrow_ang}{15}
   \fmf{fermion,label=p,label.d=0.06w}{v1,o1}
   \fmf{fermion,label=q,label.d=0.06w}{v3,v1}
  \end{fmfgraph*}
 }
\quad = \quad
\parbox{1cm}{
 \begin{fmfgraph*}(1,2)
   \fmfforce{(0,h)}{v2}
   \fmfforce{(0,.5h)}{v3}
   \fmfforce{(w,.5h)}{v1}
   \fmf{boson,label=k,label.d=0.04h}{v2,v1}
   \fmf{fermion,label=q,label.d=0.06h}{v3,v1}
  \end{fmfgraph*}
}
\quad \mbox{\LARGE -} \quad
\parbox{1cm}{
 \begin{fmfgraph*}(1,2)
   \fmfoutgoing{o1}
   \fmfforce{(w,h)}{v2}
   \fmfforce{(0,.5h)}{v1}
   \fmf{boson,label=k,label.d=0.04h}{v1,v2}
   \fmf{fermion,label=p,label.d=0.06h}{v1,o1}
  \end{fmfgraph*}
}\quad . 
\label{femrul}
\ee

In the last step of eq.~(\ref{fermionrule}), we have used $k+q=p$.

\subsection{The Basic Identities}
\label{sec2.1.2}

In this subsection we give rules which allow to relate gauge dependent parts 
of diagrams with different topology with each other. This will be used to show
that the gauge dependent parts of different diagrams cancel each other. The 
first rule relates a diagram with a 4-gluon-vertex Q to those with contraction
arrows. It is obtained by applying (\ref{maindiagram}) to diagrams with a 
second three-gluon vertex $\Gamma$:

\bc
\parbox{2.5cm}{\bc
 \begin{fmfgraph*}(2.5,1.2)
  \fmfforce{(0,h)}{i1}
  \fmfforce{(w,h)}{o1}
  \fmfforce{(.3w,0)}{v1}
  \fmfforce{(.7w,0)}{v2}
  \fmfforce{(0,0)}{v3}
  \fmfforce{(w,0)}{v4}
  \fmfforce{.4w,.1h}{v20}
  \fmfforce{.6w,.1h}{v21}
  \fmf{plains}{v21,v20}
  \fmf{arrowr}{v21,v20}
  \fmfv{l=$\nx\mu$a}{i1}
  \fmfv{l=$\nx\nu$b}{v3}
  \fmfv{l=$\nx\gamma$c}{v4}
  \fmfv{l=$\nx\delta$d}{o1}
  \fmfv{label=$\nx\phi$e,label.a=-150}{v2}
  \fmf{boson,label=$k$,l.d=.04w,l.s=right}{v3,v1}
  \fmf{boson,label=$-k'$,l.d=.04w,l.s=right}{v2,v4}
  \fmf{boson,label=$p$,l.d=.04w,l.s=left}{i1,v1}
  \fmf{boson}{v2,o1}
  \fmf{boson}{v1,v2}
  \fmfdot{v2}
 \end{fmfgraph*}
\ec}\qquad
{\LARGE -} \quad
\parbox{2.5cm}{\bc
 \begin{fmfgraph*}(2.5,1.2)
  \fmfforce{(0,h)}{i1}
  \fmfforce{(w,h)}{o1}
  \fmfforce{(.3w,0)}{v1}
  \fmfforce{(.7w,0)}{v2}
  \fmfforce{(0,0)}{v3}
  \fmfforce{(w,0)}{v4}
  \fmfforce{.41w,.1h}{v20}
  \fmfforce{.59w,.1h}{v21}
  \fmf{plains}{v20,v21}
  \fmf{arrowr}{v20,v21}
  \fmfv{l=$\nx\mu$a}{i1}
  \fmfv{l=$\nx\nu$b}{v3}
  \fmfv{l=$\nx\gamma$c}{v4}
  \fmfv{l=$\nx\delta$d}{o1}
  \fmfv{label=$\nx\phi$e,label.a=-40,label.d=0.04w}{v1}
  \fmf{boson,label=$k$,l.d=.04w,l.s=right}{v3,v1}
  \fmf{boson,label=$-k'$,l.d=.04w,l.s=right}{v2,v4}
  \fmf{boson}{o1,v1}
  \fmf{boson,label=$p$,l.d=.04w,l.s=right}{v2,i1}
  \fmf{boson}{v1,v2}
  \fmfdot{v1}
 \end{fmfgraph*}
\ec}\qquad
{\LARGE -} \quad
\parbox{2.5cm}{\bc
 \begin{fmfgraph*}(2.5,1.2)
  \fmfforce{(0,h)}{i1}
  \fmfforce{(w,h)}{o1}
  \fmfforce{(.5w,0)}{v1}
  \fmfforce{(.5w,.7h)}{v2}
  \fmfforce{(0,0)}{v3}
  \fmfforce{(w,0)}{v4}
  \fmfforce{.4w,.15h}{v20}
  \fmfforce{.4w,.55h}{v21}
  \fmf{plains}{v20,v21}
  \fmf{arrowl}{v20,v21}
  \fmfv{l=$\nx\mu$a}{i1}
  \fmfv{l=$\nx\nu$b}{v3}
  \fmfv{l=$\nx\gamma$c}{v4}
  \fmfv{l=$\nx\delta$d}{o1}  
  \fmfv{label=$\nx\phi$e,label.a=60}{v1}
  \fmf{boson,label=$-k'$,l.d=.04w,l.s=right}{v1,v4}
  \fmf{boson,label=$k$,l.d=.04w,l.s=right}{v3,v1}
  \fmf{boson,label=$p$,l.d=.04w,l.s=right}{i1,v2}
  \fmf{boson}{v2,o1}
  \fmf{boson}{v1,v2}
  \fmfdot{v1}
 \end{fmfgraph*}
\ec}\qquad
{\Large -} \quad
\parbox{2.5cm}{\bc
 \begin{fmfgraph*}(2,1.2)
  \fmfforce{(0,0)}{v1}
  \fmfforce{(.5w,0)}{v2}
  \fmfforce{(w,0)}{v3}
  \fmf{boson,label=$k$,l.d=.04w,l.s=right}{v1,v2}
  \fmf{boson,label=$-k'$,l.d=.04w,l.s=right}{v2,v3}
  \fmfforce{(0,h)}{i1}
  \fmfforce{(w,h)}{o1}
  \fmfv{l=$\nx\mu$a}{i1}
  \fmfv{l=$\nx\nu$b}{v1}
  \fmfv{l=$\nx\gamma$c}{v3}
  \fmfv{l=$\nx\delta$d}{o1}
  \fmf{boson,label=$p$,l.d=.04w,l.s=right}{i1,v2}
  \fmf{boson}{v2,o1}
  \fmf{projc}{v2,i1}
  \fmfdot{v2}
 \end{fmfgraph*}
 \ec}
= 0.
\ec

\be
i\mbox{\em{g}}\,g^{\nu\phi}f^{abe}\Gamma_{\delta\phi\gamma}^{dec}(-p-k+k', p+k, -k')
\mbox{\hspace{10cm}}\nonumber\\
 -i\mbox{\em{g}}\,g^{\gamma\phi}f^{cae}\Gamma_{\delta\nu\phi}^{dbe}(-p-k+k', k, p-k')
\mbox{\hspace{6.5cm}}\nonumber\\
 -i\mbox{\em{g}}\,g^{\delta\phi}f^{dae}\Gamma_{\phi\nu\gamma}^{ebc}(k'-k, k, -k')
\mbox{\hspace{2.5cm}}\nonumber\\
 - p^{\mu} Q_{\mu\nu\gamma\delta}^{abcd}=0\; .
\label{rel1}
\ee
The second identity addresses the case that the contraction arrow points at a 
four-gluon vertex:

\bc
 \parbox{2.5cm}{\bc
  \begin{fmfgraph*}(2.5,2.5)
   \fmfforce{(0,0)}{v1}
   \fmfforce{(0,h)}{v2}
   \fmfforce{(w,h)}{v3}
   \fmfforce{(w,0)}{v4}
   \fmfforce{(.3w,.7h)}{v5}
   \fmfforce{(.3w,h)}{v6}
   \fmfforce{(.5w,.5h)}{v7}
   \fmfforce{.4w,.7h}{v20}
   \fmfforce{.5w,.6h}{v21}
   \fmf{plains}{v21,v20}
   \fmf{arrowr}{v21,v20}
   \fmfv{l=$\nx\mu$a}{v2}
   \fmfv{l=$\nx\nu$b}{v1}
   \fmfv{l=$\nx\gamma$c}{v4}
   \fmfv{l=$\nx\delta$d}{v3}
   \fmfv{l=g}{v6}
   \fmfv{l=$\nx\footnotesize\nx\phi${\nx\small h},l.a=160,l.d=0.1w}{v7}
   \fmf{boson}{v1,v7,v4}
   \fmf{boson}{v2,v5,v7,v3}
   \fmf{boson}{v6,v5}
   \fmf{boson}{v5,v7}
   \fmfdot{v7}
  \end{fmfgraph*}
 \ec}
\quad {\Large -} \quad
\parbox{2.5cm}{\bc
  \begin{fmfgraph*}(2.5,2.5)
   \fmfforce{(0,0)}{v1}
   \fmfforce{(0,h)}{v2}
   \fmfforce{(w,h)}{v3}
   \fmfforce{(w,0)}{v4}
   \fmfforce{(.7w,.7h)}{v5}
   \fmfforce{(w,.7h)}{v6}
   \fmfforce{(.5w,.5h)}{v7}
   \fmfforce{.6w,.7h}{v20}
   \fmfforce{.5w,.6h}{v21}
   \fmf{plains}{v21,v20}
   \fmf{arrowr}{v21,v20}
   \fmfv{l=$\nx\mu$a}{v2}
   \fmfv{l=$\nx\nu$b}{v1}
   \fmfv{l=$\nx\gamma$c}{v4}
   \fmfv{l=$\nx\delta$d}{v3}
   \fmfv{l=g}{v6}
   \fmfv{l=$\nx\footnotesize\nx\phi${\nx\small h},label.a=20,l.d=0.1w}{v7}
   \fmf{boson}{v1,v7,v4}
   \fmf{boson}{v2,v7,v5,v3}
   \fmf{boson}{v6,v5}
   \fmf{boson}{v5,v7}
   \fmfdot{v7}
  \end{fmfgraph*}
 \ec}
\quad {\large -} \quad
\parbox{2.5cm}{\bc
  \begin{fmfgraph*}(2.5,2.5)
   \fmfforce{(0,0)}{v1}
   \fmfforce{(0,h)}{v2}
   \fmfforce{(w,h)}{v3}
   \fmfforce{(w,0)}{v4}
   \fmfforce{(.7w,.3h)}{v5}
   \fmfforce{(w,.3h)}{v6}
   \fmfforce{(.5w,.5h)}{v7}
   \fmfforce{.6w,.3h}{v20}
   \fmfforce{.5w,.4h}{v21}
   \fmf{plains}{v21,v20}
   \fmf{arrowr}{v21,v20}
   \fmfv{l=$\nx\mu$a}{v2}
   \fmfv{l=$\nx\nu$b}{v1}
   \fmfv{l=$\nx\gamma$c}{v4}
   \fmfv{l=$\nx\delta$d}{v3}
   \fmfv{l=g}{v6}
   \fmfv{l=$\nx\footnotesize\nx\phi${\nx\small h},label.a=-20,l.d=0.1w}{v7}
   \fmf{boson}{v1,v7,v5,v4}
   \fmf{boson}{v2,v7,v3}
   \fmf{boson}{v6,v5}
   \fmf{boson}{v5,v7}
   \fmfdot{v7}
  \end{fmfgraph*}
 \ec}
\quad + \quad
\parbox{2.5cm}{\bc
  \begin{fmfgraph*}(2.5,2.5)
   \fmfforce{(0,0)}{v1}
   \fmfforce{(0,h)}{v2}
   \fmfforce{(w,h)}{v3}
   \fmfforce{(w,0)}{v4}
   \fmfforce{(.3w,.3h)}{v5}
   \fmfforce{(0,.3h)}{v6}
   \fmfforce{(.5w,.5h)}{v7}
   \fmfforce{.4w,.3h}{v20}
   \fmfforce{.5w,.4h}{v21}
   \fmf{plains}{v21,v20}
   \fmf{arrowr}{v21,v20}
   \fmfv{l=$\nx\mu$a}{v2}
   \fmfv{l=$\nx\nu$b}{v1}
   \fmfv{l=$\nx\gamma$c}{v4}
   \fmfv{l=$\nx\delta$d}{v3}
   \fmfv{l=g}{v6}
   \fmfv{l=$\nx\footnotesize\nx\phi${\nx\small h},l.a=-160,l.d=0.1w}{v7}
   \fmf{boson}{v1,v5,v7,v4}
   \fmf{boson}{v2,v7,v3}
   \fmf{boson}{v6,v5}
   \fmf{boson}{v5,v7}
   \fmfdot{v7}
  \end{fmfgraph*}
 \ec} = 0.
\ec

\be
\qquad 
i\mbox{\em g}f^{ahg}g^{\mu\phi}Q^{hdcb}_{\phi\delta\gamma\nu} \quad - \qquad
i\mbox{\em g}f^{gdh}g^{\delta\phi}Q^{ahcb}_{\mu\phi\gamma\nu} \quad - \qquad
i\mbox{\em g}f^{ghc}g^{\gamma\phi}Q^{adhb}_{\mu\delta\phi\nu} \quad + \qquad
i\mbox{\em g}f^{bhg}g^{\nu\phi}Q^{adch}_{\mu\delta\gamma\phi}       = 0.
\label{rel2}
\ee

The third identity (which we have not found in the literature, but which will 
play an important role in Sec.~\ref{sec5}) concerns the case that the 
contraction arrow points at a vertex which carries a $G$. This identity 
follows from the Bianchi identity for the color structure coefficients 
$f^{abc}$:\\

\bc
\parbox{2.5cm}{
 \begin{fmfgraph*}(2.5,1.2)
  \fmfforce{(0,h)}{i1}
  \fmfforce{(w,h)}{o1}
  \fmfforce{(.3w,0)}{v1}
  \fmfforce{(.3w,0)}{v5}
  \fmfforce{(.7w,0)}{v2}
  \fmfforce{(0,0)}{v3}
  \fmfforce{(w,0)}{v4}
  \fmfv{l=a}{o1}
  \fmfv{l=$\nx\nu$b}{i1}
  \fmfv{l=c}{v3}
  \fmfv{l=d}{v4}
  \fmfv{label=$\nx\phi$e,label.a=30}{v5}
  \fmf{boson}{i1,v1,v2,v4}
  \fmf{boson,label=$k\quad$,label.d=0.04w,label.s=right}{v3,v1}
  \fmf{boson}{v2,o1}
  \fmfv{l=\nx\G,l.a=-140,label.d=0.04w}{v1}
  \fmfv{l={\nx\bf 1},l.a=-140}{v2}
 \end{fmfgraph*}
}\qquad
- \qquad
\parbox{2.5cm}{
 \begin{fmfgraph*}(2.5,1.2)
  \fmfforce{(0,h)}{i1}
  \fmfforce{(w,h)}{o1}
  \fmfforce{(.3w,0)}{v1}
  \fmfforce{(.3w,0)}{v5}
  \fmfforce{(.7w,0)}{v2}
  \fmfforce{(0,0)}{v3}
  \fmfforce{(w,0)}{v4}
  \fmfforce{.41w,.1h}{v20}
  \fmfforce{.59w,.1h}{v21}
  \fmf{plains}{v20,v21}
  \fmf{arrowr}{v20,v21}
  \fmfv{l=a}{o1}
  \fmfv{l=$\nx\nu$b}{i1}
  \fmfv{l=c}{v3}
  \fmfv{l=d}{v4}
  \fmfv{label=$\nx\phi$e,label.a=-40,label.d=0.04w}{v5}
  \fmf{boson,label=$k\quad$,label.d=0.04w,label.s=right}{v3,v1}
  \fmf{boson}{v2,v4}
  \fmf{boson}{o1,v1}
  \fmf{boson}{v2,i1}
  \fmf{boson}{v1,v2}
  \fmfv{l=\nx\G,l.a=-140,label.d=0.04w}{v1}
 \end{fmfgraph*}
}\qquad
- \qquad
\parbox{2.5cm}{
 \begin{fmfgraph*}(2.5,1.2)
  \fmfforce{(0,h)}{i1}
  \fmfforce{(w,h)}{o1}
  \fmfforce{(.5w,0)}{v1}
  \fmfforce{(.5w,0)}{v5}
  \fmfforce{(.5w,.7h)}{v2}
  \fmfforce{(0,0)}{v3}
  \fmfforce{(w,0)}{v4}
  \fmfforce{.4w,.15h}{v20}
  \fmfforce{.4w,.55h}{v21}
  \fmf{plains}{v20,v21}
  \fmf{arrowl}{v20,v21}
  \fmfv{l=a}{o1}
  \fmfv{l=$\nx\nu$b}{i1}
  \fmfv{l=c}{v3}
  \fmfv{l=d}{v4}
  \fmfv{label=$\nx\phi$e,label.a=60}{v5}
  \fmf{boson}{v4,v1}
  \fmf{boson,label=$k$,label.d=0.04w,label.s=right}{v3,v1}
  \fmf{boson}{i1,v2}
  \fmf{boson}{v2,o1}
  \fmf{boson}{v1,v2}
  \fmfv{l=\nx\G,l.a=-140,label.d=0.04w}{v1}
 \end{fmfgraph*}
} \qquad = 0. 
\ec
\vspace{1em}

\be
  \qquad \qquad
  i\mbox{\em g}f^{aed}G_{\nu}^{bce} (k) \quad - \quad 
  i\mbox{\em g}f^{bed}G^{\phi ,ace} (k)g_{\nu\phi} \quad - \quad
  i\mbox{\em g}f^{abe}G^{\phi ,ecd} (k)g_{\nu\phi} \qquad =  0.
\label{rel3}
\ee
The fourth identity treats the case that the contraction arrow points at a 
fermion-gluon vertex:

\bc
\parbox{2.5cm}{\bc
 \begin{fmfgraph*}(2.5,1.2)
  \fmfforce{(0,h)}{i1}
  \fmfforce{(w,h)}{o1}
  \fmfforce{(.3w,0)}{v1}
  \fmfforce{(.7w,0)}{v2}
  \fmfforce{(0,0)}{v3}
  \fmfforce{(w,0)}{v4}
  \fmfforce{.4w,.1h}{v20}
  \fmfforce{.6w,.1h}{v21}
  \fmf{plains}{v21,v20}
  \fmf{arrowr}{v21,v20}
  \fmfv{l=a}{i1}
  \fmfv{l=$\nx\nu$b}{v3}
  \fmfv{l=j}{v4}
  \fmfv{l=i}{o1}
  \fmfv{label=$\nx\phi$e,label.a=-150}{v2}
  \fmf{boson,label=$k$,l.d=.04w,l.s=right}{v3,v1}
  \fmf{fermion,label=$k'$,l.d=.04w,l.s=right}{v2,v4}
  \fmf{boson,label=$p$,l.d=.04w,l.s=right}{i1,v1}
  \fmf{fermion}{o1,v2}
  \fmf{boson}{v1,v2}
  \fmfdot{v2}
 \end{fmfgraph*}
\ec}\qquad
= \qquad
\parbox{2.5cm}{\bc
 \begin{fmfgraph*}(2,1.2)
  \fmfforce{(0,0)}{v1}
  \fmfforce{(.5w,0)}{v2}
  \fmfforce{(w,0)}{v3}
  \fmf{boson,label=$k$,l.d=.04w,l.s=right}{v1,v2}
  \fmf{fermion,label=$k'$,l.d=.04w,l.s=right}{v2,v3}
  \fmfforce{(0,h)}{i1}
  \fmfforce{(w,h)}{o1}
  \fmfv{l=a}{i1}
  \fmfv{l=$\nx\nu$b}{v1}
  \fmfv{l=j}{v3}
  \fmfv{l=i}{o1}
  \fmf{boson,label=$p$,l.d=.04w,l.s=right}{i1,v2}
  \fmf{fermion}{o1,v2}
  \fmfdot{v2}
 \end{fmfgraph*}
 \ec}
\qquad
{\LARGE -}
\qquad
\parbox{2.5cm}{\bc
 \begin{fmfgraph*}(2,1.2)
  \fmfforce{(0,0)}{v1}
  \fmfforce{(.5w,0)}{v2}
  \fmfforce{(w,0)}{v3}
  \fmf{boson,left=.8,label=$k$,l.d=.04w,l.s=left}{v1,v2}
  \fmf{fermion,label=$k'$,l.d=.04w,l.s=right}{v2,v3}
  \fmfforce{(0,h)}{i1}
  \fmfforce{(0,h)}{i2}
  \fmfforce{(w,h)}{o1}
  \fmfv{l=a}{i1}
  \fmfv{l=$\nx\nu$b}{v1}
  \fmfv{l=j}{v3}
  \fmfv{l=i}{o1}
  \fmf{boson,right=.5}{i1,v2}
  \fmfv{label=$p$,label.a=-60}{i2}
  \fmf{fermion}{o1,v2}
  \fmfdot{v2}
 \end{fmfgraph*}
\ec}
\ec

\be
-i\mbox{\em{g}}f^{abe}g^{\nu\phi}F^e_{\phi, ij}(p+k) & = & \qquad  \quad
\Upsilon^{ab}_{\nu, ij} \qquad \qquad - \qquad \qquad \Upsilon^{ba}_{\nu, ij}.
\label{rel4}
\ee

Note that the gluon lines at the four-point fermion-gluon vertex created in
(\ref{femrul}) cannot be interchanged due to the color-structure 
(cf.~Appendix \ref{appc}).

\subsection{A First Example}
\label{sec2.1.3}

To illustrate the method of Cheng and Tsai, we discuss briefly the gauge 
invariance of the {$\cal O$}($\mbox{\em g}^{2}$), contributions to the QCD 
partition function. For this purpose we will from now on concentrate on 
general covariant $\alpha$ -gauge. That means that we will use the special 
structure $G_{F}=G$, which the ghost contribution satisfies (cf.~\ref{ghocov}).
This leads to the use of eq.~(\ref{ghocovz2}) in the following. The partition 
function consists of four diagrams \footnote{We indexed each diagram with a 
letter in order to refer to it later in the text (e.g.~diagram 
(\ref{grund}.\refmy{C1}) is the third diagram in (\ref{grund})). Moreover we 
give the factor (-1) for the fermion and ghost loops in the following 
explicitly in the symmetry factor.}:

\be
+\;\frac{1}{8}
\raisebox{-1.5ex}{
 \parbox{1.5cm}{\bc
 \begin{fmfgraph}(1.5,2)
  \fmfforce{(.5w,.5h)}{v1}
  \fmfforce{(.5w,h)}{v2}
  \fmfforce{(.5w,0)}{v3}
  \fmf{boson,left}{v1,v2,v1}
  \fmf{boson,left}{v1,v3,v1}
  \fmfdot{v1}
 \end{fmfgraph}\\
\refstepcounter{diagh}
(\Alph{diagh})\label{A1}
\ec}}
+\;\frac{1}{12}
\raisebox{-1.5ex}{
\parbox{2.4cm}{\bc
 \begin{fmfgraph}(2,2)
  \fmfforce{(0,.5h)}{v1}
  \fmfforce{(w,.5h)}{v2}
  \fmf{boson,left}{v1,v2,v1}
  \fmf{boson}{v1,v2}
  \fmfdot{v1,v2}
 \end{fmfgraph}\\
\refstepcounter{diagh}
(\Alph{diagh})\label{B1}
\ec}}
-\;\frac{1}{2}
\raisebox{-1.5ex}{
\parbox{2.4cm}{\bc
 \begin{fmfgraph}(2,2)
  \fmfforce{(0,.5h)}{v1}
  \fmfforce{(w,.5h)}{v2}
  \fmf{ghost,left}{v1,v2,v1}
  \fmf{boson}{v1,v2}
  \fmfdot{v1,v2}
 \end{fmfgraph}\\
\refstepcounter{diagh}
{(\Alph{diagh})\label{C1}}
\ec}}
-\;\frac{1}{2}
\raisebox{-1.5ex}{
\parbox{2.4cm}{\bc
 \begin{fmfgraph}(2,2)
  \fmfforce{(0,.5h)}{v1}
  \fmfforce{(w,.5h)}{v2}
  \fmf{fermion,left}{v1,v2,v1}
  \fmf{boson}{v1,v2}
  \fmfdot{v1,v2}
 \end{fmfgraph}\\
\refstepcounter{diagh}
{(\Alph{diagh})\label{D1}}
\ec}}\; .
\label{grund}
\ee

The dotted lines are the usual notation for ghost propagators. For our purpose
it is however more useful to express diagrams like (\ref{grund}.\refmy{C1}) in
the following way:\\

\be
\parbox{2cm}{\bc
\begin{fmfgraph*}(2,2)
  \fmfforce{(0,.5h)}{v1}
  \fmfforce{(w,.5h)}{v2}
  \fmf{boson,left}{v1,v2,v1}
  \fmfv{label=\nx\G,l.a=140}{v1}
  \fmfv{label=\nx\G,l.a=-65}{v2}
  \fmf{boson}{v1,v2}
 \end{fmfgraph*}\\
\ec}\quad .
\ee

We follow a three step procedure:\\
We select, one after the other, a diagram in (\ref{grund}) and separate the
$a_{\mu}$-dependent part of one of the internal gluon lines. E.g.~for diagram 
(\ref{grund}.\refmy{B1}) this looks like

\be
\raisebox{-1.5ex}{
\parbox{2.4cm}{\bc
 \begin{fmfgraph*}(2,2)
  \fmfforce{(0,.5h)}{v1}
  \fmfforce{(w,.5h)}{v2}
  \fmf{boson,left}{v1,v2}
  \fmf{boson,label=$D$,l.d=.1w,l.s=right,left}{v2,v1}
  \fmf{boson}{v1,v2}
  \fmfdot{v1,v2}
 \end{fmfgraph*}\\
\stepcounter{rueck}
\refstepcounter{diagh}
(\Alph{diagh})\label{A25a}
\ec}}
= 
\raisebox{-1.5ex}{
\parbox{2.4cm}{\bc
 \begin{fmfgraph*}(2,2)
  \fmfforce{(0,.5h)}{v1}
  \fmfforce{(w,.5h)}{v2}
  \fmf{boson,left}{v1,v2}
  \fmf{boson,label=$D_{F}$,l.d=.1w,l.s=right,left}{v2,v1}
  \fmf{boson}{v1,v2}
  \fmfdot{v1,v2}
 \end{fmfgraph*}\\
\refstepcounter{diagh}
(\Alph{diagh})\label{B25a}
\ec}}
+ \; 
\raisebox{-1.5ex}{
\parbox{2.4cm}{\bc
 \begin{fmfgraph*}(2,2)
  \fmfforce{(0,.5h)}{v1}
  \fmfforce{(w,.5h)}{v2}
  \fmf{boson,left}{v1,v2}
  \fmf{boson,left}{v2,v1}
  \fmf{boson}{v1,v2}
  \fmfv{label=$\nx\Delta$,l.a=-150,l.d=.05w}{v1}
  \fmf{projs,left}{v2,v1}
  \fmfdot{v1,v2}
 \end{fmfgraph*}\\
\refstepcounter{diagh}
(\Alph{diagh})\label{C25a}
\ec}}
- 
\raisebox{-1.5ex}{
\parbox{2.4cm}{\bc
 \begin{fmfgraph*}(2,2)
  \fmfforce{(0,.5h)}{v1}
  \fmfforce{(w,.5h)}{v2}
  \fmf{boson,left}{v1,v2}
  \fmf{boson,left}{v2,v1}
  \fmf{boson}{v1,v2}
  \fmfv{label=$\nx\Delta$,l.a=-30,l.d=.05w}{v2}
  \fmf{projs,right}{v1,v2}
  \fmfdot{v1,v2}
 \end{fmfgraph*}\\
\refstepcounter{diagh}
(\Alph{diagh})\label{D25a}
\ec}}
\quad .
\label{abspalt}
\ee

Because $\Delta$ is proportional to an odd power of the momentum
$k$ diagrams (\ref{abspalt}.\refmy{C25a}) and (\ref{abspalt}.\refmy{D25a}) can
be summed. So we get:

\be
\raisebox{-1.5ex}{
\parbox{2.4cm}{\bc
 \begin{fmfgraph*}(2,2)
  \fmfforce{(0,.5h)}{v1}
  \fmfforce{(w,.5h)}{v2}
  \fmf{boson,left}{v1,v2}
  \fmf{boson,label=$D$,l.d=.1w,l.s=right,left}{v2,v1}
  \fmf{boson}{v1,v2}
  \fmfdot{v1,v2}
 \end{fmfgraph*}\\
\stepcounter{rueck}
\refstepcounter{diagh}
(\Alph{diagh})\label{A25}
\ec}}
= 
\raisebox{-1.5ex}{
\parbox{2.4cm}{\bc
 \begin{fmfgraph*}(2,2)
  \fmfforce{(0,.5h)}{v1}
  \fmfforce{(w,.5h)}{v2}
  \fmf{boson,left}{v1,v2}
  \fmf{boson,label=$D_{F}$,l.d=.1w,l.s=right,left}{v2,v1}
  \fmf{boson}{v1,v2}
  \fmfdot{v1,v2}
 \end{fmfgraph*}\\
\refstepcounter{diagh}
(\Alph{diagh})\label{B25}
\ec}}
+\; 2 \quad
\raisebox{-1.5ex}{
\parbox{2.4cm}{\bc
 \begin{fmfgraph*}(2,2)
  \fmfforce{(0,.5h)}{v1}
  \fmfforce{(w,.5h)}{v2}
  \fmf{boson,left}{v1,v2}
  \fmf{boson,left}{v2,v1}
  \fmf{boson}{v1,v2}
  \fmfv{label=$\nx\Delta$,l.a=-150,l.d=.05w}{v1}
  \fmf{projs,left}{v2,v1}
  \fmfdot{v1,v2}
 \end{fmfgraph*}\\
\refstepcounter{diagh}
(\Alph{diagh})\label{C25}
\ec}}
\; .
\label{abspaltung2}
\ee
Next we use the rules of Sec.~\ref{sec2.1.1} to decompose diagram 
(\ref{abspaltung2}.\refmy{C25}): 

\be
\raisebox{-1.5ex}{
\parbox{2cm}{\bc
 \begin{fmfgraph*}(2,2)
  \fmfforce{(0,.5h)}{v1}
  \fmfforce{(w,.5h)}{v2}
  \fmf{boson,left}{v1,v2,v1}
  \fmf{boson}{v1,v2}
  \fmf{projs,left}{v2,v1}
  \fmfv{label=$\nx\Delta$, l.a=-155, l.d=0.05w}{v1} 
  \fmfdot{v1}
 \end{fmfgraph*}\\
\stepcounter{rueck}
\refstepcounter{diagh}
(\Alph{diagh})
\label{A2}
\ec
}}
\; & \stackrel{(\ref{maindiagram})}{=} & 
+ \;
\raisebox{0em}{
\parbox{2.4cm}{\bc
 \begin{fmfgraph*}(2,2)
  \fmfforce{(0,.5h)}{v1}
  \fmfforce{(w,.5h)}{v2}  
  \fmfforce{.35w,.9h}{v20}
  \fmfforce{.65w,.9h}{v21}
  \fmf{plains,left=0.15}{v20,v21}
  \fmf{arrowl,left=0.15}{v20,v21}
\fmf{boson,left}{v2,v1}
  \fmf{boson}{v1,v2}
  \fmfv{label=$\nx\Delta$, l.a=-155, l.d=0.05w}{v1}
  \fmf{boson,left}{v1,v2}
  \fmfdot{v1}
 \end{fmfgraph*}
\ec
}} 
+ \;
\raisebox{0em}{
\parbox{2.4cm}{\bc
 \begin{fmfgraph*}(2,2)
  \fmfforce{(0,.5h)}{v1}
  \fmfforce{(w,.5h)}{v2}
  \fmf{boson,left}{v2,v1}
  \fmf{boson}{v1,v2}
  \fmf{projs,left}{v1,v2}
  \fmfv{label=\nx\G,l.a=60}{v2}
  \fmfv{label=$\nx\Delta$, l.a=-155, l.d=0.05w}{v1}
  \fmf{boson,left}{v1,v2}
  \fmfdot{v1}
 \end{fmfgraph*}
\ec
}}\;
- \;
\raisebox{0em}{
\parbox{2.4cm}{\bc
 \begin{fmfgraph*}(2,2)
  \fmfforce{(0,.5h)}{v1}
  \fmfforce{(w,.5h)}{v2}
  \fmfforce{.35w,.6h}{v20}
  \fmfforce{.65w,.6h}{v21}
  \fmf{plains}{v20,v21}
  \fmf{arrowr}{v20,v21}
  \fmf{boson,left}{v2,v1,v2}
  \fmf{boson}{v1,v2}  
  \fmfv{label=$\nx\Delta$, l.a=-155, l.d=0.05w}{v1}
  \fmf{boson}{v1,v2}
  \fmfdot{v1}
 \end{fmfgraph*}
\ec
}}
- \;
\raisebox{0em}{
\parbox{2.4cm}{\bc
 \begin{fmfgraph*}(2,2)
  \fmfforce{(0,.5h)}{v1}
  \fmfforce{(w,.5h)}{v2}
  \fmf{boson,left}{v2,v1}
  \fmf{boson}{v1,v2}
  \fmf{projc}{v1,v2}
  \fmfv{label=\nx\G,l.a=155}{v2}
  \fmfv{label=$\nx\Delta$, l.a=-155, l.d=0.05w}{v1}
  \fmf{boson,left}{v1,v2}
  \fmfdot{v1}
 \end{fmfgraph*}
\ec
}}\nonumber\\
& = & + 2\;\left[ \;
\raisebox{0em}{
\parbox{2.4cm}{\bc
 \begin{fmfgraph*}(2,2)
  \fmfforce{(0,.5h)}{v1}
  \fmfforce{(w,.5h)}{v2}
  \fmfforce{.35w,.9h}{v20}
  \fmfforce{.65w,.9h}{v21}
  \fmf{plains,left=0.15}{v20,v21}
  \fmf{arrowr,left=0.15}{v20,v21}
  \fmf{boson,left}{v2,v1}
  \fmf{boson}{v1,v2}
  \fmfv{label=$\nx\Delta$, l.a=-155, l.d=0.05w}{v1}
  \fmf{boson,left}{v1,v2}
  \fmfdot{v1}
 \end{fmfgraph*}
\ec
}}
- \;
\raisebox{0em}{
\parbox{2.4cm}{\bc
 \begin{fmfgraph*}(2,2)
  \fmfforce{(0,.5h)}{v1}
  \fmfforce{(w,.5h)}{v2}
  \fmf{boson,left}{v2,v1}
  \fmf{boson}{v1,v2}
  \fmf{projc}{v1,v2}
  \fmfv{label=\nx\G,l.a=155}{v2}
  \fmfv{label=$\nx\Delta$, l.a=-155, l.d=0.05w}{v1}
  \fmf{boson,left}{v1,v2}
  \fmfdot{v1}
 \end{fmfgraph*}
\ec
}}\;
\right]\nonumber \\
& \stackrel{(\ref{maindiagram})\atop (\ref{ghocovz2})}{=} & 
+\; 2 \left[ \;
\raisebox{0em}{
\parbox{2.4cm}{\bc
 \begin{fmfgraph*}(2,2)
  \fmfforce{(0,.5h)}{v1}
  \fmfforce{(w,.5h)}{v2}
  \fmfforce{.35w,.9h}{v20}
  \fmfforce{.65w,.9h}{v21}
  \fmf{plains,left=0.15}{v20,v21}
  \fmf{arrowr,left=0.15}{v20,v21}
  \fmf{boson,left}{v2,v1}
  \fmf{boson}{v1,v2}
  \fmfv{label=$\nx\Delta$, l.a=-155, l.d=0.05w}{v1}
  \fmf{boson,left}{v1,v2}
  \fmfdot{v1}
 \end{fmfgraph*}
\ec
}}
- \;
\raisebox{0ex}{
\parbox{2.4cm}{\bc
 \begin{fmfgraph*}(2,2)
  \fmfforce{(0,.5h)}{v1}
  \fmfforce{(w,.5h)}{v2}
  \fmfforce{.35w,.9h}{v20}
  \fmfforce{.65w,.9h}{v21}
  \fmf{plains,right=0.15}{v21,v20}
  \fmf{arrowr,right=0.15}{v21,v20}
  \fmf{boson,left}{v2,v1}
  \fmf{boson}{v1,v2}
  \fmfv{label=$\nx\Delta$, l.a=-155, l.d=0.05w}{v1}
  \fmfv{label=\nx\G, l.a=155}{v2}
  \fmf{boson,left}{v1,v2}
 \end{fmfgraph*}
\ec
}}
- \;
\raisebox{0ex}{
\parbox{2.4cm}{\bc
 \begin{fmfgraph*}(2,2)
  \fmfforce{(0,.5h)}{v1}
  \fmfforce{(w,.5h)}{v2}
  \fmfforce{(0,.5h)}{v3}
  \fmf{boson,left}{v1,v2,v1}
  \fmf{boson}{v1,v2}
  \fmfv{label=$\nx\Delta$, l.a=-155, l.d=0.05w}{v1}
  \fmfv{label=\nx\G, l.a=145, l.d=0.1w}{v3}
  \fmfv{label=\nx\G, l.a=155, l.d=0.15w}{v2}
  \fmf{projs,right}{v2,v1}
 \end{fmfgraph*}
\ec
}}\;
\right]\nonumber\\
&  \stackrel{(\ref{momentumcon})}{=} &
+\; 2 \left[\;
\raisebox{-1.5ex}{
\parbox{2.4cm}{\bc
 \begin{fmfgraph*}(2,2)
  \fmfforce{(0,.5h)}{v1}
  \fmfforce{(w,.5h)}{v2}
  \fmfforce{.35w,.9h}{v20}
  \fmfforce{.65w,.9h}{v21}
  \fmf{plains,left=0.15}{v20,v21}
  \fmf{arrowr,left=0.15}{v20,v21}
  \fmf{boson,left}{v2,v1}
  \fmf{boson}{v1,v2}
  \fmfv{label=$\nx\Delta$, l.a=-155, l.d=0.05w}{v1}
  \fmf{boson,left}{v1,v2}
  \fmfdot{v1}
 \end{fmfgraph*}
\refstepcounter{diagh}
(\Alph{diagh})\label{B2}\ec
}}
- \;
\raisebox{-1.5ex}{
\parbox{2.4cm}{\bc
 \begin{fmfgraph*}(2,2)
  \fmfforce{(0,.5h)}{v1}
  \fmfforce{(w,.5h)}{v2}
  \fmfforce{.35w,.9h}{v20}
  \fmfforce{.65w,.9h}{v21}
  \fmf{plains,right=0.15}{v21,v20}
  \fmf{arrowr,right=0.15}{v21,v20}
  \fmf{boson,left}{v2,v1}
  \fmf{boson}{v1,v2}
  \fmfv{label=$\nx\Delta$, l.a=-155, l.d=0.05w}{v1}
  \fmfv{label=\nx\G, l.a=155}{v2}
  \fmf{boson,left}{v1,v2}
 \end{fmfgraph*}
\refstepcounter{diagh}
(\Alph{diagh})\label{C2}
\ec
}}
+ \;
\raisebox{-1.5ex}{
\parbox{2.4cm}{\bc
 \begin{fmfgraph*}(2,2)
  \fmfforce{(0,.5h)}{v1}
  \fmfforce{(w,.5h)}{v2}
  \fmfforce{(0,.5h)}{v3}
  \fmf{boson,left}{v1,v2,v1}
  \fmf{boson}{v1,v2}
  \fmfv{label=$\nx\Delta$, l.a=-155, l.d=0.05w}{v1}
  \fmfv{label=\nx\G, l.a=145, l.d=0.1w}{v3}
  \fmfv{label={\nx\bf 1}, l.a=155}{v2}
 \end{fmfgraph*}
\refstepcounter{diagh}
(\Alph{diagh})\label{D2}
\ec
}}
+ \;
\raisebox{-1.5ex}{
\parbox{2.4cm}{\bc
 \begin{fmfgraph*}(2,2)
  \fmfforce{(0,.5h)}{v1}
  \fmfforce{(w,.5h)}{v2}
  \fmfforce{(0,.5h)}{v3}
  \fmf{boson,left}{v1,v2,v1}
  \fmf{boson}{v1,v2}
  \fmfv{label=$\nx\Delta$, l.a=-155, l.d=0.05w}{v1}
  \fmfv{label=\nx\G, l.a=145, l.d=0.1w}{v3}
  \fmfv{label=\nx\G, l.a=155}{v2}
  \fmf{projs,left}{v2,v1}
 \end{fmfgraph*}
\refstepcounter{diagh}
(\Alph{diagh})\label{E2}
\ec}}
\;\right].
\label{zerlegung}
\ee
As a second step, we invoke the identities of Sec.~\ref{sec2.1.2} as follows. 
Connecting the external lines of eq. (\ref{rel1}), we get

\be
\raisebox{-1.5ex}{
\parbox{2.9cm}{\bc
 \begin{fmfgraph}(2.5,2.5)
  \fmfforce{(0,h)}{i1}
  \fmfforce{(w,h)}{o1}
  \fmfforce{(.3w,.5h)}{v1}
  \fmfforce{(.7w,.5h)}{v2}
  \fmfforce{(0,.5h)}{v3}
  \fmfforce{(w,.5h)}{v4}
  \fmfforce{.4w,.4h}{v20}
  \fmfforce{.6w,.4h}{v21}
  \fmf{plains}{v21,v20}
  \fmf{arrowr}{v21,v20}
  \fmf{boson,right}{v3,v4}
  \fmf{boson}{i1,o1}
  \fmf{boson}{v3,v1}
  \fmf{boson}{v2,v4}
  \fmf{boson}{i1,v1}
  \fmf{boson}{v2,o1}
  \fmf{boson}{v1,v2}
  \fmfdot{v2}
 \end{fmfgraph}\\
\stepcounter{rueck}
\refstepcounter{diagh}
(\Alph{diagh})\label{A26}
\ec}}
- 
\raisebox{-1.5ex}{
\parbox{2.9cm}{\bc
 \begin{fmfgraph}(2.5,2.5)
  \fmfforce{(0,h)}{i1}
  \fmfforce{(w,h)}{o1}
  \fmfforce{(.3w,.5h)}{v1}
  \fmfforce{(.7w,.5h)}{v2}
  \fmfforce{(0,.5h)}{v3}
  \fmfforce{(w,.5h)}{v4}
  \fmfforce{.4w,.4h}{v20}
  \fmfforce{.6w,.4h}{v21}
  \fmf{plains}{v20,v21}
  \fmf{arrowr}{v20,v21}
  \fmf{boson,right}{v3,v4}
  \fmf{boson}{i1,o1}
  \fmf{boson}{v3,v1}
  \fmf{boson}{v2,v4}
  \fmf{boson}{o1,v1}
  \fmf{boson}{v2,i1}
  \fmf{boson}{v1,v2}
  \fmfdot{v1}
 \end{fmfgraph}\\
\refstepcounter{diagh}
(\Alph{diagh})\label{B26}
\ec}}
- 
\raisebox{-1.5ex}{
\parbox{2.9cm}{\bc
 \begin{fmfgraph}(2.5,2.5)
  \fmfforce{(0,h)}{i1}
  \fmfforce{(w,h)}{o1}
  \fmfforce{(.5w,.5h)}{v1}
  \fmfforce{(.5w,.85h)}{v2}
  \fmfforce{(0,.5h)}{v3}
  \fmfforce{(w,.5h)}{v4}
  \fmfforce{.4w,.575h}{v20}
  \fmfforce{.4w,.775h}{v21}
  \fmf{plains}{v20,v21}
  \fmf{arrowr}{v20,v21}
  \fmf{boson,right}{v3,v4}
  \fmf{boson}{i1,o1}
  \fmf{boson}{v1,v4}
  \fmf{boson}{v3,v1}
  \fmf{boson}{i1,v2}
  \fmf{boson}{v2,o1}
  \fmf{boson}{v1,v2}
  \fmfdot{v1}
 \end{fmfgraph}\\
\refstepcounter{diagh}
(\Alph{diagh})\label{C26}
\ec}}
- 
\raisebox{-1.5ex}{
\parbox{2.9cm}{\bc
 \begin{fmfgraph}(2.5,2.5)
  \fmfforce{(0,.5h)}{v1}
  \fmfforce{(.5w,.5h)}{v2}
  \fmfforce{(w,.5h)}{v3}
  \fmf{boson}{v1,v2,v3}
  \fmfforce{(0,h)}{i1}
  \fmfforce{(w,h)}{o1}
  \fmf{boson,right}{v1,v3}
  \fmf{boson}{i1,o1}
  \fmf{boson}{i1,v2}
  \fmf{boson}{v2,o1}
  \fmf{projc}{v2,i1}
  \fmfdot{v2}
 \end{fmfgraph}\\
\refstepcounter{diagh}
(\Alph{diagh})\label{D26}
\ec}}
= 0.\label{ident}
\ee

First we observe that diagram (\ref{ident}.\refmy{C26}) is zero due to the
color structure. Obviously diagram (\ref{ident}.\refmy{A26}) and
(\ref{ident}.\refmy{B26}) look very similar, and after substituting $-k$ to 
$k$ in diagram (\ref{ident}.\refmy{B26}) we arrive at

\be
\raisebox{0ex}{
\parbox{2.9cm}{\bc
 \begin{fmfgraph*}(2.5,2.5)
  \fmfforce{(0,h)}{i1}
  \fmfforce{(w,h)}{o1}
  \fmfforce{(.3w,.5h)}{v1}
  \fmfforce{(.7w,.5h)}{v2}
  \fmfforce{(0,.5h)}{v3}
  \fmfforce{(w,.5h)}{v4}
  \fmfforce{.4w,.4h}{v20}
  \fmfforce{.6w,.4h}{v21}
  \fmf{plains}{v21,v20}
  \fmf{arrowr}{v21,v20}
  \fmf{boson,right,label=$k$}{v3,v4}
  \fmf{boson,label=$p$}{i1,o1}
  \fmf{boson}{v3,v1}
  \fmf{boson}{v2,v4}
  \fmf{boson}{i1,v1}
  \fmf{boson}{v2,o1}
  \fmf{boson}{v1,v2}
  \fmfdot{v2}
 \end{fmfgraph*}
\ec}}
- 
\raisebox{0ex}{
\parbox{2.9cm}{\bc
 \begin{fmfgraph*}(2.5,2.5)
  \fmfforce{(0,h)}{i1}
  \fmfforce{(w,h)}{o1}
  \fmfforce{(.3w,.5h)}{v1}
  \fmfforce{(.7w,.5h)}{v2}
  \fmfforce{(0,.5h)}{v3}
  \fmfforce{(w,.5h)}{v4}
  \fmfforce{.4w,.4h}{v20}
  \fmfforce{.6w,.4h}{v21}
  \fmf{plains}{v20,v21}
  \fmf{arrowr}{v20,v21}
  \fmf{boson,right,label=$-k$}{v3,v4}
  \fmf{boson,label=$p$}{i1,o1}
  \fmf{boson}{v3,v1}
  \fmf{boson}{v2,v4}
  \fmf{boson}{o1,v1}
  \fmf{boson}{v2,i1}
  \fmf{boson}{v1,v2}
  \fmfdot{v1}
 \end{fmfgraph*}
\ec}}
=
\; -2 
\raisebox{0ex}{
\parbox{2.4cm}{\bc
 \begin{fmfgraph*}(2,2)
  \fmfforce{(0,.5h)}{v1}
  \fmfforce{(w,.5h)}{v2}
  \fmfforce{.35w,.9h}{v20}
  \fmfforce{.65w,.9h}{v21}
  \fmf{plains,left=0.15}{v20,v21}
  \fmf{arrowr,left=0.15}{v20,v21}
  \fmf{boson,left,label=$p$}{v2,v1}
  \fmf{boson,label=$k$}{v1,v2}
  \fmf{boson,left}{v1,v2}
  \fmfdot{v1}
 \end{fmfgraph*}\\
\ec}}
.
\label{shift}
\ee

\vspace{.5cm}

Clearly momentum shifts do not affect the value of a complete Feynman diagram.
Here, however the integrand of a complete Feynman diagram is rewritten  
algebraically as a sum and the sequence of the summation and the integration
over the internal momentum loop variables is interchanged. Performing the same
change of integration variables in all terms of the sum does not effect the
result. However, changing integration variables only in some terms, as e.g.~in
eq.~(\ref{shift}), requires a careful study of the corresponding terms in the 
sum. Here, this amounts to the demand that $a_\mu (k)$ is not too singular to 
allow for this substitutions and shifts (c.f. \cite{cheMIT}). In section 
\ref{sec4}, we shall show that this subtlety does not obstruct the arguments 
made in what follows. Altogether we get from (\ref{ident}) using (\ref{shift})
and attaching a $\Delta$

\be
\raisebox{-1.5ex}{
 \parbox{2.4cm}{\bc
 \begin{fmfgraph*}(2,2)
  \fmfforce{(0,.5h)}{v1}
  \fmfforce{(w,.5h)}{v2}
  \fmfforce{.35w,.9h}{v20}
  \fmfforce{.65w,.9h}{v21}
  \fmf{plains,left=0.15}{v20,v21}
  \fmf{arrowr,left=0.15}{v20,v21}
  \fmf{boson,left}{v2,v1}
  \fmf{boson}{v1,v2}
  \fmfv{label=$\nx\Delta$, l.a=-155, l.d=0.05w}{v1}
  \fmf{boson,left}{v1,v2}
  \fmfdot{v1}
 \end{fmfgraph*}\\
\stepcounter{rueck}
\refstepcounter{diagh}
(\Alph{diagh})\label{A3}
\ec}}
=\;-\frac{1}{2}
\raisebox{-1.5ex}{
\parbox{1.5cm}{\bc
 \begin{fmfgraph*}(1.5,2)
  \fmfforce{(.5w,.5h)}{v1}
  \fmfforce{(.5w,h)}{v2}
  \fmfforce{(.5w,0)}{v3}
  \fmf{boson,left}{v1,v2,v1}
  \fmf{boson,left}{v1,v3,v1}
  \fmfv{label=$\nx\Delta$,l.a=112,l.d=0.1w}{v1}
  \fmf{projc,right}{v1,v2}
  \fmfdot{v1}
 \end{fmfgraph*}
\refstepcounter{diagh}
(\Alph{diagh})\label{B3}\ec
}}.
\label{identity1}
\ee
Similarly, eq. (\ref{rel3}) leads to the identity

\be
\raisebox{-1.5ex}{
\parbox{2.4cm}{\bc
 \begin{fmfgraph*}(2,2)
  \fmfforce{(0,.5h)}{v1}
  \fmfforce{(w,.5h)}{v2}
  \fmfforce{(0,.5h)}{v3}
  \fmf{boson,left}{v1,v2,v1}
  \fmf{boson}{v1,v2}
  \fmfv{label=$\nx\Delta$, l.a=-155, l.d=0.05w}{v1}
  \fmfv{label=\nx\G, l.a=145, l.d=0.1w}{v3}
  \fmfv{label={\nx\bf 1}, l.a=155}{v2}
 \end{fmfgraph*}\\
\stepcounter{rueck}
\refstepcounter{diagh}
(\Alph{diagh})\label{A4}\ec
}}
=\;
\raisebox{-1.5ex}{
\parbox{2.4cm}{\bc
 \begin{fmfgraph*}(2,2)
  \fmfforce{(0,.5h)}{v1}
  \fmfforce{(w,.5h)}{v2}
  \fmfforce{.35w,.9h}{v20}
  \fmfforce{.65w,.9h}{v21}
  \fmf{plains,right=0.15}{v21,v20}
  \fmf{arrowr,right=0.15}{v21,v20}
  \fmf{boson,left}{v2,v1}
  \fmf{boson}{v1,v2}
  \fmfv{label=$\nx\Delta$, l.a=-155, l.d=0.05w}{v1}
  \fmfv{label=\nx\G, l.a=155}{v2}
  \fmf{boson,left}{v1,v2}
 \end{fmfgraph*}\\
\refstepcounter{diagh}
(\Alph{diagh})\label{B4}\ec
}}
\; .
\label{1ident}
\ee
This means that (\ref{zerlegung}.\refmy{C2}) cancels 
(\ref{zerlegung}.\refmy{D2}). Using (\ref{identity1}), eq.~(\ref{zerlegung}) 
is thus simplified to

\be
\raisebox{-1.5ex}{
\parbox{2.4cm}{\bc
 \begin{fmfgraph*}(2,2)
  \fmfforce{(0,.5h)}{v1}
  \fmfforce{(w,.5h)}{v2}
  \fmf{boson,left}{v1,v2,v1}
  \fmf{boson}{v1,v2}
  \fmf{projs,left}{v2,v1}
  \fmfv{label=$\nx\Delta$, l.a=-155, l.d=0.05w}{v1} 
  \fmfdot{v1}
 \end{fmfgraph*}\\
\stepcounter{rueck}
\refstepcounter{diagh}
(\Alph{diagh})
\label{A5}
\ec
}}
=\;-
\raisebox{-1.5ex}{
\parbox{1.5cm}{\bc
 \begin{fmfgraph*}(1.5,2)
  \fmfforce{(.5w,.5h)}{v1}
  \fmfforce{(.5w,h)}{v2}
  \fmfforce{(.5w,0)}{v3}
  \fmf{boson,left}{v1,v2,v1}
  \fmf{boson,left}{v1,v3,v1}
  \fmfv{label=$\nx\Delta$,l.a=112,l.d=0.1w}{v1}
  \fmf{projc,right}{v1,v2}
  \fmfdot{v1}
 \end{fmfgraph*}
\refstepcounter{diagh}
(\Alph{diagh})\label{B5}\ec
}}
+\;2\quad
\raisebox{-1.5ex}{
\parbox{2.4cm}{\bc
 \begin{fmfgraph*}(2,2)
  \fmfforce{(0,.5h)}{v1}
  \fmfforce{(w,.5h)}{v2}
  \fmfforce{(0,.5h)}{v3}
  \fmf{boson,left}{v1,v2,v1}
  \fmf{boson}{v1,v2}
  \fmfv{label=$\nx\Delta$, l.a=-155, l.d=0.05w}{v1}
  \fmfv{label=\nx\G, l.a=145, l.d=0.1w}{v3}
  \fmfv{label=\nx\G, l.a=155}{v2}
  \fmf{projs,left}{v2,v1}
 \end{fmfgraph*}
\refstepcounter{diagh}
(\Alph{diagh})\label{C5}
\ec}}\;.
\label{zerlegung2}
\ee
Obviously this relates the three different diagrams (\ref{grund}.\refmy{A1}),
(\ref{grund}.\refmy{B1}) and (\ref{grund}.\refmy{C1}).

We are now still left with the diagram (\ref{grund}.\refmy{D1}). Using 
(\ref{fermionrule}) we can make the following decomposition (\ref{femrul}):

\be
\raisebox{-1.5ex}{
\parbox{2.4cm}{\bc
 \begin{fmfgraph*}(2,2)
  \fmfforce{(0,.5h)}{v1}
  \fmfforce{(w,.5h)}{v2}
  \fmf{fermion,left}{v1,v2,v1}
  \fmf{boson}{v1,v2}
  \fmf{projc}{v2,v1}
  \fmfv{label=$\nx\Delta$,label.a=-30}{v1}
  \fmfdot{v1,v2}
 \end{fmfgraph*}\\
\stepcounter{rueck}
\refstepcounter{diagh}
{(\Alph{diagh})\label{A27}}
\ec}}
=  
\raisebox{-1.5ex}{
 \parbox{1.5cm}{\bc
 \begin{fmfgraph*}(1.5,2)
  \fmfforce{(.5w,.5h)}{v1}
  \fmfforce{(.5w,h)}{v2}
  \fmfforce{(.5w,0)}{v3}
  \fmf{fermion,left}{v1,v2,v1}
  \fmf{boson,left}{v1,v3,v1}
  \fmfv{label=$\nx\Delta$,label.a=-120}{v1}
  \fmfdot{v1}
 \end{fmfgraph*}\\
\refstepcounter{diagh}
(\Alph{diagh})\label{B27}
\ec}}
-
\raisebox{-1.5ex}{
 \parbox{1.5cm}{\bc
 \begin{fmfgraph*}(1.5,2)
  \fmfforce{(.5w,.5h)}{v1}
  \fmfforce{(.5w,h)}{v2}
  \fmfforce{(.5w,0)}{v3}
  \fmf{fermion,left}{v1,v2,v1}
  \fmf{boson,left}{v1,v3,v1}
  \fmfv{label=$\nx\Delta$,label.a=-60}{v1}
  \fmfdot{v1}
 \end{fmfgraph*}\\
\refstepcounter{diagh}
(\Alph{diagh})\label{C27}
\ec}}\; .
\label{fermionzerl}
\ee
Closing identity (\ref{rel4}) on the right and left side to loops and 
attaching a $\Delta$ one can see that the r.h.s.~of eq.~(\ref{fermionzerl})
is zero and in this way the $a_{\mu}$-dependent part of 
(\ref{grund}.\refmy{D1}) vanishes.\\

In the third and final step, one just has to count symmetry factors in order 
to show that all $a_\mu$-dependent terms in (\ref{grund}) cancel each other.
To do so we count the different ways to obtain the diagrams in 
(\ref{zerlegung2}) from (\ref{grund}). There are 6 possibilities to assign 
$\Delta_\mu k_\nu$ to diagram (\ref{grund}.\refmy{B1}): there are 3 lines to 
choose from  and 2 ends of the line to which the factor $\Delta$ can be 
attached. (The latter factor 2 is explicitly given in (\ref{abspaltung2}).) 
Similarly, there are 4 such possibilities for diagram (\ref{grund}.\refmy{A1})
and 2 for (\ref{grund}.\refmy{C1}). This counting shows immediately that all 
terms linear in $\Delta_\mu k_\nu$ cancel each other. Similarly one can show 
the same for the terms which are quadratic and cubic in $\Delta_\mu k_\nu$.
That is all we have to show: (\ref{grund}) leads to the same result whether 
we calculate it in an arbitrary $\alpha$-gauge or in the Feynman gauge.

\subsection{The Symmetrization Argument}
\label{sec2.1.4}

Arguing separately for each order of $\Delta_\mu k_\nu$ is very hard to do
for more complicated diagram. In this subsection we outline a technique for 
solving this problem which is based on the following observation: 
In principle gauge invariance of a sum of diagrams requires only that this sum
does not change if $a_\mu$ is simultaneously changed in {\em all} lines of the
diagrams. However, we will show in the following that the same sum of diagrams 
remains unchanged even if $a_{\mu}$ is changed in only {\em one} line in 
{\em each} diagram.  To see this we start again with the sum given in 
(\ref{grund}) and symmetrize the diagrams with respect to the gluon 
propagators:  

\be
&&\frac{1}{12}
\left[
\raisebox{-1.5ex}{
\parbox{2.4cm}{\bc
 \begin{fmfgraph*}(2,2)
  \fmfforce{(0,.5h)}{v1}
  \fmfforce{(w,.5h)}{v2}
  \fmf{boson,label=$D_1$,l.d=.075w,l.s=left,left}{v1,v2}
  \fmf{boson,label=$D_3$,l.d=.075w,l.s=right,left}{v2,v1}
  \fmf{boson,label=$D_2$,l.d=.075w,l.s=left}{v1,v2}
  \fmfdot{v1,v2}
 \end{fmfgraph*}\\
\stepcounter{rueck}
\refstepcounter{diagh}
(\Alph{diagh})\label{A7}
\ec}}
+\;\frac{1}{2}
\left(
\raisebox{-1.5ex}{
 \parbox{1.5cm}{\bc
 \begin{fmfgraph*}(1.5,2)
  \fmfforce{(.5w,.5h)}{v1}
  \fmfforce{(.5w,h)}{v2}
  \fmfforce{(.5w,0)}{v3}
  \fmf{boson,left}{v1,v2,v1}
  \fmf{boson,left}{v1,v3,v1}
  \fmfv{l=$D_1$,l.d=.075w}{v2}
  \fmfv{l=$D_2$,l.d=.09w,l.a=90}{v3}
  \fmfdot{v1}
 \end{fmfgraph*}\\
\refstepcounter{diagh}
(\Alph{diagh})\label{B7}
\ec}}
+ 
\raisebox{-1ex}{
 \parbox{1.5cm}{\bc
 \begin{fmfgraph*}(1.5,2)
  \fmfforce{(.5w,.5h)}{v1}
  \fmfforce{(.5w,h)}{v2}
  \fmfforce{(.5w,0)}{v3}
  \fmf{boson,left}{v1,v2,v1}
  \fmf{boson,left}{v1,v3,v1}
  \fmfv{l=$D_1$,l.d=.075w}{v2}
  \fmfv{l=$D_3$,l.d=.09w,l.a=90}{v3}
  \fmfdot{v1}
 \end{fmfgraph*}\\
\refstepcounter{diagh}
(\Alph{diagh})\label{C7}
\ec}}
+
\raisebox{-1.5ex}{
 \parbox{1.5cm}{\bc
 \begin{fmfgraph*}(1.5,2)
  \fmfforce{(.5w,.5h)}{v1}
  \fmfforce{(.5w,h)}{v2}
  \fmfforce{(.5w,0)}{v3}
  \fmf{boson,left}{v1,v2,v1}
  \fmf{boson,left}{v1,v3,v1}
  \fmfv{l=$D_2$,l.d=.075w}{v2}
  \fmfv{l=$D_3$,l.d=.09w,l.a=90}{v3}
  \fmfdot{v1}
 \end{fmfgraph*}\\
\refstepcounter{diagh}
(\Alph{diagh})\label{D7}
\ec}} \right)\right.
\nonumber \\
&&
- \left. 2 \left(\;
\raisebox{-1.5ex}{
\parbox{2.4cm}{\bc
 \begin{fmfgraph*}(2,2)
  \fmfforce{(0,.5h)}{v1}
  \fmfforce{(w,.5h)}{v2}
  \fmf{boson,left}{v1,v2,v1}
  \fmfv{label=$\nx\G$,l.a=55}{v2}
  \fmfv{label=$\nx\G$,l.a=-125}{v1}
  \fmf{boson,label=$D_1$,l.d=.075w,l.s=left}{v1,v2}
 \end{fmfgraph*}\\
\refstepcounter{diagh}
{(\Alph{diagh})\label{E7}}
\ec}}
\; + \; 
\raisebox{-1.5ex}{
\parbox{2.4cm}{\bc
 \begin{fmfgraph*}(2,2)
  \fmfforce{(0,.5h)}{v1}
  \fmfforce{(w,.5h)}{v2}
  \fmf{boson,left}{v1,v2,v1}
  \fmfv{label=$\nx\G$,l.a=55}{v2}
  \fmfv{label=$\nx\G$,l.a=-125}{v1}
  \fmf{boson,label=$D_2$,l.d=.075w,l.s=left}{v1,v2}
 \end{fmfgraph*}\\
\refstepcounter{diagh}
{(\Alph{diagh})\label{F7}}
\ec}}
\; + \;
\raisebox{-1.5ex}{
\parbox{2.4cm}{\bc
 \begin{fmfgraph*}(2,2)
  \fmfforce{(0,.5h)}{v1}
  \fmfforce{(w,.5h)}{v2}
  \fmf{boson,left}{v1,v2,v1}
  \fmfv{label=$\nx\G$,l.a=55}{v2}
  \fmfv{label=$\nx\G$,l.a=-125}{v1}
  \fmf{boson,label=$D_3$,l.d=.075w,l.s=left}{v1,v2}
 \end{fmfgraph*}\\
\refstepcounter{diagh}
{(\Alph{diagh})\label{G7}}
\ec}}
\;\right)\right.\nonumber \\
&&
\left. -\;2\left(
\raisebox{-1.5ex}{
\parbox{2.4cm}{\bc
 \begin{fmfgraph*}(2,2)
  \fmfforce{(0,.5h)}{v1}
  \fmfforce{(w,.5h)}{v2}
  \fmf{fermion,right}{v1,v2,v1}
  \fmf{boson,label=$D_1$,l.d=.075w,l.s=left}{v1,v2}
  \fmfdot{v1,v2}
 \end{fmfgraph*}\\
\refstepcounter{diagh}
{(\Alph{diagh})\label{H7}}
\ec}}
+
\raisebox{-1.5ex}{
\parbox{2.4cm}{\bc
 \begin{fmfgraph*}(2,2)
  \fmfforce{(0,.5h)}{v1}
  \fmfforce{(w,.5h)}{v2}
  \fmf{fermion,right}{v1,v2,v1}
  \fmf{boson,label=$D_2$,l.d=.075w,l.s=left}{v1,v2}
  \fmfdot{v1,v2}
 \end{fmfgraph*}\\
\refstepcounter{diagh}
{(\Alph{diagh})\label{I7}}
\ec}}
+ 
\raisebox{-1.5ex}{
\parbox{2.4cm}{\bc
 \begin{fmfgraph*}(2,2)
  \fmfforce{(0,.5h)}{v1}
  \fmfforce{(w,.5h)}{v2}
  \fmf{fermion,right}{v1,v2,v1}
  \fmf{boson,label=$D_3$,l.d=.075w,l.s=left}{v1,v2}
  \fmfdot{v1,v2}
 \end{fmfgraph*}\\
\refstepcounter{diagh}
{(\Alph{diagh})\label{J7}}
\ec}}
\;\right)\right].
\label{grundsym}
\ee

In the spirit of Sec.~\ref{sec2.1.3}, we choose now one line, say $D_1$, and 
split it into its gauge dependent and independent parts. This means e.g.~for 
(\ref{grundsym}.\refmy{A7}):

\be
\raisebox{-1.5ex}{
\parbox{2.4cm}{\bc
 \begin{fmfgraph*}(2,2)
  \fmfforce{(0,.5h)}{v1}
  \fmfforce{(w,.5h)}{v2}
  \fmf{boson,label=$D_1$,l.d=.075w,l.s=left,left}{v1,v2}
  \fmf{boson,label=$D_3$,l.d=.075w,l.s=right,left}{v2,v1}
  \fmf{boson,label=$D_2$,l.d=.075w,l.s=left}{v1,v2}
  \fmfdot{v1,v2}
 \end{fmfgraph*}\\
\stepcounter{rueck}
\refstepcounter{diagh}
(\Alph{diagh})\label{A8}
\ec}}
=
\raisebox{-1.5ex}{
\parbox{2.4cm}{\bc
 \begin{fmfgraph*}(2,2)
  \fmfforce{(0,.5h)}{v1}
  \fmfforce{(w,.5h)}{v2}
  \fmf{boson,label=$D_{1,,F}$,l.d=.075w,l.s=left,left}{v1,v2}
  \fmf{boson,label=$D_3$,l.d=.075w,l.s=right,left}{v2,v1}
  \fmf{boson,label=$D_2$,l.d=.075w,l.s=left}{v1,v2}
  \fmfdot{v1,v2}
 \end{fmfgraph*}\\
\refstepcounter{diagh}
(\Alph{diagh})\label{B8}
\ec}}
+\; 2 \quad
\raisebox{-1.5ex}{
\parbox{2.4cm}{\bc
 \begin{fmfgraph*}(2,2)
  \fmfforce{(0,.5h)}{v1}
  \fmfforce{(w,.5h)}{v2}
  \fmf{boson,left}{v1,v2}
  \fmf{boson,label=$D_3$,l.d=.075w,l.s=right,left}{v2,v1}
  \fmf{boson,label=$D_2$,l.d=.075w,l.s=left}{v1,v2}
  \fmfv{label=$\nx\Delta $,l.a=150,l.d=.075w}{v1}
  \fmf{projs,right}{v2,v1}
  \fmfdot{v1,v2}
 \end{fmfgraph*}\\
\refstepcounter{diagh}
(\Alph{diagh})\label{C8}
\ec}}
\label{abspaltungsym}
\ee

Clearly all diagrams in (\ref{grundsym}) containing $D_1$ must be treated in 
the same way. According to (\ref{fermionzerl}) the fermion diagram cancels 
immediately. By complete analogy with our treatment in eq.~(\ref{zerlegung}), 
one observes that the gauge-dependent part arising from the propagator $D_1$ 
in (\ref{abspaltungsym}) cancels the corresponding gauge-dependent 
contributions of $D_1$ in (\ref{grundsym}.\refmy{B7}/\refmy{C7}/\refmy{E7}). 
Consequently, we can replace in (\ref{grundsym}) the propagator $D_{1}$ by its
Feynman part $D_{1,F}$ without changing $D_2$ and $D_3$. In the procedure the 
explicit form of $D_2$ and $D_3$ is never used. As we have changed $D_1$ to  
$D_{1,F}$ without changing the value of the whole sum of diagrams we can do 
the same for $D_2$ and finally also for $D_3$. Consequently, $D_2$ and $D_3$ 
can also be replaced by their Feynman parts $D_{2,F}$ and $D_{3,F}$ without 
changing the value of (\ref{grundsym}). The set of diagrams has the same value
in Feynman gauge and in all $\alpha$-gauges and in this sense it is 
gauge-invariant.

From the fact that $D_1$ can be replaced by $D_{1,F}$ without using the
explicit form of $D_2$ and $D_3$, we draw two conclusions:

\begin{enumerate}
 \item It is sufficient to check the invariance of a set of Feynman 
       diagrams under the replacement of only one propagator in each diagram 
       by its Feynman part. The rest can be handled by symmetrization.
 \item Since no assumption is made about the form of the $D_n$, each could 
       enter the symmetrized expression with a different $\Delta_\mu k_\nu$ 
       without affecting the result of the calculation.
\end{enumerate}

In the next section we will generalize the diagrammatic rules presented here to
the case of finite temperature.

\end{fmffile}

% 
% Hier ends the first extra font ``paperf1''
%
%%%%%%%%%%%%%%%%%%%%%%%%%%%%%%%%%%%%%%%%%%%%%%%%%%%%%%%%%%%%%%%%%%%%%%%
%                                                                     %
% Starting the 2.font-file and adding additional Metafont-Definitions %
%                                                                     %
%%%%%%%%%%%%%%%%%%%%%%%%%%%%%%%%%%%%%%%%%%%%%%%%%%%%%%%%%%%%%%%%%%%%%%%

\begin{fmffile}{paperf2}

%%%%%%%%%%%%%%%%%%%%%%%%%%%%%%%%%%%%%%%%%%%%%%%%%%%%%%%%%%%%%%%%%%%%%%%
%                                                                     %
%       Some Metafont-definitions in addition to feynmf.mf            %
%                                                                     %
%%%%%%%%%%%%%%%%%%%%%%%%%%%%%%%%%%%%%%%%%%%%%%%%%%%%%%%%%%%%%%%%%%%%%%%
%
% Variables:
%

\fmfcurved
\fmfpen{thin} 
\fmfset{curly_len}{3mm}
\fmfset{wiggly_len}{2mm}

%
% Styles:
%

\fmfcmd{ %
 vardef cross_bar (expr p, len, ang) =
  ((-len/2,0)--(len/2,0))
    rotated (ang + angle direction length(p)/2 of p)
    shifted point length(p)/2 of p
 enddef;
 style_def crossed expr p = 
   ccutdraw cross_bar (p, 5mm, 45);
   ccutdraw cross_bar (p, 5mm, -45)
 enddef;}

\fmfcmd{ %
 vardef proj_tarrow (expr p, frac) =
  save a, t, z;
  pair z;
  t1 = frac*length p;
  a = angle direction t1 of p;
  z = point t1 of p;
  (t2,whatever) = p intersectiontimes
    (halfcircle scaled 3arrow_len rotated (a-90) shifted z);
  arrow_head (p, t1, t2, arrow_ang)
enddef;
 style_def projs expr p  =
    cfill (proj_tarrow (reverse p, 0.85));
 enddef;
 style_def projc expr p =
    cfill (proj_tarrow (reverse p, 0.7));
 enddef;}

\fmfcmd{ 
 vardef projl_tarrow (expr p, frac) =
  save a, t, z;
  pair z;
  t1 = frac*length p;
  a = angle direction t1 of p;
  z = point t1 of p;
  (t2,whatever) = p intersectiontimes
    (halfcircle scaled 2arrow_len rotated (a-90) shifted z);
  arrow_head (p, t1, t2, arrow_ang)
 enddef;
 style_def arrowr expr p =
   cfill (projl_tarrow (reverse p, 0.6));
 enddef;
 style_def arrowl expr p  =
   cfill (projl_tarrow (reverse p, 0.6));
 enddef;
}

\fmfcmd{
style_def plains expr p =
 draw p;
 undraw subpath (0,.15) of p;
enddef;}

%%%%%%%%%%%%%%%%%%%%%%%%%%%%%%%%%%%%%%%%%%%%%%%%%%%%%%%%%%%%%%%%%%%%%%%
%                                                                     %
%                  Here the paper continues                           %
%                                                                     %
%%%%%%%%%%%%%%%%%%%%%%%%%%%%%%%%%%%%%%%%%%%%%%%%%%%%%%%%%%%%%%%%%%%%%%%

%%%%%%%%%%%%%%%%%%%%%%%%%%%%%%%%%%%%%%%%%%%%%%%%%%%%%%%%%%%%%%%%%%%%%%
\section{The Cheng-Tsai method for Imaginary Time Perturbation Theory}
\label{sec3}
%%%%%%%%%%%%%%%%%%%%%%%%%%%%%%%%%%%%%%%%%%%%%%%%%%%%%%%%%%%%%%%%%%%%%%

So far, we have reviewed and completed the basic Cheng-Tsai rules for vacuum 
QCD. Now we explain in a first step how these rules get modified in imaginary 
time perturbation theory at finite temperature (i.e.~with a compact imaginary 
time coordinate). Subsequently, the rules are extended to a resummed 
perturbation scheme in which a dynamically generated Debye screening mass is 
introduced for the static temporal gauge propagator. This paves the way for 
checking the gauge invariance of the {$\cal O$}({\em g}$\,^4$) result for 
$Z_{\text{QCD}}$.

\subsection{Diagrammatic Cheng-Tsai Rules at Finite Temperature}
\label{sec3.1}

The Feynman rules for imaginary time perturbation theory at finite temperature
differ from the vacuum ones essentially only by changing the energy component 
$k_0$ of the four momentum $k$ to discrete imaginary Matsubara frequencies, 
$k_0 = i2n\pi\beta^{-1}$ for bosons, $k_0 = i(2n+1)\pi\beta^{-1}$ for fermions
with $\beta$ being the inverse temperature. This leads to the replacement of 
loop integrals in momentum space by 

\be
  \int \frac{d^4k}{(2\pi)^4i} \rightarrow 
  \beta \sum_{k_{0}}\int \frac{d^3k}{(2\pi)^3}.
\label{loop}
\ee

The energy-momentum conservation at each vertex is modified accordingly:

\be
  i(2\pi )^4\delta (k) \rightarrow \beta (2\pi )^3\delta_{n0}\delta(\vec{k}).
\label{3.6}
\ee

Up to factors $i$ coming from the imaginary $k_0$ components, the propagators 
and vertices look formally the same as in the vacuum case, now being functions
of $\vec{k}$ and $k_0$ (cf.~Appendix \ref{appb}). They can be split up 
diagrammatically into their Feynman gauge and explicit $a_\mu$-dependent parts
in exactly the same way as before, e.g.

\be
  D_{\mu\nu}^{ab} & = & \delta^{ab}
                  \left[g_{\mu\nu} + a_\mu (-k)k_\nu - a_\nu (k)k_\mu \right]
                          \frac{1}{k^2},\nonumber\\
  &\equiv & D_{F,\mu\nu} + \Delta_{\mu}(-k)k_{\nu} - \Delta_{\nu}(k)k_{\mu};
  \qquad \mbox{with} \quad k^2=\left.\vec{k}\right.^2+k_0^2,
\label{prop2}
\ee

\be
\mbox{
 \parbox{3cm}{\bc
  \begin{fmfgraph}(2.5,2)
   \fmfforce{(0,.5h)}{v1}
   \fmfforce{(w,.5h)}{v2}
   \fmf{boson}{v1,v2}
 \end{fmfgraph}
\ec}
\quad = \quad
 \parbox{2.5cm}{\bc
  \begin{fmfgraph*}(2.5,2)
   \fmfforce{(0,.5h)}{v1}
   \fmfforce{(w,.5h)}{v2}
   \fmf{boson,label=$D_F$,l.s=left}{v1,v2}
 \end{fmfgraph*}
\ec}
\quad + \quad
 \parbox{2.5cm}{\bc
  \begin{fmfgraph*}(2.5,2)
   \fmfforce{(0,.5h)}{v1}
   \fmfforce{(w,.5h)}{v2}
   \fmfforce{(.85w,.5h)}{v4}
   \fmf{boson}{v1,v4}
   \fmfv{label=$\nx\Delta$,l.a=35}{v1}
   \fmf{projc}{v2,v1}
 \end{fmfgraph*}
 \ec}
\quad - \quad
 \parbox{2.5cm}{\bc
  \begin{fmfgraph*}(2.5,2)
   \fmfforce{(0,.5h)}{v1}
   \fmfforce{(w,.5h)}{v2}
   \fmfforce{(.15w,.5h)}{v4}
   \fmf{boson}{v4,v2}
   \fmfv{label=$\nx\Delta$,l.a=145}{v2}
   \fmf{projc}{v1,v2}
 \end{fmfgraph*}
\ec}\quad ,
}\nonumber
\ee
where now the Feynman propagator is defined by

\be
  D_{F,\mu\nu}^{ab} = \delta^{ab}g_{\mu\nu}\frac{1}{k^2} 
\ee
while $\Delta_{\mu}^{ab}$ takes the form

\be
  \Delta_{\mu}^{ab}(k) = \delta^{ab}\frac{a_{\mu}(k)}{k^{2}}.
\ee

With these changes, the diagrammatic rules
(\ref{maindiagram}), (\ref{ghocovz2}), (\ref{momentumcon}) and 
(\ref{femrul}) look the same, the ghost contribution reading now 
(c.f.~(\ref{ghost}))

\be
  G_{\mu}^{abd} (k) 
         = 
  -if^{abd}\left[\left(a\cdot k - 1\right)k_\mu - k^2a_\mu\right] \frac{1}{k^2}.
\label{ghost2}
\ee

Also, the basic identities (\ref{rel1}-\ref{rel4}) carry over without further 
modification. Note that the application of the Cheng-Tsai method in imaginary 
time perturbation theory proceeds as in the vacuum case. It is not obstructed 
by the replacement (\ref{loop}). All rules are purely algebraic, establishing 
the cancellation of $a_\mu$-dependent terms on the level of the integrand 
before any loop integrations are performed (for a detailed discussion, 
cf.~Sec.~\ref{sec4}). Consequently, they do not depend on changing the 
$k_0$-loop integral to an infinite sum.

\subsection{A Simple Resummation Scheme}
\label{sec3.2}

As already mentioned, naive calculations in finite temperature perturbation 
theory with massless degrees of freedom can lead to gauge dependent and 
infrared divergent results for physical quantities since the loop expansion 
does not coincide with an expansion in orders of the coupling constant 
\cite{bra90b}. Depending on the physical quantity one wants to calculate, this
requires the use of more or less refined resummation schemes 
\cite{bra90a,arn93,bra95}. A prominent example is the QCD partition function 
$Z_{\text{QCD}}$ for which naive perturbation theory leads to an incomplete 
two-loop result, missing the {$\cal O$}({\em g}$\,^3$)-contribution 
completely, and becomes infrared divergent at the three-loop level. At least 
up to order {$\cal O$}({\em g}$\,^5$), a consistent resummation scheme curing 
all infrared divergences is obtained by reorganizing perturbation theory with 
a massive static $A_0$-propagator \cite{arn95,arn93}.

\be
  {\cal L} = \left({\cal L} + \frac{1}{2}m^2A_0^aA_0^a\delta_{k_0,0}\right)
                            - \frac{1}{2}m^2A_0^aA_0^a\delta_{k_0,0}.
\label{lagrangien}
\ee

Here, $m$ denotes the dynamical generated Debye screening mass which takes 
into account the screening of electric fields in a polarizable plasma at 
finite temperature and density. At lowest order in the coupling constant the 
Debye mass is obtained as the infrared limit of the corresponding component 
of the one-loop self energy $\Pi_{\mu\nu}$,

\be
  m^2 = \lim_{\vec{k}\rightarrow 0} \Pi_{00}(k_0=0, \vec{k}).
\label{selfform}
\ee

\be
\mbox{\bf \Large $-\Pi_{\mu\nu}$}
\quad  = \frac{1}{2} \quad 
\parbox{2cm}{\bc
 \begin{fmfgraph}(2,2)
  \fmfincoming{i1}
  \fmfoutgoing{o1}
  \fmf{boson}{i1,v1,v1,o1}
  \fmfdot{v1}
 \end{fmfgraph}\ec
}
+ \frac{1}{2}
\parbox{2.6cm}{\bc
 \begin{fmfgraph}(2.5,2)
  \fmfincoming{i1}
  \fmfoutgoing{o1}
  \fmf{boson}{i1,v1}
  \fmf{boson}{v2,o1}
  \fmf{boson,left,tension=.3}{v1,v2,v1}
  \fmfdot{v1,v2}
 \end{fmfgraph}\ec
}
- \;
\parbox{2.5cm}{\bc
 \begin{fmfgraph*}(2.5,2)
  \fmfincoming{i1}
  \fmfoutgoing{o1}
  \fmf{boson}{i1,v1}
  \fmf{boson}{v2,o1}
  \fmfv{label=\nx\G,l.a=114}{v1}
  \fmfv{label=\nx\G,l.a=-67}{v2}
  \fmf{boson,left,tension=.3}{v1,v2,v1}
 \end{fmfgraph*}\ec
}
- \;
\parbox{2.5cm}{\bc
 \begin{fmfgraph}(2.5,2)
  \fmfincoming{i1}
  \fmfoutgoing{o1}
  \fmf{boson}{i1,v1}
  \fmf{boson}{v2,o1}
  \fmf{fermion,left,tension=.3}{v1,v2,v1}
  \fmfdot{v1,v2}
 \end{fmfgraph}\ec
}
\label{self}
\ee

In the improved imaginary time perturbation theory, resummed according to
(\ref{lagrangien}) and (\ref{self}), the gauge boson propagator reads

\be
D_{\mu\nu}^{ab}(k) & = & \delta^{ab}
                  \left[g_{\mu\nu} + a_\mu (-k)k_\nu - a_\nu (k)k_\mu 
                        -\delta_{\mu 0}\delta_{\nu 0}
                           \delta_{k_0 0}\frac{m^2}{m^2 + k^2}\right]
                          \frac{1}{k^2},\nonumber\\
  &\equiv & \underbrace{D_{F,\mu\nu}^{ab}
 -\delta^{ab}\delta_{\mu 0}\delta_{\nu 0}
   \delta_{k_0 0}\frac{m^2}{k^2\left(m^2 + k^2\right)}}_{D_{F,\mu\nu}'^{ab}}
  +\Delta_{\mu}(-k)k_{\nu} - \Delta_{\nu}(k)k_{\mu}.
\label{gluonm}
\ee

\be
\mbox{
 \parbox{3cm}{\bc
  \begin{fmfgraph}(2.5,2)
   \fmfforce{(0,.5h)}{v1}
   \fmfforce{(w,.5h)}{v2}
   \fmf{boson}{v1,v2}
 \end{fmfgraph}
\ec}
\quad = \quad
 \parbox{2.5cm}{\bc
  \begin{fmfgraph*}(2.5,2)
   \fmfforce{(0,.5h)}{v1}
   \fmfforce{(w,.5h)}{v2}
   \fmf{boson,label=$D_F'$,l.s=left}{v1,v2}
 \end{fmfgraph*}
\ec}
\quad + \quad
 \parbox{2.5cm}{\bc
  \begin{fmfgraph*}(2.5,2)
   \fmfforce{(0,.5h)}{v1}
   \fmfforce{(w,.5h)}{v2}
   \fmfforce{(.85w,.5h)}{v4}
   \fmf{boson}{v1,v4}
   \fmfv{label=$\nx\Delta$,label.a=35}{v1}
   \fmf{projc}{v2,v1}
 \end{fmfgraph*}
 \ec}
\quad - \quad
 \parbox{2.5cm}{\bc
  \begin{fmfgraph*}(2.5,2)
   \fmfforce{(0,.5h)}{v1}
   \fmfforce{(w,.5h)}{v2}
   \fmfforce{(.15w,.5h)}{v4}
   \fmf{boson}{v4,v2}
   \fmfv{label=$\nx\Delta$,label.a=145}{v2}
   \fmf{projc}{v1,v2}
 \end{fmfgraph*}
 \ec}
}
\ee
The counterterm in (\ref{lagrangien}) is treated diagrammatically as a new
two-point vertex

\be
\parbox{3cm}{\bc
 \begin{fmfgraph*}(2.3,2.3)
  \fmfincoming{i1}
  \fmfoutgoing{o1}
  \fmfv{label=a}{i1}
  \fmfv{label=b}{o1}
  \fmf{boson,label=$\nx\mu\nx\qquad\nx\nu$,l.s=right}{i1,o1}
  \fmf{crossed}{i1,o1}
 \end{fmfgraph*}\ec
}\qquad
= \qquad 
\delta^{ab}\delta_{\nu 0}\delta_{\mu 0}\delta_{k_0 0}m^2
\label{counter}
\ee

Note that the counter term in (\ref{lagrangien}) breaks the gauge invariance
explicitly. Hence, it is a priori unclear whether this reorganization of
perturbation theory preserves the gauge invariance at a fixed order in the 
coupling constant. \\ 
Due to the mass term in (\ref{gluonm}), the Cheng-Tsai rule for the action
of the ``projection'' ${\cal P}$ on the propagator $D_{\mu\nu}^{ab}$ receives 
now an extra contribution

\be
 -i\mbox{\em{g}}f^{abc}{\cal P}_{\mu}^\rho D_{\rho\nu}^{cd}(k) \quad = \quad 
  i\mbox{\em{g}}f^{abd}\delta_{\mu 0}\delta_{\nu 0}\delta_{k_00}
  \frac{m^2}{m^2+k^2}\quad-\quad i\mbox{\em{g}}f^{abd}g_{\mu\nu} 
  \quad - \quad G_{\mu}^{abd} (k)k_\nu .
\label{mainm}
\ee

\bc
 \parbox{3cm}{\bc
  \begin{fmfgraph*}(2.5,2)
   \fmfforce{(0,.5h)}{v1}
   \fmfforce{(w,.5h)}{v2}
   \fmfforce{(0,.5h)}{v3}
   \fmf{boson}{v1,v2}
   \fmfv{label=${\nx\cal P}$,l.a=35}{v3}
 \end{fmfgraph*}
\ec }
\quad = 
\quad + \quad
 \parbox{2.5cm}{\bc
  \begin{fmfgraph*}(2.5,2)
   \fmfforce{(0,.5h)}{v1}
   \fmfforce{(w,.5h)}{v2}
   \fmf{boson,label=$\nx\star$,l.d=.05w,l.s=left}{v1,v2}
   \fmf{crossed}{v2,v1}
 \end{fmfgraph*}
\ec}
\quad {\LARGE -} \quad
 \parbox{2.5cm}{\bc
  \begin{fmfgraph}(2.5,2)
   \fmfforce{(0,.5h)}{v1}
   \fmfforce{(w,.5h)}{v2}
   \fmfforce{.4w,.6h}{v20}
   \fmfforce{.6w,.6h}{v21}
   \fmf{plains}{v21,v20}
   \fmf{arrowl}{v21,v20}
   \fmf{boson}{v1,v2}
 \end{fmfgraph}
\ec}
\quad {\LARGE -} \quad
 \parbox{2.5cm}{\bc
  \begin{fmfgraph*}(2.5,2)
   \fmfforce{(0,.5h)}{v1}
   \fmfforce{(w,.5h)}{v2}
   \fmfforce{(.85w,.5h)}{v4}
   \fmf{boson}{v1,v4}
   \fmfv{label=\nx\G,l.a=35}{v1}
   \fmf{projc}{v2,v1}
 \end{fmfgraph*}
 \ec}
\ec

Thus (\ref{maindiagram}) is modified to

\be
 \parbox{2cm}{\bc
  \begin{fmfgraph}(2,2)
   \fmfoutgoing{o1}
   \fmfforce{(0,h)}{v2}
   \fmfforce{(0,0)}{v3}
   \fmfforce{(.4w,.5h)}{v1}
   \fmf{boson}{v2,v1}
   \fmfset{arrow_ang}{25}
   \fmf{projc}{v1,v2}
   \fmfset{arrow_ang}{15}
   \fmf{boson}{v3,v1,o1}
   \fmfdot{v1}
  \end{fmfgraph}
 \ec}
\quad = \quad + \quad
 \parbox{2cm}{\bc
  \begin{fmfgraph*}(2,2)
   \fmfoutgoing{o1}
   \fmfforce{(0,h)}{v2}
   \fmfforce{(0,0)}{v3}
   \fmfforce{(.4w,.5h)}{v1}
   \fmf{boson}{v2,v1}
   \fmf{boson}{v3,v1}
   \fmf{boson,label=$\nx\star$,l.d=.05w,l.s=left}{v1,o1}
   \fmf{crossed}{o1,v1}
  \end{fmfgraph*}
 \ec}
 \quad - \quad
 \parbox{2cm}{\bc
  \begin{fmfgraph}(2,2)
   \fmfoutgoing{o1}
   \fmfforce{(0,h)}{v2}
   \fmfforce{(0,0)}{v3}
   \fmfforce{(.4w,.5h)}{v1}
   \fmfforce{.6w,.4h}{v20}
   \fmfforce{.8w,.4h}{v21}
   \fmf{plains}{v21,v20}
   \fmf{arrowr}{v21,v20}
   \fmf{boson}{v2,v1}
   \fmf{boson}{v3,v1}
   \fmf{boson}{v1,o1}
  \end{fmfgraph}
 \ec}
\quad - \quad
 \parbox{2cm}{\bc
  \begin{fmfgraph*}(2,2)
   \fmfoutgoing{o1}
   \fmfforce{(0,h)}{v2}
   \fmfforce{(0,0)}{v3}
   \fmfforce{(.9w,.5h)}{v4}
   \fmfforce{(.4w,.5h)}{v1}
   \fmfv{label=\nx\G, l.a=35, l.d=0.05w}{v1}
   \fmf{boson}{v2,v1}
   \fmf{boson}{v3,v1}
   \fmf{boson}{v1,v4}
   \fmfset{arrow_ang}{30}
   \fmf{projc}{o1,v1}
   \fmfset{arrow_ang}{15}
  \end{fmfgraph*}
 \ec}\nonumber \\
- \quad
 \parbox{2cm}{\bc
  \begin{fmfgraph*}(2,2)
   \fmfoutgoing{o1}
   \fmfforce{(0,h)}{v2}
   \fmfforce{(0,0)}{v3}
   \fmfforce{(.4w,.5h)}{v1}
   \fmf{boson}{v2,v1}
   \fmf{boson}{o1,v1}
   \fmf{boson,label=$\nx\star$,l.d=.05w,l.s=right}{v1,v3}
   \fmf{crossed}{v3,v1}
  \end{fmfgraph*}
 \ec}
\quad  + \quad
 \parbox{2cm}{\bc
  \begin{fmfgraph}(2,2)
   \fmfoutgoing{o1}
   \fmfforce{(0,h)}{v2}
   \fmfforce{(0,0)}{v3}
   \fmfforce{(.4w,.5h)}{v1}
   \fmf{boson}{v2,v1}
   \fmf{boson}{v1,o1}
   \fmfforce{.322w,.253h}{v20}
   \fmfforce{.198w,.097h}{v21}
   \fmf{plains}{v21,v20}
   \fmf{arrowr}{v21,v20}
   \fmf{boson}{v2,v1}
   \fmf{boson}{v1,o1}
   \fmf{boson}{v1,v3}
  \end{fmfgraph}
 \ec}
\quad  + \quad
 \parbox{2cm}{\bc
  \begin{fmfgraph*}(2,2)
   \fmfoutgoing{o1}
   \fmfforce{(0,h)}{v2}
   \fmfforce{(0,0)}{v3}
   \fmfforce{(.08w,.1h)}{v4}
   \fmfforce{(.4w,.5h)}{v1}
   \fmfv{label=\nx\G, l.a=-85, l.d=0.1w}{v1}
   \fmf{boson}{v2,v1}
   \fmf{boson}{v4,v1}
   \fmf{boson}{v1,o1}
   \fmfset{arrow_ang}{30}
   \fmf{projc}{v3,v1}
   \fmfset{arrow_ang}{15}
  \end{fmfgraph*}
 \ec}
\label{maindiagram2}
\ee

Since eq.~(\ref{lagrangien}) does not involve resummed vertices, all 
Cheng-Tsai rules relating different vertices to each other remain unchanged.
Consequently, we have the diagrammatic reformulation of the energy-momentum 
conservation (\ref{momentumcon}), the fermion-vertex rule (\ref{femrul}) and 
the basic identities (\ref{rel1}-\ref{rel4}) at our disposal. Additionally, 
there is a rule linking the counter term (\ref{counter}) to the mass-dependent
term in (\ref{mainm}):

\be
  i\mbox{\em{g}}f^{abe}g_{\mu\sigma}\delta_{\sigma 0}\delta_{\alpha 0}
  \delta_{k_0 0}m^2D^{ed, \alpha}{}_\nu (k) 
  & = &  
  i\mbox{\em{g}}f^{abd}\delta_{\mu 0}\delta_{\nu 0}\delta_{k_0 0}
  \frac{m^2}{m^2+k^2}
\label{massrelation}
\ee

\be
\parbox{3.1cm}{
\begin{fmfgraph}(2.7,.5)
 \fmfforce{(0,.5h)}{v1}
 \fmfforce{(w,.5h)}{v2}
 \fmfforce{(.4w,.5h)}{v3}
 \fmfforce{(.6w,.5h)}{v4}
 \fmfforce{.15w,.75h}{v20}
 \fmfforce{.35w,.75h}{v21}
 \fmf{plains}{v21,v20}
 \fmf{arrowr}{v21,v20}
 \fmf{boson}{v1,v3,v4,v2}
 \fmf{crossed}{v3,v4}
 \fmf{boson}{v1,v3}
\end{fmfgraph}
}
=
\parbox{3.1cm}{
\begin{fmfgraph*}(2.7,.5)
 \fmfforce{(0,.5h)}{v1}
 \fmfforce{(w,.5h)}{v2}
 \fmf{boson}{v1,v2}
 \fmf{crossed}{v1,v2}
 \fmf{boson,label=$\nx\star$,l.d=.05w,l.s=left}{v1,v2}
\end{fmfgraph*}
}\nonumber
\ee

\subsection{The QCD Partition Function}
\label{sec3.3}

With the improved imaginary time perturbation scheme described in the last 
subsection, the QCD partition function has been calculated to 
{$\cal O$}({\em g}$\,^5$) \cite{arn94,arn95,zha95}. Since it is our main aim 
to check the gauge invariance of these results, let us shortly recall them.\\
$Z_{\text{QCD}}$ is given by the trace

\be
  Z_{\text{QCD}}= \mbox{Tr}e^{-\beta H}
\label{parfun}
\ee

where the Hamiltonian is defined in terms of (\ref{lagrangien}). The 
corresponding free energy is

\be
  F=-\frac{T}{V}\ln Z_{\text{QCD}}.
\ee

Its lowest order contribution is the Stefan-Boltzmann part, which is 
diagrammatically represented through one loop diagrams and  has been
calculated in many different gauges (c.f.~\cite{lan87}). The higher order 
corrections to this free part can be represented diagrammatically up to
{$\cal O$}({\em g}$\,^{5}$) as follows:

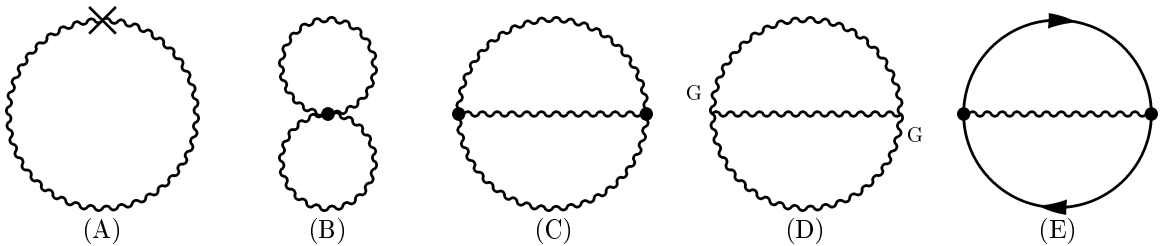
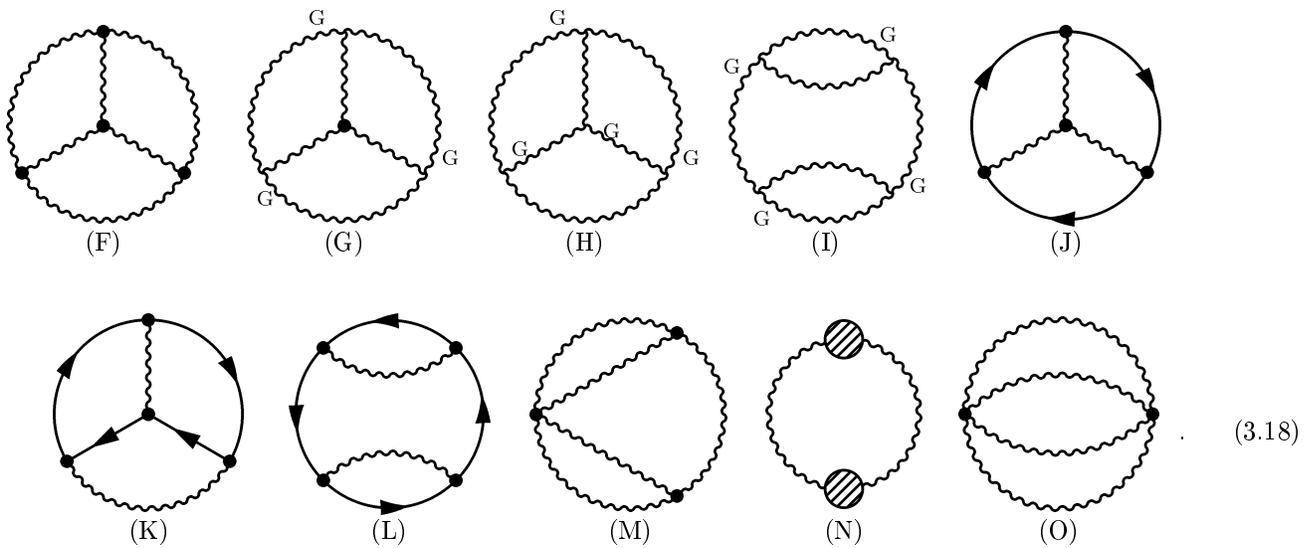
\be
\parbox{3cm}{\bc
 \begin{fmfgraph}(2.5,2.5)
  \fmfforce{(0,.5h)}{v1}
  \fmfforce{(w,.5h)}{v2}
  \fmf{boson,left}{v1,v2,v1}
  \fmf{crossed,left}{v1,v2}
 \end{fmfgraph}
\stepcounter{rueck}
\refstepcounter{diagh}
(\Alph{diagh})\label{C9}
\ec
}
\parbox{3cm}{\bc
 \begin{fmfgraph}(2.5,2.5)
  \fmfforce{(.5w,.5h)}{v1}
  \fmfforce{(.5w,h)}{v2}
  \fmfforce{(.5w,0)}{v3}
  \fmf{boson,left}{v1,v2,v1}
  \fmf{boson,left}{v1,v3,v1}
  \fmfdot{v1}
 \end{fmfgraph}
\refstepcounter{diagh}
(\Alph{diagh})\label{D9}
\ec
}
\parbox{3cm}{\bc
 \begin{fmfgraph}(2.5,2.5)
  \fmfforce{(0,.5h)}{v1}
  \fmfforce{(w,.5h)}{v2}
  \fmf{boson,left}{v1,v2,v1}
  \fmf{boson}{v1,v2}
  \fmfdot{v1,v2}
 \end{fmfgraph}
\refstepcounter{diagh}
(\Alph{diagh})\label{E9}
\ec
}\quad
\parbox{3cm}{\bc
 \begin{fmfgraph*}(2.5,2.5)
  \fmfforce{(0,.5h)}{v1}
  \fmfforce{(w,.5h)}{v2}
  \fmf{boson,left}{v1,v2,v1}
  \fmfv{label=\nx\G,label.a=120}{v1}
  \fmfv{label=\nx\G,label.a=-60}{v2}
  \fmf{boson}{v1,v2}
 \end{fmfgraph*}
\refstepcounter{diagh}
(\Alph{diagh})\label{F9}
\ec
}\quad
\parbox{3cm}{\bc
 \begin{fmfgraph}(2.5,2.5)
  \fmfforce{(0,.5h)}{v1}
  \fmfforce{(w,.5h)}{v2}
  \fmf{fermion,left}{v1,v2,v1}
  \fmf{boson}{v1,v2}
  \fmfdot{v1,v2}
 \end{fmfgraph}
\refstepcounter{diagh}
(\Alph{diagh})\label{F9f}
\ec
}\nonumber\\
\parbox{3.2cm}{\bc 
\begin{fmfgraph*}(2.5,2.5)
 \stargl
 \starplace{}{}{}{}{}{}{}{}
 \starplacedummy{}{}{}{}{}{}{}{}
 \startypes{boson}{}{boson}{}{boson}{}
 \startypec{boson}{}{boson}{}{boson}{}
 \starproj{}{}{}{}{}{}
 \fmfdot{v1,v2,v3,v4}
\end{fmfgraph*}\\
\refstepcounter{diagh}
(\Alph{diagh})\label{G9}
\ec
}
\parbox{3.2cm}{\bc
\begin{fmfgraph*}(2.5,2.5)
 \stargl
 \starplace{\nx\G}{157}{\nx\G}{-80}{\nx\G}{25}{}{}
 \starplacedummy{}{}{}{}{}{}{}{}
 \startypes{boson}{}{boson}{}{boson}{}
 \startypec{boson}{}{boson}{}{boson}{}
 \starproj{}{}{}{}{}{}
 \fmfdot{v4}
\end{fmfgraph*}\\
\refstepcounter{diagh}
(\Alph{diagh})\label{H9}
\ec
}
\parbox{3.2cm}{\bc
\begin{fmfgraph*}(2.5,2.5)
 \stargl
 \starplace{\nx\G}{157}{\nx\G}{65}{\nx\G}{25}{\nx\G}{-14}
 \starplacedummy{}{}{}{}{}{}{}{}
 \startypes{boson}{}{boson}{}{boson}{}
 \startypec{boson}{}{boson}{}{boson}{}
 \starproj{}{}{}{}{}{}
\end{fmfgraph*}\\
\refstepcounter{diagh}
(\Alph{diagh})\label{I9}
\ec
}
\parbox{3.2cm}{\bc
 \begin{fmfgraph*}(2.5,2.5)
   \selfgl
   \selfplace{\nx\G}{100}{\nx\G}{-160}{\nx\G}{-80}{\nx\G}{22}
   \selfplacedummy{}{}{}{}{}{}{}{}
   \selftypeb{boson}{}{boson}{}{boson}{}{boson}{}
   \selftypes{boson}{}{boson}{}
   \selfproj{}{}{}{}{}{}
 \end{fmfgraph*}\\
\refstepcounter{diagh}
(\Alph{diagh})\label{J9}
\ec
}
\parbox{3.2cm}{\bc
\begin{fmfgraph*}(2.5,2.5)
 \stargl
 \starplace{}{}{}{}{}{}{}{}
 \starplacedummy{}{}{}{}{}{}{}{}
 \startypes{fermion}{}{fermion}{}{fermion}{}
 \startypec{boson}{}{boson}{}{boson}{}
 \starproj{}{}{}{}{}{}
 \fmfdot{v1,v2,v3,v4}
\end{fmfgraph*}\\
\refstepcounter{diagh}
(\Alph{diagh})\label{H9f}
\ec
}\nonumber\\
\parbox{3.2cm}{\bc
\begin{fmfgraph*}(2.5,2.5)
 \stargl
 \starplace{}{}{}{}{}{}{}{}
 \starplacedummy{}{}{}{}{}{}{}{}
 \startypes{fermion}{}{fermion}{}{boson}{}
 \startypec{boson}{}{fermion}{}{fermion}{}
 \starproj{}{}{}{}{}{}
 \fmfdot{v1,v2,v3,v4}
\end{fmfgraph*}\\
\refstepcounter{diagh}
(\Alph{diagh})\label{I9f}
\ec
}
\parbox{3.2cm}{\bc
 \begin{fmfgraph*}(2.5,2.5)
   \selfgl
   \selfplace{}{}{}{}{}{}{}{}
   \selfplacedummy{}{}{}{}{}{}{}{}
   \selftypeb{fermion}{}{boson}{}{fermion}{}{boson}{}
   \selftypes{fermion}{}{fermion}{}
   \selfproj{}{}{}{}{}{}
   \fmfdot{v1,v2,v3,v4}
 \end{fmfgraph*}\\
\refstepcounter{diagh}
(\Alph{diagh})\label{J9f}
\ec
}
\parbox{3.2cm}{\bc
\begin{fmfgraph*}(2.5,2.5)
 \pacgl
 \pacplace{}{}{}{}{}{}
 \pacplacedummy{}{}{}{}{}{}
 \pactypes{boson}{}{boson}{}{boson}{}
 \pactypec{boson}{}{boson}{}
 \pacproj{}{}{}{}{}
 \fmfdot{v1,v2,v3}
\end{fmfgraph*}\\
\refstepcounter{diagh}
(\Alph{diagh})\label{K9}
\ec
}
\parbox{2.5cm}{\bc
\begin{fmfgraph*}(2.5,2.5)
 \fmfforce{(.5w,.9h)}{v1}
 \fmfforce{(.5w,.1h)}{v2}
 \fmfblob{0.2w}{v1,v2}
 \fmf{boson,left}{v2,v1}
 \fmf{boson,left}{v1,v2}
\end{fmfgraph*}\\
\refstepcounter{diagh}
(\Alph{diagh})\label{L9}
\ec
}
\parbox{3.2cm}{\bc
 \begin{fmfgraph*}(2.5,2.5)
  \fmfforce{(0,.5h)}{v1}
  \fmfforce{(w,.5h)}{v2}
  \fmf{boson,left}{v1,v2,v1}
  \fmf{boson,left=.414}{v1,v2,v1}
  \fmfdot{v1,v2}
 \end{fmfgraph*}\\
\refstepcounter{diagh}
(\Alph{diagh})\label{M9}
\ec
}
.\label{alldia}
\ee
Here, the Feynman rules of section \ref{sec3.2} are understood, and the 
blob is defined in terms of the self-energy of (\ref{self}) as

\be
\parbox{3cm}{\bc
 \begin{fmfgraph}(2.5,2)
  \fmfincoming{i1}
  \fmfoutgoing{o1}
  \fmf{boson}{i1,v1,o1}
  \fmfblob{0.2w}{v1}
 \end{fmfgraph}\ec
}
= -\;\mbox{\large $\Pi_{\mu\nu}$}\; +
\parbox{3cm}{\bc
 \begin{fmfgraph}(2.5,2)
  \fmfincoming{i1}
  \fmfoutgoing{o1}
  \fmf{boson}{i1,o1}
  \fmf{crossed}{i1,o1}
 \end{fmfgraph}\ec
}.
\label{self2}
\ee

The complete {$\cal O$}({\em g}$\,^3$) result of these contributions has been 
obtained by Kapusta \cite{kap79} nearly two decades ago. It is given via the 
resummed two-loop calculation including the diagrams 
(\ref{alldia}.\refmy{C9}-\refmy{F9f}), and was obtained in \cite{kap79} in 
Feynman gauge. Only recently, Arnold and Zhai have calculated the 
{$\cal O$}({\em g}$\,^4$) contributions \cite{arn94,arn95}, being able to 
reduce all occurring loop integrals to standard ones and to the so-called 
scalar basketball diagram. At this order all mass terms can be neglected in the
3-loop diagrams. With essentially the same methods but including the mass terms
in the three-loop diagrams, the calculation was pushed to 
{$\cal O$}({\em g}$\,^5$) by Zhai and Kastening \cite{zha95}. Beyond 
{$\cal O$}({\em g}$\,^5$), simple resummation based on (\ref{lagrangien}) is 
known to lead to infrared divergences. This is the Linde argument 
\cite{lin80,kap89} that diagrams with arbitrarily high numbers of loops all 
contribute to {$\cal O$}({\em g}$\,^6$). However it is possible to organize 
the 6th order contribution into the two terms $c_1${\em g}$\,^6$ and 
$c_2${\em g}$\,^6\ln g$ as it is done in lattice calculations \cite{kar96}; 
then other resummation techniques \cite{bra95} may give access to $c_2$, while
$c_{1}$ must be computed numerically. For a more extensive review of the 
resummation techniques in order to calculate quantities as 
e.g.~$Z_{\text{QCD}}$, we refer to \cite{tho96}.

\subsection{From Two Loop Vacuum Diagrams to Free Energy at 
{$\cal O$}({\em g}$\,^3$)}
\label{3.4}

The gauge invariance of diagrams (\ref{alldia}.\refmy{D9}-\refmy{F9f}) for
the vacuum case has already been discussed in Sec.~\ref{sec2}.
Since the decomposition rule (\ref{main}) has changed to
(\ref{mainm}), additional diagrams appear, namely

\be
-\quad
\parbox{2.4cm}{\bc
 \begin{fmfgraph*}(2,2)
  \fmfforce{(0,.5h)}{v1}
  \fmfforce{(w,.5h)}{v2}
  \fmf{boson,left}{v2,v1}
  \fmf{boson}{v1,v2}
  \fmfv{label=$\nx\Delta$, l.a=-155, l.d=0.05w}{v1}
  \fmf{crossed,left}{v1,v2}
  \fmf{boson,left,label=$\nx\star$,l.d=0.1w,l.s=left}{v1,v2}
  \fmfdot{v1}
 \end{fmfgraph*}
\ec
}\qquad
\mbox{and} \quad\;\;\;
+\;\;
\parbox{2.4cm}{\bc
 \begin{fmfgraph*}(2,2)
  \fmfforce{(0,.5h)}{v1}
  \fmfforce{(w,.5h)}{v2}
  \fmf{boson,left}{v2,v1}
  \fmf{boson}{v1,v2}
  \fmfv{label=$\nx\Delta$,label.a=-155,label.d=0.05w}{v1}
  \fmfv{label=\nx\G,label.a=155,label.d=0.15w}{v2}
  \fmf{boson}{v1,v2}
  \fmf{crossed,left}{v1,v2}
  \fmf{boson,left,label=$\nx\star$,l.d=0.1w,l.s=left}{v1,v2}
 \end{fmfgraph*}
\ec
}\; .
\label{restdiag}
\ee

Both diagrams are of {$\cal O$}({\em g}$\,^4$) and, as we will see in section 
\ref{sec5}, cancel diagrams stemming from {$\cal O$}({\em g}$\,^4$) counter 
terms. Clearly these diagrams should be neglected in a calculation which is 
only accurate to {$\cal O$}({\em g}$\,^3$). Hence the obtained result for 
$Z_{\text{QCD}}$ up to {$\cal O$}({\em g}$\,^3$) is $\alpha$-independent.

\section{About the problem of shifting momenta}
\label{sec4}

In the previous Sections we have shown how powerful the diagrammatic method of
Cheng and Tsai is. However, we have not discussed one crucial technicality, 
which is the subject of the present Section.
Namely if various diagrams are related to each other (like in (\ref{shift})), 
shifts in the momenta have to be performed before
the diagrams are really algebraically identical. This requires that the used
propagators are not too singular. Otherwise shifting momenta might change the
result of a loop integral and thus diagrams like (\ref{identity1}.\refmy{A3})
and (\ref{identity1}.\refmy{B3}) would {\it not} cancel each other. 
Therefore naively using the Cheng-Tsai method and ignoring this 
problem can lead to serious inconsistencies. Indeed, one of the major results 
of the original Cheng/Tsai work was to point out the subtlety that the 
widely used naive  versions of the different sets of temporal and Coulomb 
gauge Feynman rules are inconsistent with basic principles. This can be traced
back to inappropriate shifts of loop momenta \cite{cheMIT,che86a}. 

First we recall that momentum shifts are 
necessary to connect the gauge dependent parts of different diagrams
(c.f.~(\ref{shift}), (\ref{identity1})).
As long as the involved propagators have no singularities shifting momenta in
loop integrals causes no trouble. This might change if the propagators
have poles. To see this we study the temporal gauge with its gauge condition

\be
A_0 = 0  \,.
\ee
Naively one expects the ghost fields to decouple and the gluon propagator to be

\be
D^{ij}(k) &=&
-\frac{i}{k^2 + i\varepsilon} \left( g^{ij}+{k^i k^j \over \vec k^2} \right)
+ {i\over k_0^2} {k^i k^j \over \vec k^2} \\
D^{\mu 0}(k) &=& D^{0\mu}(k) = 0  
\ee
This is of course a special case of the general form (\ref{prop}) with the
choice 

\be
a_\mu(k) = + {g_{\mu 0}\over k_0} - {k_\mu \over 2 k_0^2}  \,.
\ee

However the spatially longitudinal part of the free temporal 
propagator has a double pole at $k_0=0$ making momentum shifts 
difficult to control.
To perform loop calculations one must introduce a pole prescription which 
obviously changes $a_\mu$. We choose

\be
a_\mu(k) = + g_{\mu 0}{1\over 2}
\left[ {1\over k_0+i\eta} +{1\over k_0-i\eta}  \right]
- k_\mu {1\over 4}
\left[ {1\over (k_0+i\eta)^2} +{1\over (k_0-i\eta)^2}  \right].
\label{hawebe}
\ee
Thus the longitudinal part of the gauge propagator becomes

\be
D_L(k) &:=& {k_i k_j \over \vec k^2} D_{ij}(k) 
\nonumber \\ &=& 
{i\over 2} 
\left[ {1\over (k_0+i\eta)^2} 
        \left( 1+{2ik_0 \eta -\eta^2 \over k^2 +i\epsilon} \right)
      +{1\over (k_0-i\eta)^2}
        \left( 1+{-2ik_0 \eta -\eta^2 \over k^2 +i\epsilon} \right)
\right]
\nonumber \\ &=& 
{i\over 2}      
\left[ {1\over (k_0+i\eta)^2} +{1\over (k_0-i\eta)^2} \right] + {\cal O}(\eta)  \,.
\label{longmode}
\ee

Note that this yields the principal value prescription of the 
$k_0$-pole, which was frequently used for loop calculations in temporal 
gauge until it turned out that the results disagree with Feynman gauge 
calculation \cite{car82}. It reintroduces the ghost field

\be
G^{abc}_\mu = +i\mbox{\em{g}} f^{abc}
{\eta^2 k_\mu + g_{\mu 0} k^2 k_0 \over (k^2+i\epsilon)(k_0^2 + \eta^2)} \, 
\label{ghosttemp}
\ee
and the temporal mode

\be
D_{00} = -{i\over k^2+i\epsilon}
\left[-{\eta^2\over 2(k_0+i\eta)^2} 
      -{\eta^2\over 2(k_0-i\eta)^2} \right].
\label{tempmode}
\ee
Note that the expressions (\ref{longmode}), (\ref{ghosttemp}) and 
(\ref{tempmode}) are easily calculated by inserting (\ref{hawebe}) in 
(\ref{prop}) and (\ref{ghost}). 

Naively one expects that all the contributions involving ghosts and/or
temporal gauge modes vanish for $\eta \to 0$ due to the appearance of
$\eta^2$ in the numerators of (\ref{ghosttemp}) and (\ref{tempmode}).
But this is not true as we will show now.
Let us concentrate on a two-loop vacuum diagram with two longitudinal (L) and
one temporal (O) propagator, as shown in fig.~\ref{bldl2} \cite{che86a}.\\ 

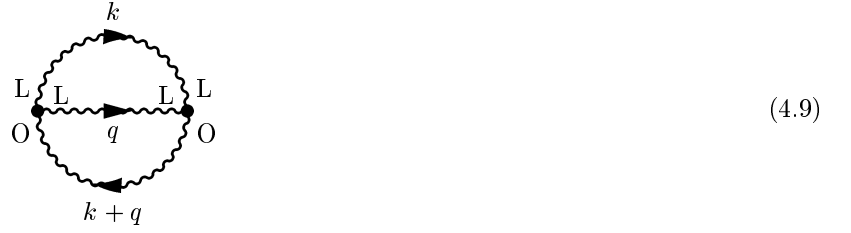
\be
\parbox{2.4cm}{\bc
 \begin{fmfgraph*}(2,2)
  \fmfforce{(0,.5h)}{v1}
  \fmfforce{(w,.5h)}{v2}
  \fmfforce{(0,.5h)}{v3}
  \fmfforce{(w,.5h)}{v4}
  \fmfforce{(0,.5h)}{v5}
  \fmfforce{(w,.5h)}{v6}
  \fmf{phantom_arrow,left}{v3,v4}
  \fmf{phantom_arrow,left}{v4,v3}
  \fmf{phantom_arrow}{v3,v4}
  \fmf{boson,left,label=$k$}{v1,v2}
  \fmf{boson,left,label=$k+q$}{v2,v1}
  \fmf{boson,label=$q$}{v1,v2}
  \fmfdot{v1,v2}
  \fmfv{l=L,l.a=120}{v1}
  \fmfv{l=L,l.a=25}{v3}
  \fmfv{l=O,l.a=-120}{v5}
  \fmfv{l=L,l.a=60}{v2}
  \fmfv{l=L,l.a=155}{v4}
  \fmfv{l=O,l.a=-60}{v6}
 \end{fmfgraph*}\\
\ec}
\label{bldl2}
\ee
\vspace*{.5ex}

If $\eta$ is taken to zero {\em before} the loop integrations are performed 
there would be no contribution from this diagram since the temporal part 
(\ref{tempmode}) of the gluon propagator vanishes in this limit. But $\eta$ 
serves as a regulator for the $k_0$-pole in (\ref{longmode}) and should be 
taken to zero only {\em after} the energy integrations are performed. To check
whether there is a contribution in this limit we restrict ourselves to the 
leading term in $\eta$, i.e. we take into account only the first contribution 
of (\ref{longmode}). Thus we have to calculate (c.f.~\ref{bldl2})

\be
\lefteqn{ \int\!\! dk_0\,dp_0 \, (k-p)_0 g^{ii'} (k-p)_0 g^{jj'} 
{k_i k_j \over \vec k^2} D_L(k)  {p_{i'} p_{j'} \over \vec p^2} D_L(p)
D_{00}(k+p) }
\nonumber \\ 
&\sim & \int\!\! dk_0\,dp_0 \, (k_0-p_0)^2
\left[ {1\over (k_0+i\eta)^2} +{1\over (k_0-i\eta)^2} \right]
\left[ {1\over (p_0+i\eta)^2} +{1\over (p_0-i\eta)^2} \right]
\nonumber \\ && \times
{1\over (k+p)^2+i\epsilon}
\left[{\eta^2\over (k_0+p_0+i\eta)^2} 
      +{\eta^2\over (k_0+p_0-i\eta)^2} \right]  \,.
\ee
By scaling $k_0=\kappa_0 \eta$, $p_0 = \pi_0 \eta$ we get

\be
\lefteqn{\int\!\! d\kappa_0\,d\pi_0 \, (\kappa_0-\pi_0)^2
\left[ {1\over (\kappa_0+i)^2} +{1\over (\kappa_0-i)^2} \right]
\left[ {1\over (\pi_0+i)^2} +{1\over (\pi_0-i)^2} \right] }
\nonumber \\ &&\times
{1\over \eta^2(\kappa+\pi)^2-(\vec k+\vec p)^2+i\epsilon}
\left[{1\over (\kappa_0+\pi_0+i)^2} 
      +{1\over (\kappa_0+\pi_0-i)^2} \right] 
\nonumber \\
& \stackrel{\eta \to 0}{\to}& {-1 \over (\vec k+\vec p)^2} 
\int\!\! d\kappa_0\,d\pi_0 \, (\kappa_0-\pi_0)^2
\left[ {1\over (\kappa_0+i)^2} +{1\over (\kappa_0-i)^2} \right]
\left[ {1\over (\pi_0+i)^2} +{1\over (\pi_0-i)^2} \right]
\nonumber \\ && \times 
\left[{1\over (\kappa_0+\pi_0+i)^2} 
      +{1\over (\kappa_0+\pi_0-i)^2} \right] \\
& & =  \frac{16}{9}\pi^2 {1 \over (\vec k+\vec p)^2}\nonumber
\ee
which might be calculated with the contour method 
and does not vanish in contrast to the naive expectation that temporal modes
should not contribute in calculations carried out in the temporal gauge. 
Whether temporal gauge modes and/or ghost
fields yield non-vanishing contributions to observable quantities strongly
depends on the pole prescription one uses for the $k_0$-pole in the 
longitudinal part of the gauge propagator.

This shows that infrared singularities in the propagator must be handled with
extreme care. In principle the same holds true for UV-singularities caused by
propagators which do not vanish at infinity. A prominent example is the 
temporal mode of the Coulomb propagator \cite{che86a}

\be
D^{00}(k) = {i\over \left.\vec{k}\right.^2}  \to \quad\mbox{finite} \qquad
\mbox{for} \quad k_0 \to \infty .
\ee

In the case at hand we should worry about infrared problems since they
appear in the naive perturbation theory at finite temperature and cause the
necessity of resummation. Indeed in our decomposition (c.f. (\ref{mainm}))

\be
-i\mbox{\em{g}}f^{abc}\left( g_{\mu\nu} - {k_\mu k_\nu \over k^2} \right) 
   D_{\rm resum}^{\nu\rho,cd}(k)
= 
-i\mbox{\em{g}}f^{abd}\frac{1}{k^2}g_{\mu\nu}
+i\mbox{\em{g}}f^{abd}
 \delta_{\mu 0}\delta_{\nu 0}\delta_{k_0 0}\frac{m^2}{k^2\left(m^2+k^2\right)}
- G_{\mu}^{abd}(k)\frac{k_\nu}{k^2} 
\label{main2}
\ee
the terms at the r.h.s.~have IR singularities. As we have shown in the 
previous sections the three terms are treated differently in further 
diagrammatic manipulations, e.g.~the necessary momentum shifts for the first 
term of the r.h.s. of (\ref{main2}) is different from the ones for the third 
term. Thus loop integrals over the sum on the r.h.s. of (\ref{main2}) which 
are well-defined since the sum is IR safe have to be split in sums over loop 
integrals which might cause trouble. Fortunately these considerations are 
purely academic since the transverse projector 
$\left( g_{\mu\nu} - {k_\mu k_\nu \over k^2} \right)$ in (\ref{main2}) is 
{\em always} accompanied by a factor $k^2$ which makes all terms at the 
r.h.s.~of (\ref{main2}) IR save (c.f.~(\ref{mainm})). Thus in our calculations
performed in the previous sections no IR singular terms ever appear. Of course
this holds only true for covariant $\alpha$-gauges. Once axial gauges or the 
Coulomb gauge are used the considerations of this section might become 
relevant again.\\
Now we turn to the three loop diagrams and prove the gauge invariance 
(more precisely the $\alpha$-independence) of the QCD partition function as 
calculated up to {$\cal O$}({\em g}$\,^4$) using the resummation scheme 
outlined in subsection \ref{sec3.1}.

%%%%%%%%%%%%%%%%%%%%%%%%%%%%%%%%%%%%%%%%%%%%%%%%%%%%%%%%%%%%%%%%%%%%%%
\section{Gauge invariance of $Z_{\text{QCD}}$ at 
{$\cal O$}(\lowercase{{\em g}}$\,^{4}$)}
\label{sec5}

%%%%%%%%%%%%%%%%%%%%%%%%%%%%%%%%%%%%%%%%%%%%%%%%%%%%%%%%%%%%%%%%%%%%%%

In this Section we prove the gauge invariance, more precisely the 
$\alpha$-independence of the {$\cal O$}({\em g}$\,^4$)
contributions to the partition function. Decomposing the corresponding
diagrams of (\ref{alldia}) with the help of the extended Cheng-Tsai rule
(\ref{maindiagram2}), we obtain two different sets of diagrams:
\begin{enumerate}
 \item[I.]diagrams with the same graphical representation as the vacuum
       contributions but which have an implicit mass dependence stemming
       solely from the massive static gluon propagator $D_{F}'$ in
       (\ref{gluonm}).
 \item[II.] diagrams which contain explicit mass terms, i.e., they contain 
       ``crosses'' or ``cross-stars''. These again split into two sets:
       \begin{enumerate}
         \item diagrams in which the crosses arise from the mass counterterms
               (\ref{counter}/\ref{self2}) to the three-loop diagrams.
         \item diagrams in which the crosses arise from the additional terms
               in the identity (\ref{alldia}) compared to (\ref{maindiagram}),
               where an arrow pointing to a three-gluon vertex is decomposed.
       \end{enumerate}
\end{enumerate}

To check the gauge-invariance of the set of diagrams I, one follows closely
that of the vacuum case. It is a straightforward application of the original
Cheng-Tsai rules discussed in Sec.~\ref{sec2}. The most important technical
steps are given in Appendix \ref{secapa}. This leaves us with the 
gauge-invariance check of the set II. Here, we analyse in a first Subsection
the structure of IIa and we find that the $\alpha$-dependence of this
set is given by the diagrams (\ref{massg5}.\refmy{E22}-\refmy{G22}) below. In 
a second subsection, we turn then to the set IIb showing that its 
{$\cal O$}({\em g}$\,^4$) contribution cancels exactly the remaining terms
of IIa.

\subsection{Mass Counterterms to Three-Loop}
\label{sec5.1}

We discuss now the two-loop diagrams of the perturbative expansion 
(\ref{alldia}) of Z$_{\text{QCD}}$ which occur as counterterms to the
three-loop contribution:

\be
\raisebox{-1.5ex}{
\parbox{2.8cm}{\bc
 \begin{fmfgraph}(2.27,2.27)
 \fmfforce{(.5w,0)}{v3}
 \fmfforce{(0,.5h)}{v2}
 \fmfforce{(w,.5h)}{v1}
 \fmfblob{0.2w}{v3}
 \fmf{boson,right=.5}{v2,v3,v1}
 \fmf{boson,right}{v1,v2}
 \fmf{crossed,right}{v1,v2}
\end{fmfgraph}\\[.22cm]
\stepcounter{rueck}
\refstepcounter{diagh}
(\Alph{diagh})\label{A21}
\ec
}}
& = & +\;\frac{1}{2} 
\raisebox{-1.5ex}{
\parbox{3cm}{\bc
 \begin{fmfgraph}(2.5,2.5)
  \selfgl
  \fmf{boson,right=0.414}{v3,v4,v3}
  \fmf{boson,right=0.414}{v4,v1,v2,v3}
  \fmf{crossed,right=.414}{v1,v2}
  \fmfdot{v3,v4}
 \end{fmfgraph}\\
\refstepcounter{diagh}
(\Alph{diagh})\label{B21}
\ec
}}
-
\raisebox{-1.5ex}{
\parbox{3cm}{\bc
 \begin{fmfgraph*}(2.5,2.5)
  \selfgl
  \fmf{boson,right=0.414}{v3,v4,v3}
  \fmf{boson,right=0.414}{v4,v1,v2,v3}
  \fmf{crossed,right=.414}{v1,v2}
  \fmfv{label=\nx\G,label.a=-70}{v3}
  \fmfv{label=\nx\G,label.a=110}{v4}
 \end{fmfgraph*}\\
\refstepcounter{diagh}
(\Alph{diagh})\label{C21}
\ec
}} +\;\frac{1}{2}
\raisebox{-1.5ex}{
\parbox{2.5cm}{\bc 
 \begin{fmfgraph}(2.5,2.5)
  \fmfforce{(.5w,.250h)}{v2}
  \fmfforce{(.125w,.625h)}{v1}
  \fmfforce{(.875w,.625h)}{v3}
  \fmfforce{(.5w,0)}{v4}
  \fmf{boson,right}{v3,v1}
  \fmf{crossed,right}{v3,v1}
  \fmf{boson,right=.5}{v1,v2,v3}
  \fmf{boson,right}{v2,v4,v2}
  \fmfdot{v2}
 \end{fmfgraph}\\
\refstepcounter{diagh}
(\Alph{diagh})\label{D21}
\ec
}}\nonumber \\ 
& & -
\raisebox{-1.5ex}{
\parbox{3cm}{\bc
 \begin{fmfgraph}(2.5,2.5)
  \selfgl
  \fmf{fermion,right=0.414}{v3,v4,v3}
  \fmf{boson,right=0.414}{v4,v1,v2,v3}
  \fmf{crossed,right=.414}{v1,v2}
  \fmfdot{v3,v4}
 \end{fmfgraph}\\
\refstepcounter{diagh}
(\Alph{diagh})\label{F21}
\ec
}} +
\raisebox{-1.5ex}{
\parbox{3cm}{\bc
\begin{fmfgraph}(2.5,2.5)
 \selfgl
 \fmf{boson,right=.414}{v1,v2,v3,v4,v1}
 \fmf{crossed,right=.414}{v1,v2}
 \fmf{crossed,right=.414}{v3,v4}
\end{fmfgraph}\\
\refstepcounter{diagh}
(\Alph{diagh})\label{E21}
\ec
}}
\label{diaOg3mass2}
\ee

Diagrams (\ref{diaOg3mass2}.\refmy{C21},\refmy{F21},\refmy{E21}) contain the
$\alpha$-independent, static gluon propagator only. The  $\alpha$-independent 
part of the remaining two diagrams can be analysed with the help of
(\ref{momentumcon}) following the procedure of Sec.~\ref{sec2.1.3}. Decomposing
the $\alpha$-dependent parts of diagram (\ref{diaOg3mass2}.\refmy{B21}), we
obtain:

\be
\raisebox{-1.5ex}{
\parbox{3cm}{\bc
 \begin{fmfgraph*}(2.5,2.5)
  \selfgl
  \fmf{boson,right=0.414}{v3,v4,v3}
  \fmf{boson,right=0.414}{v4,v1,v2,v3}
  \fmf{crossed,right=.414}{v1,v2}
  \fmfv{label=$\nx\Delta$,label.a=-110}{v4}
  \fmf{projs,right=.414}{v3,v4}
  \fmfdot{v3,v4}
 \end{fmfgraph*}\\
\stepcounter{rueck}
\refstepcounter{diagh}
(\Alph{diagh})\label{A22}
\ec
}}
& = & - \;
\raisebox{0ex}{
\parbox{3cm}{\bc
 \begin{fmfgraph*}(2.5,2.5)
  \selfgl
  \fmfforce{(.1w,.4h)}{v20}
  \fmfforce{(.1w,.6h)}{v21}
  \fmf{plains,right=0.1}{v21,v20}
  \fmf{arrowr,right=0.1}{v21,v20}
  \fmf{boson,right=0.414}{v3,v4,v3}
  \fmf{boson,right=0.414}{v4,v1,v2,v3}
  \fmf{crossed,right=.414}{v1,v2}
  \fmfv{label=$\nx\Delta$,l.a=-110}{v4}
  \fmf{boson,left=.414}{v3,v2}
  \fmfdot{v4}
 \end{fmfgraph*}\\
\ec
}} 
+\;
\raisebox{0ex}{
\parbox{3cm}{\bc
 \begin{fmfgraph*}(2.5,2.5)
  \selfgl
  \fmf{boson,right=0.414}{v3,v4,v3}
  \fmf{boson,right=0.414}{v4,v1,v2,v3}
  \fmf{crossed,right=.414}{v1,v2}
  \fmfv{label=$\nx\Delta$,label.a=-110}{v4}
  \fmf{boson,right=.414,label=$\nx\star$,l.d=.05w}{v2,v3}
  \fmf{crossed,right=.414}{v2,v3}
  \fmfdot{v4}
 \end{fmfgraph*}\\
\ec
}}
+
\raisebox{0ex}{
\parbox{3cm}{\bc
 \begin{fmfgraph*}(2.5,2.5)
  \selfgl
  \fmfforce{(.4w,.192h)}{v20}
  \fmfforce{(.6w,.192h)}{v21}
  \fmf{plains,right=0.1}{v21,v20}
  \fmf{arrowr,right=0.1}{v21,v20}
  \fmf{boson,right=0.414}{v3,v4,v3}
  \fmf{boson,right=0.414}{v4,v1,v2,v3}
  \fmf{crossed,right=.414}{v1,v2}
  \fmfv{label=$\nx\Delta$,label.a=-110}{v4}
  \fmf{boson,right=.414}{v4,v3}
  \fmfdot{v4}
 \end{fmfgraph*}\\
\ec
}}\nonumber\\
& &  +
\raisebox{0ex}{
\parbox{3cm}{\bc
 \begin{fmfgraph*}(2.5,2.5)
  \selfgl
  \fmf{boson,right=0.414}{v3,v4,v3}
  \fmf{boson,right=0.414}{v4,v1,v2,v3}
  \fmf{crossed,right=.414}{v1,v2}
  \fmfv{label=$\nx\Delta$,l.a=-110}{v4}
  \fmf{projs,right=.414}{v4,v3}
  \fmfv{label=\nx\G,l.a=77,l.d=.1w}{v3}
  \fmfdot{v4}
 \end{fmfgraph*}\\
\ec
}}
-
\raisebox{0ex}{
\parbox{3cm}{\bc
 \begin{fmfgraph*}(2.5,2.5)
  \selfgl
  \fmf{boson,right=0.414}{v3,v4,v3}
  \fmf{boson,right=0.414}{v4,v1,v2,v3}
  \fmf{crossed,right=.414}{v1,v2}
  \fmfv{label=$\nx\Delta$,l.a=-110}{v4}
  \fmf{boson,right=.414,label=$\nx\star$,l.d=.05w}{v4,v3}
  \fmf{crossed,right=.414}{v4,v3}
  \fmfdot{v4}
 \end{fmfgraph*}\\[1em]
\ec
}}\nonumber \\
& = & 
+
\raisebox{-1.5ex}{
\parbox{3cm}{\bc
 \begin{fmfgraph*}(2.5,2.5)
  \selfgl
  \fmfforce{(.4w,.192h)}{v20}
  \fmfforce{(.6w,.192h)}{v21}
  \fmf{plains,right=0.1}{v21,v20}
  \fmf{arrowr,right=0.1}{v21,v20}
  \fmf{boson,right=0.414}{v3,v4,v3}
  \fmf{boson,right=0.414}{v4,v1,v2,v3}
  \fmf{crossed,right=.414}{v1,v2}
  \fmfv{label=$\nx\Delta$,l.a=-110}{v4}
  \fmf{boson,right=.414}{v4,v3}
  \fmfdot{v4}
 \end{fmfgraph*}\\
\refstepcounter{diagh}
(\Alph{diagh})\label{B22}
\ec
}}
- \;
\raisebox{-1.5ex}{
\parbox{3cm}{\bc
 \begin{fmfgraph*}(2.5,2.5)
  \selfgl
  \fmfforce{(.1w,.4h)}{v20}
  \fmfforce{(.1w,.6h)}{v21}
  \fmf{plains,right=0.1}{v21,v20}
  \fmf{arrowr,right=0.1}{v21,v20}
  \fmf{boson,right=0.414}{v3,v4,v3}
  \fmf{boson,right=0.414}{v4,v1,v2,v3}
  \fmf{crossed,right=.414}{v1,v2}
  \fmfv{label=$\nx\Delta$,l.a=-110}{v4}
  \fmf{boson,left=.414}{v3,v2}
  \fmfdot{v4}
 \end{fmfgraph*}\\
\refstepcounter{diagh}
(\Alph{diagh})\label{C22}
\ec}}
-
\raisebox{-1.5ex}{
\parbox{3cm}{\bc
 \begin{fmfgraph*}(2.5,2.5)
  \selfgl
  \fmfforce{(.9w,.4h)}{v20}
  \fmfforce{(.9w,.6h)}{v21}
  \fmf{plains,left=0.1}{v21,v20}
  \fmf{arrowr,left=0.1}{v21,v20}
  \fmf{boson,right=0.414}{v3,v4,v3}
  \fmf{boson,right=0.414}{v4,v1,v2,v3}
  \fmf{crossed,right=.414}{v1,v2}
  \fmf{boson,left=.414}{v1,v4}
 \fmfv{label=$\nx\Delta$,l.a=-110}{v4}
  \fmfv{label=\nx\G,l.a=77,l.d=.1w}{v3}
 \end{fmfgraph*}\\
\refstepcounter{diagh}
(\Alph{diagh})\label{D22}
\ec}}\nonumber\\
& & + \;
\raisebox{-1.5ex}{
\parbox{3cm}{\bc
 \begin{fmfgraph*}(2.5,2.5)
  \selfgl
  \fmf{boson,right=0.414}{v3,v4,v3}
  \fmf{boson,right=0.414}{v4,v1,v2,v3}
  \fmf{crossed,right=0.414}{v1,v2}
  \fmfv{label=$\nx\Delta$,l.a=-110}{v4}
  \fmf{boson,right=.414,label=$\nx\star$,l.d=.05w}{v2,v3}
  \fmf{crossed,right=.414}{v2,v3}
  \fmfdot{v4}
 \end{fmfgraph*}\\
\refstepcounter{diagh}
(\Alph{diagh})\label{E22}
\ec}}
-
\raisebox{-1.5ex}{
\parbox{3cm}{\bc
 \begin{fmfgraph*}(2.5,2.5)
  \selfgl
  \fmf{boson,right=0.414}{v3,v4,v3}
  \fmf{boson,right=0.414}{v4,v1,v2,v3}
  \fmf{crossed,right=.414}{v1,v2}
  \fmfv{label=$\nx\Delta$,l.a=-110}{v4}
  \fmf{boson,right=.414,label=$\nx\star$,l.d=.05w}{v4,v3}
  \fmf{crossed,right=.414}{v4,v3}
  \fmfdot{v4}
 \end{fmfgraph*}\\
\refstepcounter{diagh}
(\Alph{diagh})\label{F22}
\ec}}
+
\raisebox{-1.5ex}{
\parbox{3cm}{\bc
 \begin{fmfgraph*}(2.5,2.5)
  \selfgl
  \fmf{boson,right=0.414}{v3,v4,v3}
  \fmf{boson,right=0.414}{v4,v1,v2,v3}
  \fmf{crossed,right=.414}{v1,v2}
  \fmf{crossed,right=.414}{v4,v1}
  \fmf{boson,right=.414,label=$\nx\star$,l.d=.05w}{v4,v1}
  \fmfv{label=$\nx\Delta$,l.a=-110}{v4}
  \fmfv{label=\nx\G,l.a=77,l.d=.1w}{v3}
 \end{fmfgraph*}\\
\refstepcounter{diagh}
(\Alph{diagh})\label{G22}
\ec}}
\label{massg5}
\ee

This allows for the following cancellations:
\begin{enumerate}
  \item[(i)] Diagram (\ref{massg5}.\refmy{B22}) cancels the $\alpha$-dependent
        part of (\ref{diaOg3mass2}.\refmy{D21}). This is a consequence of the 
        identity (\ref{rel1}), relating a four-gluon vertex to two
        contracted  three gluon vertices.
  \item[(ii)] The diagrams (\ref{massg5}.\refmy{C22}) and 
        (\ref{massg5}.\refmy{D22}) cancel the $\alpha$-dependent 
        {$\cal O$}({\em g}$\,^4$) contributions (\ref{restdiag}) of the 
        two-loop part 
        (\ref{alldia}.\refmy{E9}-\refmy{G9}) of Z$_{\text{QCD}}$. That both
        sets of diagrams have exactly the same structure follows from the
        identity (\ref{massrelation}) which relates a contraction arrow on a
        cross to a cross-star. It remains to show that the prefactors of
        (\ref{massg5}.\refmy{C22}), (\ref{massg5}.\refmy{D22}) and
        (\ref{restdiag}) allow for cancellation.\\
        To determine these prefactors, we first specifiy the symmetry factors
        attached to each diagram. Then, we symmetrize all diagrams according
        to the arguments of Sec.~\ref{sec2.1.4}. For the three-loop 
        contributions discussed here, this amounts to six gluon propagators to
        be symmetrized. Let us start with diagram (\ref{restdiag}). Their 
        combinatorical factor is $\frac{1}{12}$ and they represent an 
        $\alpha$-dependent part of the basic diagram (\ref{grund}.\refmy{B1}).
        Symmetrizing the latter with respect to its six different gluon 
        propagators gives another factor $\frac{1}{20}$. The total prefactor 
        of (\ref{restdiag}) is 
        $4\cdot\frac{1}{12}\cdot\frac{1}{20}=\frac{1}{60}$, since always four 
        differently symmetrized versions of (\ref{grund}.\refmy{B1}) lead to 
        the same version of (\ref{restdiag}) upon decomposition, e.g.,

\be
\raisebox{-1.5ex}{
\parbox{2.4cm}{\bc
 \begin{fmfgraph*}(2,2)
  \fmfforce{(0,.5h)}{v1}
  \fmfforce{(w,.5h)}{v2}
  \fmf{boson,label=$D_3/D_4/D_5/D_6$,l.d=.075w,l.s=right,right}{v2,v1}
  \fmf{boson,label=$D_2$,l.d=.075w,l.s=left}{v1,v2}
  \fmfv{label=$\nx\Delta $,l.a=-150,l.d=.075w}{v1}
  \fmf{projs,left}{v2,v1}
  \fmf{boson,left}{v2,v1}
  \fmfdot{v1,v2}
 \end{fmfgraph*}\\
\stepcounter{rueck}
\refstepcounter{diagh}
(\Alph{diagh})\label{A28}
\ec}}
\quad
\rightarrow
\quad
\raisebox{-1.5ex}{
\parbox{2.4cm}{\bc
 \begin{fmfgraph*}(2,2)
  \fmfforce{(0,.5h)}{v1}
  \fmfforce{(w,.5h)}{v2}
  \fmf{boson,left}{v2,v1}
  \fmf{boson,label=$D_2$}{v1,v2}
  \fmfv{label=$\nx\Delta$, l.a=-155, l.d=0.05w}{v1}
  \fmf{crossed,left}{v1,v2}
  \fmf{boson,left,label=$\nx\star$,l.d=0.1w,l.s=left}{v1,v2}
  \fmfdot{v1}
 \end{fmfgraph*}
\refstepcounter{diagh}
(\Alph{diagh})\label{B28}
\ec
}}.
\ee

       On the other hand, diagram (\ref{diaOg3mass2}.\refmy{B21}) whose 
       decomposition leads to (\ref{massg5}.\refmy{C22}) and
       (\ref{massg5}.\refmy{D22}), carries a combinatorical factor
       $\frac{1}{4}$. Symmetrization with respect to the possible gluon
       propagators gives a factor $\frac{1}{15}$. Consequently, the terms 
       (\ref{massg5}.\refmy{C22}) and (\ref{massg5}.\refmy{D22}) have a total
       prefactor $\frac{1}{60}$, too, and cancel the contribution from
       (\ref{restdiag}). We are left with the diagrams
       (\ref{massg5}.\refmy{E22}-\refmy{G22}) which must be treated together
       with the set of diagrams IIb.
\end{enumerate}

\subsection{Three-Loop Diagrams with Explicit Mass Dependence}
\label{sec5.2}

The complete set IIb of $\alpha$-dependent three-loop contributions which
have an explicit mass dependence is given in the following
eqs.~(\ref{diamassg5}) and (\ref{diamassg4}):

\end{fmffile}

% 
% Hier ends the second extra font ``paperf2''
%
%%%%%%%%%%%%%%%%%%%%%%%%%%%%%%%%%%%%%%%%%%%%%%%%%%%%%%%%%%%%%%%%%%%%%%%
%                                                                     %
% Starting the 3.font-file and adding additional Metafont-Definitions %
%                                                                     %
%%%%%%%%%%%%%%%%%%%%%%%%%%%%%%%%%%%%%%%%%%%%%%%%%%%%%%%%%%%%%%%%%%%%%%%

\begin{fmffile}{paperf3}

%%%%%%%%%%%%%%%%%%%%%%%%%%%%%%%%%%%%%%%%%%%%%%%%%%%%%%%%%%%%%%%%%%%%%%%
%                                                                     %
%       Some Metafont-definitions in addition to feynmf.mf            %
%                                                                     %
%%%%%%%%%%%%%%%%%%%%%%%%%%%%%%%%%%%%%%%%%%%%%%%%%%%%%%%%%%%%%%%%%%%%%%%
%
% Variables:
%

\fmfcurved
\fmfpen{thin} 
\fmfset{curly_len}{3mm}
\fmfset{wiggly_len}{2mm}

%
% Styles:
%

\fmfcmd{ %
 vardef cross_bar (expr p, len, ang) =
  ((-len/2,0)--(len/2,0))
    rotated (ang + angle direction length(p)/2 of p)
    shifted point length(p)/2 of p
 enddef;
 style_def crossed expr p = 
   ccutdraw cross_bar (p, 5mm, 45);
   ccutdraw cross_bar (p, 5mm, -45)
 enddef;}

\fmfcmd{ %
 vardef proj_tarrow (expr p, frac) =
  save a, t, z;
  pair z;
  t1 = frac*length p;
  a = angle direction t1 of p;
  z = point t1 of p;
  (t2,whatever) = p intersectiontimes
    (halfcircle scaled 3arrow_len rotated (a-90) shifted z);
  arrow_head (p, t1, t2, arrow_ang)
enddef;
 style_def projs expr p  =
    cfill (proj_tarrow (reverse p, 0.85));
 enddef;
 style_def projc expr p =
    cfill (proj_tarrow (reverse p, 0.7));
 enddef;}

\fmfcmd{ 
 vardef projl_tarrow (expr p, frac) =
  save a, t, z;
  pair z;
  t1 = frac*length p;
  a = angle direction t1 of p;
  z = point t1 of p;
  (t2,whatever) = p intersectiontimes
    (halfcircle scaled 2arrow_len rotated (a-90) shifted z);
  arrow_head (p, t1, t2, arrow_ang)
 enddef;
 style_def arrowr expr p =
   cfill (projl_tarrow (reverse p, 0.6));
 enddef;
 style_def arrowl expr p  =
   cfill (projl_tarrow (reverse p, 0.6));
 enddef;
}

\fmfcmd{
style_def plains expr p =
 draw p;
 undraw subpath (0,.15) of p;
enddef;}

%%%%%%%%%%%%%%%%%%%%%%%%%%%%%%%%%%%%%%%%%%%%%%%%%%%%%%%%%%%%%%%%%%%%%%%
%                                                                     %
%                  Here the paper continues                           %
%                                                                     %
%%%%%%%%%%%%%%%%%%%%%%%%%%%%%%%%%%%%%%%%%%%%%%%%%%%%%%%%%%%%%%%%%%%%%%%

\be
-
\raisebox{-1.5ex}{ 
 \parbox{3cm}{\bc 
 \begin{fmfgraph*}(2.5,2.5)
  \stargl
  \starplace{}{}{}{}{}{}{$\nx\Delta$}{55}
  \starplacedummy{}{}{}{}{}{}{}{}
  \startypes{boson}{}{boson}{}{boson}{}
  \startypec{boson}{}{boson}{}{boson}{}
  \starproj{}{}{}{}{}{}
  \fmf{crossed,left=0.578,label=$\nx\star$,l.s=left,l.d=.05w}{v1,v3}
  \fmfdot{v2,v3,v4}
 \end{fmfgraph*}\\
\stepcounter{rueck}
\refstepcounter{diagh}
(\Alph{diagh})
\label{A41}
\ec
}}
-
\raisebox{-1.5ex}{
 \parbox{3cm}{\bc 
 \begin{fmfgraph*}(2.5,2.5)
  \stargl
  \starplace{\nx\G}{25}{}{}{}{}{$\nx\Delta$}{55}
  \starplacedummy{}{}{}{}{}{}{}{}
  \startypes{boson}{}{boson}{}{boson}{}
  \startypec{boson}{}{boson}{}{boson}{}
  \starproj{}{}{}{}{}{}
  \fmf{crossed,label=$\nx\star$,l.s=right,l.d=0.05w}{v3,v4}
  \fmfdot{v2,v4}
 \end{fmfgraph*}\\
\refstepcounter{diagh}
(\Alph{diagh})
\label{B41}
\ec
}}
+
\raisebox{-1.5ex}{
\parbox{3cm}{\bc 
\begin{fmfgraph*}(2.5,2.5)
 \stargl
 \starplace{\nx\G}{25}{}{}{}{}{$\nx\Delta$}{55}
 \starplacedummy{}{}{}{}{}{}{}{}
 \startypes{boson}{}{boson}{}{boson}{}
 \startypec{boson}{}{boson}{}{boson}{}
 \starproj{}{}{}{}{}{}
 \fmf{crossed,left=0.578,label=$\nx\star$,l.s=right,l.d=0.05w}{v3,v2}
 \fmfdot{v2,v4}
\end{fmfgraph*}\\
\refstepcounter{diagh}
(\Alph{diagh})
\label{C41}
\ec
}}
+
\raisebox{-1.5ex}{
\parbox{3cm}{\bc 
\begin{fmfgraph*}(2.5,2.5)
 \stargl
 \starplace{\nx\G}{25}{}{}{\nx\G}{115}{$\nx\Delta$}{55}
 \starplacedummy{}{}{}{}{}{}{}{}
 \startypes{boson}{}{boson}{}{boson}{}
 \startypec{boson}{}{boson}{}{boson}{}
 \starproj{}{}{}{}{}{}
 \fmf{crossed,label=$\nx\star$,l.s=right,l.d=0.05w}{v4,v2}
 \fmfdot{v2}
\end{fmfgraph*}\\
\refstepcounter{diagh}
(\Alph{diagh})
\label{D41}
\ec
}}\nonumber\\
+
\raisebox{-1.5ex}{
\parbox{3cm}{\bc 
\begin{fmfgraph*}(2.5,2.5)
 \stargl
 \starplace{\nx\G}{25}{}{}{\nx\G}{115}{$\nx\Delta$}{55}
 \starplacedummy{}{}{}{}{}{}{\nx\G}{-112}
 \startypes{boson}{}{boson}{}{boson}{}
 \startypec{boson}{}{boson}{}{boson}{}
 \starproj{}{}{}{}{}{}
 \fmf{crossed,left=0.6,label=$\nx\star$,l.s=left,l.d=0.05w}{v2,v1}
\end{fmfgraph*}\\
\refstepcounter{diagh}
(\Alph{diagh})
\label{E41}
\ec
}}
-
\raisebox{-1.5ex}{
\parbox{3cm}{\bc 
\begin{fmfgraph*}(2.5,2.5)
 \stargl
 \starplace{\nx\G}{25}{}{}{\nx\G}{115}{$\nx\Delta$}{55}
 \starplacedummy{}{}{}{}{}{}{\nx\G}{-112}
 \startypes{boson}{}{boson}{}{boson}{}
 \startypec{boson}{}{boson}{}{boson}{}
 \starproj{}{}{}{}{}{}
 \fmf{crossed,left=0.578,label=$\nx\star$,l.s=right,l.d=0.05w}{v3,v2}
\end{fmfgraph*}\\
\refstepcounter{diagh}
(\Alph{diagh})
\label{F41}
\ec
}}
+
\raisebox{-1.5ex}{
\parbox{3cm}{\bc 
\begin{fmfgraph*}(2.5,2.5)
 \stargl
 \starplace{\nx\G}{25}{}{}{\nx\G}{-90}{$\nx\Delta$}{55}
 \starplacedummy{}{}{}{}{}{}{}{}
 \startypes{boson}{}{boson}{}{boson}{}
 \startypec{boson}{}{boson}{}{boson}{}
 \starproj{}{}{}{}{}{}
 \fmf{crossed,label=$\nx\star$,l.s=right,l.d=0.05w}{v4,v2} 
\end{fmfgraph*}\\
\refstepcounter{diagh}
(\Alph{diagh})
\label{G41}
\ec
}}
-
\raisebox{-1.5ex}{
\parbox{3cm}{\bc 
\begin{fmfgraph*}(2.5,2.5)
 \stargl
 \starplace{\nx\G}{25}{}{}{\nx\G}{-90}{$\nx\Delta$}{55}
 \starplacedummy{}{}{}{}{}{}{}{}
 \startypes{boson}{}{boson}{}{boson}{}
 \startypec{boson}{}{boson}{}{boson}{}
 \starproj{}{}{}{}{}{}
 \fmf{crossed,left=0.578,label=$\nx\star$,l.s=left,l.d=0.05w}{v2,v1}
\end{fmfgraph*}\\
\refstepcounter{diagh}
(\Alph{diagh})
\label{H41}
\ec
}}\nonumber \\
+
\raisebox{-1.5ex}{
\parbox{3cm}{\bc 
\begin{fmfgraph*}(2.5,2.5)
 \stargl
 \starplace{\nx\G}{25}{\nx\G}{10}{\nx\G}{-90}{$\nx\Delta$}{55}
 \starplacedummy{}{}{}{}{}{}{}{}
 \startypes{boson}{}{boson}{}{boson}{}
 \startypec{boson}{}{boson}{}{boson}{}
 \starproj{}{}{}{}{}{}
 \fmf{crossed,label=$\nx\star$,l.s=right,l.d=0.05w}{v3,v4}
\end{fmfgraph*}\\
\refstepcounter{diagh}
(\Alph{diagh})
\label{I41}
\ec
}}
+
\raisebox{-1.5ex}{
\parbox{3cm}{\bc
\begin{fmfgraph*}(2.5,2.5)
 \stargl
 \starplace{\nx\G}{25}{\nx\G}{157}{\nx\G}{-90}{}{}
 \starplacedummy{$\nx\Delta$}{-57}{}{}{}{}{}{}
 \startypes{boson}{}{boson}{}{boson}{}
 \startypec{boson}{}{boson}{}{boson}{}
 \starproj{}{}{}{}{}{}
 \fmf{crossed,label=$\nx\star$,l.s=right,l.d=0.05w}{v4,v2}
\end{fmfgraph*}\\
\refstepcounter{diagh}
(\Alph{diagh})
\label{J41}
\ec}}
- 
\raisebox{-1.5ex}{
\parbox{3cm}{\bc
\begin{fmfgraph*}(2.5,2.5)
 \stargl
 \starplace{\nx\G}{25}{\nx\G}{157}{\nx\G}{-90}{}{}
 \starplacedummy{$\nx\Delta$}{-57}{}{}{}{}{}{}
 \startypes{boson}{}{boson}{}{boson}{}
 \startypec{boson}{}{boson}{}{boson}{}
 \starproj{}{}{}{}{}{}
 \fmf{crossed,label=$\nx\star$,l.s=right,l.d=0.05w}{v3,v4}
\end{fmfgraph*}\\
\refstepcounter{diagh}
(\Alph{diagh})
\label{K41}
\ec
}}
+
\raisebox{-1.5ex}{
\parbox{3cm}{\bc
\begin{fmfgraph*}(2.5,2.5)
 \stargl
 \starplace{}{}{}{}{}{}{}{}
 \starplacedummy{$\nx\Delta$}{-57}{}{}{}{}{}{}
 \startypes{fermion}{}{fermion}{}{fermion}{}
 \startypec{boson}{}{boson}{}{boson}{}
 \starproj{}{}{}{}{}{}
 \fmf{crossed,label=$\nx\star$,l.s=right,l.d=0.05w}{v4,v2}
 \fmfdot{v1,v2,v3}
\end{fmfgraph*}\\
\refstepcounter{diagh}
(\Alph{diagh})
\label{J41f}
\ec}}
\nonumber\\
- 
\raisebox{-1.5ex}{
\parbox{3cm}{\bc
\begin{fmfgraph*}(2.5,2.5)
 \stargl
 \starplace{}{}{}{}{}{}{}{}
 \starplacedummy{$\nx\Delta$}{-57}{}{}{}{}{}{}
 \startypes{fermion}{}{fermion}{}{fermion}{}
 \startypec{boson}{}{boson}{}{boson}{}
 \starproj{}{}{}{}{}{}
 \fmf{crossed,label=$\nx\star$,l.s=right,l.d=0.05w}{v3,v4}
 \fmfdot{v1,v2,v3}
\end{fmfgraph*}\\
\refstepcounter{diagh}
(\Alph{diagh})
\label{K41f}
\ec
}}
-\;
\frac{1}{2}\;
\raisebox{-1.5ex}{
\parbox{3cm}{\bc
\begin{fmfgraph*}(2.5,2.5)
 \pacgl
 \pacplace{$\nx\Delta$}{115}{}{}{}{}
 \pacplacedummy{}{}{}{}{}{}
 \pactypes{boson}{}{boson}{}{boson}{}
 \pactypec{boson}{}{boson}{}
 \pacproj{}{}{}{}{}
 \fmf{crossed,label=$\nx\star$,l.s=left,l.d=0.05w}{v1,v3}
 \fmfdot{v1,v2}
\end{fmfgraph*}\\
\refstepcounter{diagh}
(\Alph{diagh})
\label{L41}
\ec}}
+\;
\frac{1}{2}\;
\raisebox{-1.5ex}{
\parbox{3cm}{\bc
\begin{fmfgraph*}(2.5,2.5)
 \pacgl
 \pacplace{$\nx\Delta$}{115}{}{}{}{}
 \pacplacedummy{}{}{}{}{}{}
 \pactypes{boson}{}{boson}{}{boson}{}
 \pactypec{boson}{}{boson}{}
 \pacproj{}{}{}{}{}
 \fmf{crossed,right=0.578,label=$\nx\star$,l.s=right,l.d=0.05w}{v2,v3}
 \fmfdot{v1,v2}
\end{fmfgraph*}\\
\refstepcounter{diagh}
(\Alph{diagh})
\label{M41}
\ec
}}\;
-\;
\frac{1}{2}\;
\raisebox{-1.5ex}{
\parbox{3cm}{\bc
\begin{fmfgraph*}(2.5,2.5)
 \pacgl
 \pacplace{$\nx\Delta$}{115}{}{}{\nx\G}{-19}
 \pacplacedummy{}{}{}{}{}{}
 \pactypes{boson}{}{boson}{}{boson}{}
 \pactypec{boson}{}{boson}{}
 \pacproj{}{}{}{}{}
 \fmf{crossed,left=0.578,label=$\nx\star$,l.s=right,l.d=0.05w}{v2,v1}
 \fmfdot{v1}
\end{fmfgraph*}\\
\refstepcounter{diagh}
(\Alph{diagh})
\label{N41}
\ec
}}\nonumber\\
+\;
\frac{1}{2}\;
\raisebox{-1.5ex}{
\parbox{3cm}{\bc
\begin{fmfgraph*}(2.5,2.5)
 \pacgl
 \pacplace{$\nx\Delta$}{115}{}{}{\nx\G}{-19}
 \pacplacedummy{}{}{}{}{}{}
 \pactypes{boson}{}{boson}{}{boson}{}
 \pactypec{boson}{}{boson}{}
 \pacproj{}{}{}{}{}
 \fmf{crossed,label=$\nx\star$,l.s=left,l.d=0.05w}{v1,v2}
 \fmfdot{v1}
\end{fmfgraph*}\\
\refstepcounter{diagh}
(\Alph{diagh})
\label{O41}
\ec
}}
-\;
\frac{1}{4}
\raisebox{-1.5ex}{
\parbox{3cm}{\bc
\begin{fmfgraph*}(2.5,2.5)
 \pacgl
 \pacplace{}{}{$\nx\Delta$}{15}{}{}
 \pacplacedummy{}{}{}{}{}{}
 \pactypes{boson}{}{boson}{}{boson}{}
 \pactypec{boson}{}{boson}{}
 \pacproj{}{}{}{}{}
 \fmf{crossed,right=0.578,label=$\nx\star$,l.s=right,l.d=0.05w}{v3,v1}
 \fmfdot{v1,v2}
\end{fmfgraph*}\\
\refstepcounter{diagh}
(\Alph{diagh})
\label{S41}
\ec
}}
+\;
\frac{1}{4}
\raisebox{-1.5ex}{
\parbox{3cm}{\bc
\begin{fmfgraph*}(2.5,2.5)
 \pacgl
 \pacplace{}{}{$\nx\Delta$}{15}{}{}
 \pacplacedummy{}{}{}{}{}{}
 \pactypes{boson}{}{boson}{}{boson}{}
 \pactypec{boson}{}{boson}{}
 \pacproj{}{}{}{}{}
 \fmf{crossed,label=$\nx\star$,l.s=left,l.d=.05w}{v1,v3}
 \fmfdot{v1,v2}
\end{fmfgraph*}\\
\refstepcounter{diagh}
(\Alph{diagh})
\label{T41}
\ec
}}\label{diamassg5}
\ee
All the diagrams in (\ref{diamassg5}) are at least of 
{$\cal O$}({\em g}$\,^5$) as can be seen by splitting each loop integral into
hard {$\cal O$}(T) and soft {$\cal O$}(m) momenta.\\

In contrast, some contributions from the ``self-energy''-diagram
(\ref{alldia}.\refmy{L9}) are {$\cal O$}({\em g}$\,^4$). In the following
equation, all diagrams containing such  {$\cal O$}({\em g}$\,^4$)-parts are
labelled with letters. The remaining diagrams, being {$\cal O$}({\em g}$\,^5$)
and higher are left unlabelled

\be
&\displaystyle
-\;\frac{1}{2}
\raisebox{-1.5ex}{
\parbox{3cm}{\bc
 \begin{fmfgraph*}(2.5,2.5)
   \selfgl
   \selfplace{}{}{$\nx\Delta$}{90}{}{}{}{}
   \selfplacedummy{}{}{}{}{}{}{}{}
   \selftypeb{boson}{}{boson}{}{boson}{}{boson}{}
   \selftypes{boson}{}{boson}{}
   \selfproj{}{}{}{}{}{}
   \fmf{crossed,right=0.414,label=$\nx\star$,l.s=left,l.d=.05w}{v2,v1}
   \fmfdot{v2,v3,v4}
 \end{fmfgraph*}\\
\stepcounter{rueck}
\refstepcounter{diagh}
(\Alph{diagh})
\label{A42}
\ec
}}
+\;\frac{1}{2}
\raisebox{-1.5ex}{
\parbox{3cm}{\bc
 \begin{fmfgraph*}(2.5,2.5)
   \selfgl
   \selfplace{}{}{$\nx\Delta$}{90}{}{}{}{}
   \selfplacedummy{}{}{}{}{}{}{}{}
   \selftypeb{boson}{}{boson}{}{boson}{}{boson}{}
   \selftypes{boson}{}{boson}{}
   \selfproj{}{}{}{}{}{}
   \fmf{crossed,right=0.414,label=$\nx\star$,l.s=right,l.d=0.05w}{v4,v1}
   \fmfdot{v2,v3,v4}
 \end{fmfgraph*}\\
\refstepcounter{diagh}
(\Alph{diagh})
\label{B42}
\ec
}}\;
+\frac{1}{2}\;
\raisebox{-1.5ex}{
\parbox{3cm}{\bc
 \begin{fmfgraph*}(2.5,2.5)
   \selfgl
   \selfplace{\nx\G}{-107}{$\nx\Delta$}{90}{}{}{}{}
   \selfplacedummy{}{}{}{}{}{}{}{}
   \selftypeb{boson}{}{boson}{}{boson}{}{boson}{}
   \selftypes{boson}{}{boson}{}
   \selfproj{}{}{}{}{}{}
   \fmf{crossed,right=0.414,label=$\nx\star$,l.s=right,l.d=.05w}{v2,v3}
   \fmfdot{v3,v4}
 \end{fmfgraph*}\\
\refstepcounter{diagh}
(\Alph{diagh})
\label{C42}
\ec
}}
+
\raisebox{0ex}{
\parbox{3cm}{\bc
 \begin{fmfgraph*}(2.5,2.5)
   \selfgl
   \selfplace{\nx\G}{-107}{$\nx\Delta$}{90}{}{}{}{}
   \selfplacedummy{}{}{\nx\G}{-156}{}{}{}{}
   \selftypeb{boson}{}{boson}{}{boson}{}{boson}{}
   \selftypes{boson}{}{boson}{}
   \selfproj{}{}{}{}{}{}
   \fmf{crossed,right=0.414,label=$\nx\star$,l.s=left,l.d=0.05w}{v3,v4}
   \fmfdot{v4}
 \end{fmfgraph*}
\ec
}}&\nonumber\\
&\displaystyle+
\raisebox{0ex}{
\parbox{3cm}{\bc
 \begin{fmfgraph*}(2.5,2.5)
   \selfgl
   \selfplace{\nx\G}{-107}{$\nx\Delta$}{90}{\nx\G}{-93}{}{}
   \selfplacedummy{}{}{\nx\G}{-156}{}{}{}{}
   \selftypeb{boson}{}{boson}{}{boson}{}{boson}{}
   \selftypes{boson}{}{boson}{}
   \selfproj{}{}{}{}{}{}
   \fmf{crossed,right=0.414,label=$\nx\star$,l.s=right,l.d=0.05w}{v4,v1}
 \end{fmfgraph*}
\ec
}}\;
-
\raisebox{0ex}{
\parbox{3cm}{\bc
 \begin{fmfgraph*}(2.5,2.5)
   \selfgl
   \selfplace{\nx\G}{-107}{$\nx\Delta$}{90}{\nx\G}{-93}{}{}
   \selfplacedummy{}{}{\nx\G}{-156}{}{}{}{}
   \selftypeb{boson}{}{boson}{}{boson}{}{boson}{}
   \selftypes{boson}{}{boson}{}
   \selfproj{}{}{}{}{}{}
   \fmf{crossed,right=0.414,label=$\nx\star$,l.s=right,l.d=0.05w}{v4,v3}
 \end{fmfgraph*}
\ec
}}
+
\raisebox{0ex}{
\parbox{3cm}{\bc
 \begin{fmfgraph*}(2.5,2.5)
   \selfgl
   \selfplace{\nx\G}{-23}{$\nx\Delta$}{90}{}{}{}{}
   \selfplacedummy{}{}{}{}{}{}{}{}
   \selftypeb{boson}{}{boson}{}{boson}{}{boson}{}
   \selftypes{boson}{}{boson}{}
   \selfproj{}{}{}{}{}{}
   \fmf{crossed,right=0.414,label=$\nx\star$,l.s=right,l.d=0.05w}{v4,v3}
   \fmfdot{v2,v3}
 \end{fmfgraph*}
\ec
}}
+
\raisebox{0ex}{
\parbox{3cm}{\bc
 \begin{fmfgraph*}(2.5,2.5)
   \selfgl
   \selfplace{\nx\G}{-23}{$\nx\Delta$}{90}{}{}{\nx\G}{106}
   \selfplacedummy{}{}{}{}{}{}{}{}
   \selftypeb{boson}{}{boson}{}{boson}{}{boson}{}
   \selftypes{boson}{}{boson}{}
   \selfproj{}{}{}{}{}{}
   \fmf{crossed,right=0.414,label=$\nx\star$,l.s=right,l.d=.05w}{v2,v3}
   \fmfdot{v2}
 \end{fmfgraph*}
\ec
}}&\nonumber\\
&\displaystyle-
\raisebox{0ex}{
\parbox{3cm}{\bc
 \begin{fmfgraph*}(2.5,2.5)
   \selfgl
   \selfplace{\nx\G}{-23}{$\nx\Delta$}{90}{}{}{\nx\G}{106}
   \selfplacedummy{}{}{}{}{}{}{}{}
   \selftypeb{boson}{}{boson}{}{boson}{}{boson}{}
   \selftypes{boson}{}{boson}{}
   \selfproj{}{}{}{}{}{}
   \fmf{crossed,right=0.414,label=$\nx\star$,l.s=left,l.d=0.05w}{v3,v4}
   \fmfdot{v2}
 \end{fmfgraph*}
\ec}}
+
\raisebox{0ex}{
\parbox{3cm}{\bc
 \begin{fmfgraph*}(2.5,2.5)
   \selfgl
   \selfplace{\nx\G}{-23}{$\nx\Delta$}{90}{\nx\G}{157}{\nx\G}{106}
   \selfplacedummy{}{}{}{}{}{}{}{}
   \selftypeb{boson}{}{boson}{}{boson}{}{boson}{}
   \selftypes{boson}{}{boson}{}
   \selfproj{}{}{}{}{}{}
   \fmf{crossed,right=0.414,label=$\nx\star$,l.s=left,l.d=.05w}{v2,v1}
 \end{fmfgraph*}
\ec
}}
-\;\frac{1}{2}
\raisebox{-1.5ex}{
\parbox{3cm}{\bc
 \begin{fmfgraph*}(2.5,2.5)
   \selfgl
   \selfplace{}{}{}{}{$\nx\Delta$}{155}{}{}
   \selfplacedummy{}{}{}{}{}{}{}{}
   \selftypeb{boson}{}{boson}{}{boson}{}{boson}{}
   \selftypes{boson}{}{boson}{}
   \selfproj{}{}{}{}{}{}
   \fmf{crossed,right=0.414,label=$\nx\star$,l.s=left,l.d=.05w}{v2,v1}
   \fmfdot{v1,v3,v4}
 \end{fmfgraph*}\\
\refstepcounter{diagh}
(\Alph{diagh})
\label{D42}
\ec
}}
-
\;\frac{1}{2}
\raisebox{-1.5ex}{
\parbox{3cm}{\bc
 \begin{fmfgraph*}(2.5,2.5)
   \selfgl
   \selfplace{}{}{\nx\G}{-75}{$\nx\Delta$}{155}{}{}
   \selfplacedummy{}{}{}{}{}{}{}{}
   \selftypeb{boson}{}{boson}{}{boson}{}{boson}{}
   \selftypes{boson}{}{boson}{}
   \selfproj{}{}{}{}{}{}
   \fmf{crossed,right=0.414,label=$\nx\star$,l.s=right,l.d=0.05w}{v4,v1}
   \fmfdot{v3,v4}
 \end{fmfgraph*}\\
\refstepcounter{diagh}
(\Alph{diagh})
\label{E42}
\ec
}}&\nonumber\\
&\displaystyle+\;\frac{1}{2}
\raisebox{-1.5ex}{
\parbox{3cm}{\bc
 \begin{fmfgraph*}(2.5,2.5)
   \selfgl
   \selfplace{}{}{\nx\G}{-74}{$\nx\Delta$}{155}{}{}
   \selfplacedummy{}{}{}{}{}{}{}{}
   \selftypeb{boson}{}{boson}{}{boson}{}{boson}{}
   \selftypes{boson}{}{boson}{}
   \selfproj{}{}{}{}{}{}
   \fmf{crossed,right=0.414,label=$\nx\star$,l.s=right,l.d=0.05w}{v1,v2}
   \fmfdot{v3,v4}
 \end{fmfgraph*}\\
\refstepcounter{diagh}
(\Alph{diagh})
\label{F42}
\ec
}}
-
\raisebox{0ex}{
\parbox{3cm}{\bc
 \begin{fmfgraph*}(2.5,2.5)
   \selfgl
   \selfplace{\nx\G}{-23}{\nx\G}{-74}{$\nx\Delta$}{155}{}{}
   \selfplacedummy{}{}{}{}{}{}{}{}
   \selftypeb{boson}{}{boson}{}{boson}{}{boson}{}
   \selftypes{boson}{}{boson}{}
   \selfproj{}{}{}{}{}{}
   \fmf{crossed,right=0.414,label=$\nx\star$,l.s=right,l.d=0.05w}{v4,v3}
   \fmfdot{v3}
 \end{fmfgraph*}
\ec
}}
+
\raisebox{0ex}{
\parbox{3cm}{\bc
 \begin{fmfgraph*}(2.5,2.5)
   \selfgl
   \selfplace{\nx\G}{-23}{\nx\G}{-74}{$\nx\Delta$}{155}{\nx\G}{105}
   \selfplacedummy{}{}{}{}{}{}{}{}
   \selftypeb{boson}{}{boson}{}{boson}{}{boson}{}
   \selftypes{boson}{}{boson}{}
   \selfproj{}{}{}{}{}{}
   \fmf{crossed,right=0.414,label=$\nx\star$,l.s=left,l.d=0.05w}{v3,v4}
 \end{fmfgraph*}
\ec
}}
+
\raisebox{-1.5ex}{
\parbox{3cm}{\bc
 \begin{fmfgraph*}(2.5,2.5)
   \selfgl
   \selfplace{\nx\G}{-107}{\nx\G}{75}{}{}{$\nx\Delta$}{-100}
   \selfplacedummy{}{}{}{}{}{}{}{}
   \selftypeb{boson}{}{boson}{}{boson}{}{boson}{}
   \selftypes{boson}{}{boson}{}
   \selfproj{}{}{}{}{}{}
   \fmf{crossed,right=0.414,label=$\nx\star$,l.s=right,l.d=0.05w}{v4,v3}
   \fmfdot{v4}
 \end{fmfgraph*}\\
\refstepcounter{diagh}
(\Alph{diagh})
\label{G42}
\ec
}}&\nonumber\\
&\displaystyle-\;
\raisebox{-1.5ex}{
\parbox{3cm}{\bc
 \begin{fmfgraph*}(2.5,2.5)
   \selfgl
   \selfplace{\nx\G}{-107}{\nx\G}{75}{}{}{$\nx\Delta$}{-100}
   \selfplacedummy{}{}{}{}{}{}{}{}
   \selftypeb{boson}{}{boson}{}{boson}{}{boson}{}
   \selftypes{boson}{}{boson}{}
   \selfproj{}{}{}{}{}{}
   \fmf{crossed,right=0.414,label=$\nx\star$,l.s=right,l.d=.05w}{v2,v3}
   \fmfdot{v4}
 \end{fmfgraph*}\\
\refstepcounter{diagh}
(\Alph{diagh})
\label{H42}
\ec
}}
-
\raisebox{-1.5ex}{
\parbox{3cm}{\bc
 \begin{fmfgraph*}(2.5,2.5)
   \selfgl
   \selfplace{\nx\G}{-107}{\nx\G}{75}{\nx\G}{74}{$\nx\Delta$}{-100}
   \selfplacedummy{}{}{}{}{}{}{}{}
   \selftypeb{boson}{}{boson}{}{boson}{}{boson}{}
   \selftypes{boson}{}{boson}{}
   \selfproj{}{}{}{}{}{}
   \fmf{crossed,right=0.414,label=$\nx\star$,l.s=right,l.d=0.05w}{v4,v1}
 \end{fmfgraph*}\\
\refstepcounter{diagh}
(\Alph{diagh})
\label{I42}
\ec
}}
+
\raisebox{-1.5ex}{
\parbox{3cm}{\bc
 \begin{fmfgraph*}(2.5,2.5)
   \selfgl
   \selfplace{\nx\G}{-107}{\nx\G}{75}{}{}{}{}
   \selfplacedummy{}{}{$\nx\Delta$}{-156}{}{}{}{}
   \selftypeb{boson}{}{boson}{}{boson}{}{boson}{}
   \selftypes{boson}{}{boson}{}
   \selfproj{}{}{}{}{}{}
   \fmf{crossed,right=0.414,label=$\nx\star$,l.s=left,l.d=0.05w}{v3,v4}
 \end{fmfgraph*}\\
\refstepcounter{diagh}
(\Alph{diagh})
\label{J42}
\ec
}}
+
\raisebox{-1.5ex}{
\parbox{3cm}{\bc
 \begin{fmfgraph*}(2.5,2.5)
   \selfgl
   \selfplace{\nx\G}{-107}{\nx\G}{75}{\nx\G}{-85}{}{}
   \selfplacedummy{}{}{$\nx\Delta$}{-156}{}{}{}{}
   \selftypeb{boson}{}{boson}{}{boson}{}{boson}{}
   \selftypes{boson}{}{boson}{}
   \selfproj{}{}{}{}{}{}
   \fmf{crossed,right=0.414,label=$\nx\star$,l.s=right,l.d=0.05w}{v4,v1}
 \end{fmfgraph*}\\
\refstepcounter{diagh}
(\Alph{diagh})
\label{K42}
\ec
}}&\nonumber\\
&\displaystyle-
\raisebox{-1.5ex}{
\parbox{3cm}{\bc
 \begin{fmfgraph*}(2.5,2.5)
   \selfgl
   \selfplace{\nx\G}{-107}{\nx\G}{75}{\nx\G}{-85}{}{}
   \selfplacedummy{}{}{$\nx\Delta$}{-156}{}{}{}{}
   \selftypeb{boson}{}{boson}{}{boson}{}{boson}{}
   \selftypes{boson}{}{boson}{}
   \selfproj{}{}{}{}{}{}
   \fmf{crossed,right=0.414,label=$\nx\star$,l.s=right,l.d=0.05w}{v4,v3}
 \end{fmfgraph*}\\
\refstepcounter{diagh}
(\Alph{diagh})
\label{L42}
\ec
}}
+
\raisebox{-1.5ex}{
\parbox{3cm}{\bc
 \begin{fmfgraph*}(2.5,2.5)
   \selfgl
   \selfplace{}{}{}{}{}{}{$\nx\Delta$}{-100}
   \selfplacedummy{}{}{}{}{}{}{}{}
   \selftypeb{fermion}{}{fermion}{}{boson}{}{boson}{}
   \selftypes{boson}{}{boson}{}
   \selfproj{}{}{}{}{}{}
   \fmf{crossed,right=0.414,label=$\nx\star$,l.s=right,l.d=0.05w}{v4,v3}
   \fmfdot{v4,v1,v2}
 \end{fmfgraph*}\\
\refstepcounter{diagh}
(\Alph{diagh})
\label{G42f}
\ec
}}
-\;
\raisebox{-1.5ex}{
\parbox{3cm}{\bc
 \begin{fmfgraph*}(2.5,2.5)
   \selfgl
   \selfplace{}{}{}{}{}{}{$\nx\Delta$}{-100}
   \selfplacedummy{}{}{}{}{}{}{}{}
   \selftypeb{fermion}{}{fermion}{}{boson}{}{boson}{}
   \selftypes{boson}{}{boson}{}
   \selfproj{}{}{}{}{}{}
   \fmf{crossed,right=0.414,label=$\nx\star$,l.s=right,l.d=.05w}{v2,v3}
   \fmfdot{v4,v1,v2}
 \end{fmfgraph*}\\
\refstepcounter{diagh}
(\Alph{diagh})
\label{H42f}
\ec
}}
-
\raisebox{-1.5ex}{
\parbox{3cm}{\bc
 \begin{fmfgraph*}(2.5,2.5)
   \selfgl
   \selfplace{}{}{}{}{\nx\G}{75}{$\nx\Delta$}{-100}
   \selfplacedummy{}{}{}{}{}{}{}{}
   \selftypeb{fermion}{}{fermion}{}{boson}{}{boson}{}
   \selftypes{boson}{}{boson}{}
   \selfproj{}{}{}{}{}{}
   \fmf{crossed,right=0.414,label=$\nx\star$,l.s=right,l.d=0.05w}{v4,v1}
   \fmfdot{v1,v2}
 \end{fmfgraph*}\\
\refstepcounter{diagh}
(\Alph{diagh})
\label{I42f}
\ec
}}&\nonumber\\
&\displaystyle+
\raisebox{-1.5ex}{
\parbox{3cm}{\bc
 \begin{fmfgraph*}(2.5,2.5)
   \selfgl
   \selfplace{}{}{}{}{}{}{}{}
   \selfplacedummy{}{}{$\nx\Delta$}{-156}{}{}{}{}
   \selftypeb{fermion}{}{fermion}{}{boson}{}{boson}{}
   \selftypes{boson}{}{boson}{}
   \selfproj{}{}{}{}{}{}
   \fmf{crossed,right=0.414,label=$\nx\star$,l.s=left,l.d=0.05w}{v3,v4}
   \fmfdot{v1,v2}
 \end{fmfgraph*}\\
\refstepcounter{diagh}
(\Alph{diagh})
\label{J42f}
\ec
}}
+
\raisebox{-1.5ex}{
\parbox{3cm}{\bc
 \begin{fmfgraph*}(2.5,2.5)
   \selfgl
   \selfplace{}{}{}{}{\nx\G}{-85}{}{}
   \selfplacedummy{}{}{$\nx\Delta$}{-156}{}{}{}{}
   \selftypeb{fermion}{}{fermion}{}{boson}{}{boson}{}
   \selftypes{boson}{}{boson}{}
   \selfproj{}{}{}{}{}{}
   \fmf{crossed,right=0.414,label=$\nx\star$,l.s=right,l.d=0.05w}{v4,v1}
   \fmfdot{v1,v2}
 \end{fmfgraph*}\\
\refstepcounter{diagh}
(\Alph{diagh})
\label{K42f}
\ec
}}
-
\raisebox{-1.5ex}{
\parbox{3cm}{\bc
 \begin{fmfgraph*}(2.5,2.5)
   \selfgl
   \selfplace{}{}{}{}{\nx\G}{-85}{}{}
   \selfplacedummy{}{}{$\nx\Delta$}{-156}{}{}{}{}
   \selftypeb{fermion}{}{fermion}{}{boson}{}{boson}{}
   \selftypes{boson}{}{boson}{}
   \selfproj{}{}{}{}{}{}
   \fmf{crossed,right=0.414,label=$\nx\star$,l.s=right,l.d=0.05w}{v4,v3}
   \fmfdot{v1,v2}
 \end{fmfgraph*}\\
\refstepcounter{diagh}
(\Alph{diagh})
\label{L42f}
\ec
}}
-
\;\frac{1}{2}
\raisebox{-1.5ex}{ 
\parbox{2.5cm}{\bc
 \begin{fmfgraph*}(2.5,2.5)
  \selftwogl
  \selftwoplace{$\nx\Delta$}{-120}{}{}{}{}{}{}
  \selftwoplacedummy{}{}{}{}{}{}
  \selftwotypes{boson}{}{boson}{}{boson}{}{boson}{}
  \selftwotypeb{boson}{}
  \selftwoproj{}{}{}{}{}
  \fmf{crossed,right=0.578,label=$\nx\star$,l.s=left,l.d=0.05w}{v2,v3}
  \fmfdot{v1,v3}
 \end{fmfgraph*}\\
\refstepcounter{diagh}
(\Alph{diagh})
\label{M42}
\ec}}
&\nonumber\\
&\displaystyle-\;
\frac{1}{2}\raisebox{-1.5ex}{
 \parbox{2.5cm}{\bc
  \begin{fmfgraph*}(2.5,2.5)
   \selftwogl
   \selftwoplace{$\nx\Delta$}{-120}{\nx\G}{-85}{}{}{}{}
   \selftwoplacedummy{}{}{}{}{}{}
   \selftwotypes{boson}{}{boson}{}{boson}{}{boson}{}
   \selftwotypeb{boson}{}
   \selftwoproj{}{}{}{}{}
   \fmf{crossed,left=0.578,label=$\nx\star$,l.s=left,l.d=0.05w}{v1,v3}
   \fmfdot{v1}
  \end{fmfgraph*}\\
 \refstepcounter{diagh}
 (\Alph{diagh})
 \label{N42}
 \ec}}
+\;
\frac{1}{2}\raisebox{-1.5ex}{
 \parbox{2.5cm}{\bc
  \begin{fmfgraph*}(2.5,2.5)
   \selftwogl
   \selftwoplace{$\nx\Delta$}{-120}{\nx\G}{-85}{}{}{}{}
   \selftwoplacedummy{}{}{}{}{}{}
   \selftwotypes{boson}{}{boson}{}{boson}{}{boson}{}
   \selftwotypeb{boson}{}
   \selftwoproj{}{}{}{}{}
   \fmf{crossed,right=0.578,label=$\nx\star$,l.s=right,l.d=0.05w}{v3,v2}
   \fmfdot{v1}
  \end{fmfgraph*}\\
 \refstepcounter{diagh}
 (\Alph{diagh})
 \label{O42}
 \ec}}
-\;
\frac{1}{2}\raisebox{-1.5ex}{
 \parbox{2.5cm}{\bc
  \begin{fmfgraph*}(2.5,2.5)
   \selftwogl
   \selftwoplace{}{}{$\nx\Delta$}{-85}{}{}{}{}
   \selftwoplacedummy{}{}{}{}{}{}
   \selftwotypes{boson}{}{boson}{}{boson}{}{boson}{}
   \selftwotypeb{boson}{}
   \selftwoproj{}{}{}{}{}
   \fmf{crossed,left=0.578,label=$\nx\star$,l.s=left,l.d=0.05w}{v1,v3}
   \fmfdot{v1,v2}
  \end{fmfgraph*}\\
 \refstepcounter{diagh}
 (\Alph{diagh})
 \label{P42}
 \ec}}
+\;
\frac{1}{2}\raisebox{-1.5ex}{
 \parbox{2.5cm}{\bc
  \begin{fmfgraph*}(2.5,2.5)
   \selftwogl
   \selftwoplace{}{}{$\nx\Delta$}{-85}{}{}{}{}
   \selftwoplacedummy{}{}{}{}{}{}
   \selftwotypes{boson}{}{boson}{}{boson}{}{boson}{}
   \selftwotypeb{boson}{}
   \selftwoproj{}{}{}{}{}
   \fmf{crossed,right=0.578,label=$\nx\star$,l.s=right,l.d=0.05w}{v3,v2}
   \fmfdot{v1,v2}
  \end{fmfgraph*}\\
 \refstepcounter{diagh}
 (\Alph{diagh})
 \label{Q42}
 \ec}}&\nonumber\\
&\displaystyle+\;
\frac{1}{2}\raisebox{-1.5ex}{
 \parbox{2.5cm}{\bc
  \begin{fmfgraph*}(2.5,2.5)
   \selftwogl
   \selftwoplace{}{}{$\nx\Delta$}{-85}{\nx\G}{170}{}{}
   \selftwoplacedummy{}{}{}{}{}{}
   \selftwotypes{boson}{}{boson}{}{boson}{}{boson}{}
   \selftwotypeb{boson}{}
   \selftwoproj{}{}{}{}{}
   \fmf{crossed,left=0.578,label=$\nx\star$,l.s=left,l.d=0.05w}{v2,v1}
   \fmfdot{v1}
  \end{fmfgraph*}\\
 \refstepcounter{diagh}
 (\Alph{diagh})
 \label{R42}
\ec}}
\label{diamassg4}
\ee

\end{fmffile}

%
% Here ends the third font ``paperf3''
%
%%%%%%%%%%%%%%%%%%%%%%%%%%%%%%%%%%%%%%%%%%%%%%%%%%%%%%%%%%%%%%%%%%%%%%%
%                                                                     %
% Starting the 4.font-file and adding additional Metafont-Definitions %
%                                                                     %
%%%%%%%%%%%%%%%%%%%%%%%%%%%%%%%%%%%%%%%%%%%%%%%%%%%%%%%%%%%%%%%%%%%%%%%

\begin{fmffile}{paperf4}

%%%%%%%%%%%%%%%%%%%%%%%%%%%%%%%%%%%%%%%%%%%%%%%%%%%%%%%%%%%%%%%%%%%%%%%
%                                                                     %
%       Some Metafont-definitions in addition to feynmf.mf            %
%                                                                     %
%%%%%%%%%%%%%%%%%%%%%%%%%%%%%%%%%%%%%%%%%%%%%%%%%%%%%%%%%%%%%%%%%%%%%%%
%
% Variables:
%

\fmfcurved
\fmfpen{thin} 
\fmfset{curly_len}{3mm}
\fmfset{wiggly_len}{2mm}

%
% Styles:
%

\fmfcmd{ %
 vardef cross_bar (expr p, len, ang) =
  ((-len/2,0)--(len/2,0))
    rotated (ang + angle direction length(p)/2 of p)
    shifted point length(p)/2 of p
 enddef;
 style_def crossed expr p = 
   ccutdraw cross_bar (p, 5mm, 45);
   ccutdraw cross_bar (p, 5mm, -45)
 enddef;}

\fmfcmd{ %
 vardef proj_tarrow (expr p, frac) =
  save a, t, z;
  pair z;
  t1 = frac*length p;
  a = angle direction t1 of p;
  z = point t1 of p;
  (t2,whatever) = p intersectiontimes
    (halfcircle scaled 3arrow_len rotated (a-90) shifted z);
  arrow_head (p, t1, t2, arrow_ang)
enddef;
 style_def projs expr p  =
    cfill (proj_tarrow (reverse p, 0.85));
 enddef;
 style_def projc expr p =
    cfill (proj_tarrow (reverse p, 0.7));
 enddef;}

\fmfcmd{ 
 vardef projl_tarrow (expr p, frac) =
  save a, t, z;
  pair z;
  t1 = frac*length p;
  a = angle direction t1 of p;
  z = point t1 of p;
  (t2,whatever) = p intersectiontimes
    (halfcircle scaled 2arrow_len rotated (a-90) shifted z);
  arrow_head (p, t1, t2, arrow_ang)
 enddef;
 style_def arrowr expr p =
   cfill (projl_tarrow (reverse p, 0.6));
 enddef;
 style_def arrowl expr p  =
   cfill (projl_tarrow (reverse p, 0.6));
 enddef;
}

\fmfcmd{
style_def plains expr p =
 draw p;
 undraw subpath (0,.15) of p;
enddef;}

%%%%%%%%%%%%%%%%%%%%%%%%%%%%%%%%%%%%%%%%%%%%%%%%%%%%%%%%%%%%%%%%%%%%%%%
%                                                                     %
%                  Here the paper continues                           %
%                                                                     %
%%%%%%%%%%%%%%%%%%%%%%%%%%%%%%%%%%%%%%%%%%%%%%%%%%%%%%%%%%%%%%%%%%%%%%%

The ${\cal O}(\mbox{\em g}^4)$-parts of 
(\ref{diamassg4}.\refmy{A42}-\refmy{R42}) are characterized by special
contributions of hard (h) and soft (s) momenta in each diagram. For these 
combinations one  can cancel all ${\cal O}(\mbox{\em g}^4)$-parts with the 
help of two relations:

\begin{enumerate}
  \item[(i)] the definition of the Debye-mass (\ref{selfform}/\ref{self}) and
  \item[(ii)] the transversality of the one-loop self-energy. 
\end{enumerate}

Let us first focus on (i). As an example we show the 
${\cal O}(\mbox{\em g}^4)$-parts of the diagrams 
(\ref{diamassg4}.\refmy{B42}/\refmy{H42}/\refmy{H42f}/\refmy{P42}) 
(note that the propagators with an explicit mass-dependence (the 
``cross-stars'') are always soft):

\be
\frac{1}{2}\;
\raisebox{-1.5ex}{
\parbox{3cm}{\bc
 \begin{fmfgraph*}(2.5,2.5)
   \selfgl
   \selfplace{}{}{}{}{}{}{$\nx\Delta$}{-100}
   \selfplacedummy{}{}{}{}{}{}{}{}
   \selftypeb{boson}{\nx\h}{boson}{\nx\h}{boson}{\nx\s}{boson}{\nx\s}
   \selftypes{boson}{\nx\s}{boson}{}
   \selfproj{}{}{}{}{}{}
   \fmf{crossed,right=0.4,label=$\nx\star$,l.s=right,l.d=.05w}{v2,v3}
   \fmfdot{v4,v1,v2}
 \end{fmfgraph*}\\
\stepcounter{rueck}
\refstepcounter{diagh}
(\Alph{diagh})
\label{A43}
\ec
}}
-\;
\raisebox{-1.5ex}{
\parbox{3cm}{\bc
 \begin{fmfgraph*}(2.5,2.5)
   \selfgl
   \selfplace{\nx\G}{-107}{\nx\G}{75}{}{}{$\nx\Delta$}{-100}
   \selfplacedummy{}{}{}{}{}{}{}{}
   \selftypeb{boson}{\nx\h}{boson}{\nx\h}{boson}{\nx\s}{boson}{\nx\s}
   \selftypes{boson}{\nx\s}{boson}{}
   \selfproj{}{}{}{}{}{}
   \fmf{crossed,right=0.4,label=$\nx\star$,l.s=right,l.d=.05w}{v2,v3}
   \fmfdot{v4}
 \end{fmfgraph*}\\
\refstepcounter{diagh}
(\Alph{diagh})
\label{B43}
\ec
}}
-\;
\raisebox{-1.5ex}{
\parbox{3cm}{\bc
 \begin{fmfgraph*}(2.5,2.5)
   \selfgl
   \selfplace{}{}{}{}{}{}{$\nx\Delta$}{-100}
   \selfplacedummy{}{}{}{}{}{}{}{}
   \selftypeb{fermion}{\nx\h}{fermion}{\nx\h}{boson}{\nx\s}{boson}{\nx\s}
   \selftypes{boson}{\nx\s}{boson}{}
   \selfproj{}{}{}{}{}{}
   \fmf{crossed,right=0.4,label=$\nx\star$,l.s=right,l.d=.05w}{v2,v3}
   \fmfdot{v4,v1,v2}
 \end{fmfgraph*}\\
\refstepcounter{diagh}
(\Alph{diagh})
\label{B43f}
\ec
}}
+
\frac{1}{2}\raisebox{-1.5ex}{
 \parbox{2.5cm}{\bc
  \begin{fmfgraph*}(2.5,2.5)
   \selftwogl
   \selftwoplace{}{}{}{}{$\nx\Delta$}{-100}{}{}
   \selftwoplacedummy{}{}{}{}{}{}
   \selftwotypes{boson}{}{boson}{\nx\s}{boson}{\nx\s}{boson}{\nx\s}
   \selftwotypeb{boson}{\nx\h}
   \selftwoproj{}{}{}{}{}
   \fmf{crossed,left=0.6,label=$\nx\star$,l.s=left,l.d=0.05w}{v2,v1}
   \fmfdot{v1,v3}
  \end{fmfgraph*}\\
 \refstepcounter{diagh}
 (\Alph{diagh})
 \label{C43}
 \ec}}
\ee

The definition of the one-loop Debye-mass (\ref{selfform}/\ref{self}) assures 
that the sum of the four diagrams above is just minus the diagram 
(\ref{massg5}.\refmy{E22}):

\be
\raisebox{0ex}{
\parbox{3cm}{\bc
 \begin{fmfgraph*}(2.5,2.5)
  \selfgl
  \fmf{boson,right=0.414,label=\nx\s,label.s=left,l.d=0.05w}{v3,v4}
  \fmf{boson,right=0.414,label=\nx\s,label.s=left,l.d=0.05w}{v4,v3} 
  \fmf{boson,right=0.414}{v1,v2,v3}
 \fmf{boson,right=0.414,label=\nx\s,label.s=left,l.d=0.05w}{v4,v1}
  \fmf{crossed,right=0.414}{v1,v2}
  \fmfv{label=$\nx\Delta$,l.a=-110}{v4}
  \fmf{boson,right=.414,label=$\nx\star$,l.d=.05w}{v2,v3}
  \fmf{crossed,right=.414}{v2,v3}
  \fmfdot{v4}
 \end{fmfgraph*}\\
\ec}}
\ee

Consequently, these five diagrams cancel. In the same way we can identify the 
${\cal O}(\mbox{\em g}^4)$-parts of the diagrams 
(\ref{diamassg4}.\refmy{A42}/\refmy{G42}/\refmy{G42f}/\refmy{Q42})
with (\ref{massg5}.\refmy{F22}) and 
(\ref{diamassg4}.\refmy{C42}/\refmy{I42}/\refmy{I42f}/\refmy{R42}) with
(\ref{massg5}.\refmy{G22}). Note that it is not surprising that the actual 
definition of the Debye mass enters at this stage of the calculation. If the
mass $m$ in the counterterm (\ref{counter}) is not defined properly via the 
static limit of the self energy, the result for the physical quantity of 
interest might be incorrect and gauge dependent.\\
Now we turn to argument (ii). We group the diagrams which are still
left together to four-diagram blocks so that the sum can be depicted as one 
diagram which has a one-loop self-energy insertion. This insertion 
is always Lorentz-contracted with $\Delta_{\mu}$, which contains 
$k_{\mu}$ for covariant gauges. Due to the known relation for the one-loop 
self-energy,

\be
k_{\mu}\Pi^{\mu\nu} = 0,
\ee
these blocks of four diagrams are zero. To give an example, we look at the
diagrams (\ref{diamassg4}.\refmy{D42}/\refmy{J42}/\refmy{J42f}/\refmy{M42}):

\be
-\;\frac{1}{2}
\raisebox{-1.5ex}{
\parbox{3cm}{\bc
 \begin{fmfgraph*}(2.5,2.5)
   \selfgl
   \selfplace{}{}{}{}{}{}{}{}
   \selfplacedummy{}{}{$\nx\Delta$}{-156}{}{}{}{}
   \selftypeb{boson}{}{boson}{}{boson}{}{boson}{\nx\s}
   \selftypes{boson}{\nx\s}{boson}{\nx\s}
   \selfproj{}{}{}{}{}{}
   \fmf{crossed,right=0.414,label=$\nx\star$,l.s=left,l.d=0.05w}{v3,v4}
   \fmfdot{v4}
 \end{fmfgraph*}\\
\stepcounter{rueck}
\refstepcounter{diagh}
(\Alph{diagh})
\label{A44}
\ec
}}
+
\raisebox{-1.5ex}{
\parbox{3cm}{\bc
 \begin{fmfgraph*}(2.5,2.5)
   \selfgl
   \selfplace{\nx\G}{-107}{\nx\G}{75}{}{}{}{}
   \selfplacedummy{}{}{$\nx\Delta$}{-156}{}{}{}{}
   \selftypeb{boson}{}{boson}{}{boson}{}{boson}{\nx\s}
   \selftypes{boson}{\nx\s}{boson}{\nx\s}
   \selfproj{}{}{}{}{}{}
   \fmf{crossed,right=0.414,label=$\nx\star$,l.s=left,l.d=0.05w}{v3,v4}
   \fmfdot{v4}
 \end{fmfgraph*}\\
\refstepcounter{diagh}
(\Alph{diagh})
\label{B44}
\ec
}}
+
\raisebox{-1.5ex}{
\parbox{3cm}{\bc
 \begin{fmfgraph*}(2.5,2.5)
   \selfgl
   \selfplace{}{}{}{}{}{}{}{}
   \selfplacedummy{}{}{$\nx\Delta$}{-156}{}{}{}{}
   \selftypeb{fermion}{}{fermion}{}{boson}{}{boson}{\nx\s}
   \selftypes{boson}{\nx\s}{boson}{\nx\s}
   \selfproj{}{}{}{}{}{}
   \fmf{crossed,right=0.414,label=$\nx\star$,l.s=left,l.d=0.05w}{v3,v4}
   \fmfdot{v4,v1,v2}
 \end{fmfgraph*}\\
\refstepcounter{diagh}
(\Alph{diagh})
\label{B44f}
\ec
}}
-\;
\frac{1}{2}
\raisebox{-1.5ex}{ 
\parbox{2.5cm}{\bc
 \begin{fmfgraph*}(2.5,2.5)
  \selftwogl
  \selftwoplace{$\nx\Delta$}{-120}{}{}{}{}{}{}
  \selftwoplacedummy{}{}{}{}{}{}
  \selftwotypes{boson}{\nx\s}{boson}{\nx\s}{boson}{\nx\s}{boson}{}
  \selftwotypeb{boson}{}
  \selftwoproj{}{}{}{}{}
  \fmf{crossed,right=0.578,label=$\nx\star$,l.s=left,l.d=0.05w}{v2,v3}
  \fmfdot{v1,v3}
 \end{fmfgraph*}\\
\refstepcounter{diagh}
(\Alph{diagh})
\label{C44}
\ec}}
= 
-
\raisebox{-1.5ex}{ 
\parbox{2.5cm}{\bc
 \begin{fmfgraph*}(2.5,2.5)
  \selftwogl
  \selftwoplace{$\nx\Delta$}{-120}{}{}{}{}{}{}
  \selftwoplacedummy{}{}{}{}{}{}
  \selftwotypes{boson}{\nx\s}{boson}{\nx\s}{boson}{\nx\s}{boson}{}
  \selftwotypeb{phantom}{}
  \selftwoproj{}{}{}{}{}
  \fmf{crossed,right=0.578,label=$\nx\star$,l.s=left,l.d=0.05w}{v2,v3}
  \fmfblob{0.2w}{v1}
  \fmfdot{v3}
 \end{fmfgraph*}\\
\refstepcounter{diagh}
(\Alph{diagh})
\label{D44}
\ec}}
\ee

The same holds true for 
(\ref{diamassg4}.\refmy{E42}/\refmy{K42}/\refmy{K42f}/\refmy{N42}) and
(\ref{diamassg4}.\refmy{F42}/\refmy{L42}/\refmy{L42f}/\refmy{M42}).
In this way ${\cal O}$({\em g}$\,^{4}$)-contributions of gauge dependent parts 
of the three loop diagrams are cancelled. Thus we have proven that the 
partition function up to ${\cal O}$({\em g}$\,^{4}$) is $\alpha$-independent 
for an arbitrary covariant gauge. Of course it would be interesting to extend 
the proof to {$\cal O$}({\em g}$\,^5$). Here, however, all diagrams of
(\ref{diamassg5}) and (\ref{diamassg4}) contribute and so far we haven't found
a way to prove their cancellation diagrammatically.

%%%%%%%%%%%%%%%%%%%%%%%%%%%%%%%%%%%%%%%%%%%%%%%%%%%%%%%%%%%%%%%%%%%%%%
\section{Summary and Conclusions}\label{sec6}
%%%%%%%%%%%%%%%%%%%%%%%%%%%%%%%%%%%%%%%%%%%%%%%%%%%%%%%%%%%%%%%%%%%%%%

We briefly summarize the necessary steps to prove the $\alpha$-independence 
of a sum of loop diagrams in a general covariant gauge:

\begin{itemize}
 \item Start with the diagram which has only three-gluon vertices (like
       (\ref{alldia}.\refmy{E9}) for the two-loop and 
       (\ref{alldia}.\refmy{G9}) for the three-loop case). Decompose one
       gluon propagator according to (\ref{gluonm}). Then treat the rest
       of the diagrams in the same way.
 \item Use (\ref{maindiagram2}) and the equivalent of (\ref{ghocovz2}) as long 
       as there are $k_\mu$-arrows pointing at a three-gluon vertex,
       (\ref{femrul}) as long as there are such arrows pointing at a
       fermion-gluon vertex.
 \item Use (\ref{momentumcon}) to pull arrows out of closed ghost loops
       (c.f.~the last equality in (\ref{zerlegung})).
 \item At this stage all $a_\mu$-dependent diagrams which have
       neither contraction arrows nor four-point vertices should drop out.
 \item Next, diagrams with different topology are related by the rules
       of Sec.~\ref{sec2}. To get an idea which diagrams are related 
       it is instructive to contract the lines with contraction arrows.
 \item In all remaining diagrams the momentum integrations have to be split
       up in hard and soft modes to figure out which diagrams contribute
       to the order of {\em g} one is interested in. The current definition 
       of the Debye mass and the transversality of the one-loop self-energy 
       has to be taken into account.
 \item All diagrams which have no counterparts so far ought to be of
       higher order in the coupling constant.
 \item Finally symmetry factors have to be counted as outlined in 
       Secs.~\ref{sec2.1.4} and \ref{sec5.1}.
\end{itemize}

In this article we have presented a method to check whether a 
resummation scheme 
preserves gauge invariance. This method is diagrammatic and relates 
systematically diagrams with different topology. This avoids tedious analytical
calculations since it can be seen 
already at the diagrammatic level which algebraic expressions cancel
each other. (This means that in a calculation the integrands cancels
before performing the integrations.)
Of course this diagrammatic method is less elegant than formal
proofs of gauge invariance using Ward identities. On the other hand
formal proofs are sometimes not straightforward or ambiguous due to
singularities. Originally this diagrammatic method was invented by
Cheng and Tsai \cite{cheMIT} to check the consistency of different sets
of Feynman rules for vacuum QCD. Here we have given an example that this 
method can be extended to check gauge invariance of
physical quantities calculated in improved perturbation schemes. We have
applied the diagrammatic method to the calculation of the free energy of
finite temperature QCD up to O({\em g}$\,^4$).     

Strictly speaking we have only shown that the result for the partition 
function/free energy is independent of $\alpha$ for arbitrary covariant
$\alpha$-gauges. In principle it should also be checked that the result
remains the same if one performs the calculation in a different class of
gauges like e.g.~axial or Coulomb gauge. Thus from a rigorous point of view 
further investigations are required 
to prove  the gauge invariance of $Z_{\text{QCD}}$.
Of course the diagrammatic method outlined here is not restricted to 
covariant gauges but can be used for arbitrary gauges \cite{cheMIT,leu96}.
Note that then one has to deal with 
the momentum shift problems discussed in Sec.~\ref{sec4}. 
In practice however
$\alpha$-independence is often used synonymously for gauge invariance. Indeed 
e.g..~the famous ``plasmon puzzle'' \cite{bra90b}, i.e., the fact that 
the plasmon damping rate calculated without resummation turns out to be 
gauge dependent, already shows up for different values of $\alpha$.\\
We expect that the method presented here can be extended 
to more complicated resummation schemes
(like e.g.~``hard thermal loops'' \cite{bra90a} where also vertex 
resummation is taken into account) and to electro-weak gauge
theory (where the partition function in the vicinity of the phase transition
is of special interest \cite{buc95}). For the latter case we also refer 
to \cite{fen96a} where a similar diagrammatic method is used to minimize 
the computational effort for calculating S-matrix elements in vacuum quantum
field theory.

%%%%%%%%%%%%%%%%%%%%%%%%%%%%%%%%%%%%%%%%%%%%%%%%%%%%%%%%%%%%%%%%%%%%%%

\acknowledgments

We acknowledge enlightening discussions with 
Anton Rebhan and useful remarks by Markus Thoma. 
U.H.~and S.L.~would like to thank Prof.~B.~M\"uller and the
Duke University Physics Department, where part of this work was done, for 
their kind hospitality. U.H.~would like to thank CERN for warm hospitality and
a stimulating atmosphere during the final stages of this work, too.
This work was supported in part by BMBF, DFG and GSI.  

\appendix

%%%%%%%%%%%%%%%%%%%%%%%%%%%%%%%%%%%%%%%%%%%%%%%%%%%%%%%%%%%%%%%%%%%%%
\section{Some Diagrammatics for $O(\lowercase{g}^4)$ - gauge invariance}
\label{secapa}
%%%%%%%%%%%%%%%%%%%%%%%%%%%%%%%%%%%%%%%%%%%%%%%%%%%%%%%%%%%%%%%%%%%%%%

In this Appendix, we give further details of how to prove the gauge invariance
of the `` vacuum like '' three loop contributions to the QCD partition 
function. \\
The explicit mass dependent diagrams are discussed in Sec.\ref{sec5}. In 
addition, the straightforward application of the diagrammatic rules of 
Sec.~\ref{sec2} to (\ref{alldia}) results in approximately 100
a$_\mu$-dependent diagrams. We have checked that the sum of these contributions
cancels. For lack of space, we restrict our presentation here to a special 
example: The cancellation of all a$_\mu$-dependent contributions to star 
diagrams.\\[1ex]

\centerline{\bf Cancellation of the Star}
\vspace*{1ex}

Instead of presenting all diagrams we demonstrate for the star diagrams how 
the Cheng-Tsai rules work for three-loop diagrams. Then we outline briefly how
all the other diagrams drop out. 

\begin{enumerate}
  \item The star diagrams allow the following decompositions:

\be
\raisebox{-1.5ex}{
 \parbox{3cm}{\bc 
  \begin{fmfgraph*}(2.5,2.5)
   \stargl
   \starplace{}{}{}{}{}{}{}{}
   \starplacedummy{}{}{}{}{}{}{}{}
   \startypes{boson}{}{boson}{}{boson}{}
   \startypec{boson}{$D$}{boson}{}{boson}{}
   \starproj{}{}{}{}{}{}
   \fmfdot{v1,v2,v3,v4}
  \end{fmfgraph*}\\
 \stepcounter{rueck}
 \refstepcounter{diagh}
 (\Alph{diagh})
 \label{A12}
 \ec}
}
=
\raisebox{-1.5ex}{
 \parbox{3cm}{\bc 
  \begin{fmfgraph*}(2.5,2.5)
   \stargl
   \starplace{}{}{}{}{}{}{}{}
   \starplacedummy{}{}{}{}{}{}{}{}
   \startypes{boson}{}{boson}{}{boson}{}
   \startypec{boson}{$D_F'$}{boson}{}{boson}{}
   \starproj{}{}{}{}{}{}
   \fmfdot{v1,v2,v3,v4}
  \end{fmfgraph*}\\
 \refstepcounter{diagh}
 (\Alph{diagh})
 \label{B12}
 \ec}
}
+\;2
\raisebox{-1.5ex}{
 \parbox{3cm}{\bc 
  \begin{fmfgraph*}(2.5,2.5)
   \stargl
   \starplace{}{}{}{}{}{}{$\nx\Delta$}{55}
   \starplacedummy{}{}{}{}{}{}{}{}
   \startypes{boson}{}{boson}{}{boson}{}
   \startypec{boson}{}{boson}{}{boson}{}
   \starproj{}{}{}{proj_v}{}{}
   \fmfdot{v1,v2,v3,v4}
  \end{fmfgraph*}\\
 \refstepcounter{diagh}
 (\Alph{diagh})
 \label{C12}
 \ec}
}
\label{decompos}
\ee

\be
-
\raisebox{-1.5ex}{
\parbox{3cm}{\bc 
 \begin{fmfgraph*}(2.5,2.5)
  \stargl
  \starplace{\nx\G}{150}{\nx\G}{-80}{\nx\G}{20}{}{}
  \starplacedummy{}{}{}{}{}{}{}{}
  \startypes{boson}{}{boson}{}{boson}{}
  \startypec{boson}{$D$}{boson}{}{boson}{}
  \starproj{}{}{}{}{}{}
  \fmfdot{v4}
 \end{fmfgraph*}\\
\stepcounter{rueck}
\refstepcounter{diagh}
(\Alph{diagh})
\label{A15}
\ec
}}
=
-\raisebox{-1.5ex}{
\parbox{3cm}{\bc 
 \begin{fmfgraph*}(2.5,2.5)
  \stargl
  \starplace{\nx\G}{150}{\nx\G}{-80}{\nx\G}{20}{}{}
  \starplacedummy{}{}{}{}{}{}{}{}
  \startypes{boson}{}{boson}{}{boson}{}
  \startypec{boson}{$D_F'$}{boson}{}{boson}{}
  \starproj{}{}{}{}{}{}
  \fmfdot{v4}
 \end{fmfgraph*}\\
\refstepcounter{diagh}
(\Alph{diagh})
\label{B15}
\ec
}}
-
\raisebox{-1.5ex}{
\parbox{3cm}{\bc
 \begin{fmfgraph*}(2.5,2.5)
  \stargl
  \starplace{\nx\G}{150}{\nx\G}{-80}{\nx\G}{20}{$\nx\Delta$}{55}
  \starplacedummy{}{}{}{}{}{}{}{}
  \startypes{boson}{}{boson}{}{boson}{}
  \startypec{boson}{}{boson}{}{boson}{}
  \starproj{}{}{}{proj_v}{}{}
  \fmfdot{v4}
 \end{fmfgraph*}\\
\refstepcounter{diagh}
(\Alph{diagh})
\label{C15}
\ec}}
-
\raisebox{-1.5ex}{
\parbox{3cm}{\bc
 \begin{fmfgraph*}(2.5,2.5)
  \stargl
  \starplace{\nx\G}{150}{\nx\G}{-80}{\nx\G}{20}{}{}
  \starplacedummy{$\nx\Delta$}{-55}{}{}{}{}{}{}
  \startypes{boson}{}{boson}{}{boson}{}
  \startypec{boson}{}{boson}{}{boson}{}
  \starproj{}{}{}{proj_h}{}{}
  \fmfdot{v4}
 \end{fmfgraph*}\\
\refstepcounter{diagh}
(\Alph{diagh})
\label{D15}
\ec
}}
\label{decompos1}
\ee

\be
-\raisebox{-1.5ex}{
\parbox{3cm}{\bc 
\begin{fmfgraph*}(2.5,2.5)
 \stargl
 \starplace{\nx\G}{150}{\nx\G}{68}{\nx\G}{20}{\nx\G}{-70}
 \starplacedummy{}{}{}{}{}{}{}{}
 \startypes{boson}{}{boson}{}{boson}{}
 \startypec{boson}{$D$}{boson}{}{boson}{}
 \starproj{}{}{}{}{}{}
\end{fmfgraph*}\\
\stepcounter{rueck}
\refstepcounter{diagh}
(\Alph{diagh})
\label{A16}
\ec
}}
=
-\raisebox{-1.5ex}{
\parbox{3cm}{\bc 
\begin{fmfgraph*}(2.5,2.5)
 \stargl
 \starplace{\nx\G}{150}{\nx\G}{68}{\nx\G}{20}{\nx\G}{-70}
 \starplacedummy{}{}{}{}{}{}{}{}
 \startypes{boson}{}{boson}{}{boson}{}
 \startypec{boson}{$D_F'$}{boson}{}{boson}{}
 \starproj{}{}{}{}{}{}
\end{fmfgraph*}\\
\refstepcounter{diagh}
(\Alph{diagh})
\label{B16}
\ec
}}
-\;2
\raisebox{-1.5ex}{
\parbox{3cm}{\bc 
\begin{fmfgraph*}(2.5,2.5)
 \stargl
 \starplace{\nx\G}{150}{\nx\G}{68}{\nx\G}{20}{\nx\G}{-70}
 \starplacedummy{}{}{}{}{}{}{$\nx\Delta$}{55}
 \startypes{boson}{}{boson}{}{boson}{}
 \startypec{boson}{}{boson}{}{boson}{}
 \starproj{}{}{}{proj_v}{}{}
\end{fmfgraph*}\\
\refstepcounter{diagh}
(\Alph{diagh})
\label{C16}
\ec
}}
\label{decompos2}
\ee

\be
\raisebox{-1.5ex}{
\parbox{3cm}{\bc 
\begin{fmfgraph*}(2.5,2.5)
 \stargl
 \starplace{}{}{}{}{}{}{}{}
 \starplacedummy{}{}{}{}{}{}{}{}
 \startypes{fermion}{}{fermion}{}{fermion}{}
 \startypec{boson}{D}{boson}{}{boson}{}
 \starproj{}{}{}{}{}{}
 \fmfdot{v1,v2,v3,v4}
\end{fmfgraph*}\\
\stepcounter{rueck}
\refstepcounter{diagh}
(\Alph{diagh})
\label{A15f}
\ec
}}
=
\raisebox{-1.5ex}{
\parbox{3cm}{\bc 
\begin{fmfgraph*}(2.5,2.5)
 \stargl
 \starplace{}{}{}{}{}{}{}{}
 \starplacedummy{}{}{}{}{}{}{}{}
 \startypes{fermion}{}{fermion}{}{fermion}{}
 \startypec{boson}{$D_F'$}{boson}{}{boson}{}
 \starproj{}{}{}{}{}{}
 \fmfdot{v1,v2,v3,v4}
\end{fmfgraph*}\\
\refstepcounter{diagh}
(\Alph{diagh})
\label{B15f}
\ec
}}
+
\raisebox{-1.5ex}{
\parbox{3cm}{\bc
\begin{fmfgraph*}(2.5,2.5)
 \stargl
 \starplace{}{}{}{}{}{}{$\nx\Delta$}{55}
 \starplacedummy{}{}{}{}{}{}{}{}
 \startypes{fermion}{}{fermion}{}{fermion}{}
 \startypec{boson}{}{boson}{}{boson}{}
 \starproj{}{}{}{proj_v}{}{}
 \fmfdot{v1,v2,v3,v4}
\end{fmfgraph*}\\
\refstepcounter{diagh}
(\Alph{diagh})
\label{C15f}
\ec}}
+
\raisebox{-1.5ex}{
\parbox{3cm}{\bc
\begin{fmfgraph*}(2.5,2.5)
 \stargl
 \starplace{$\nx\Delta$}{-55}{}{}{}{}{}{}
 \starplacedummy{}{}{}{}{}{}{}{}
 \startypes{fermion}{}{fermion}{}{fermion}{}
 \startypec{boson}{}{boson}{}{boson}{}
 \starproj{}{}{}{proj_h}{}{}
 \fmfdot{v1,v2,v3,v4}
\end{fmfgraph*}\\
\refstepcounter{diagh}
(\Alph{diagh})
\label{D15f}
\ec
}}\label{decompos4}
\ee

\be
\raisebox{-1.5ex}{
\parbox{3cm}{\bc 
\begin{fmfgraph*}(2.5,2.5)
 \stargl
 \starplace{}{}{}{}{}{}{}{}
 \starplacedummy{}{}{}{}{}{}{}{}
 \startypes{fermion}{}{fermion}{}{boson}{}
 \startypec{boson}{D}{fermion}{}{fermion}{}
 \starproj{}{}{}{}{}{}
 \fmfdot{v1,v2,v3,v4}
\end{fmfgraph*}\\
\stepcounter{rueck}
\refstepcounter{diagh}
(\Alph{diagh})
\label{A16f}
\ec
}}
=
\raisebox{-1.5ex}{
\parbox{3cm}{\bc 
\begin{fmfgraph*}(2.5,2.5)
 \stargl
 \starplace{}{}{}{}{}{}{}{}
 \starplacedummy{}{}{}{}{}{}{}{}
 \startypes{fermion}{}{fermion}{}{boson}{}
 \startypec{boson}{$D_F'$}{fermion}{}{fermion}{}
 \starproj{}{}{}{}{}{}
 \fmfdot{v1,v2,v3,v4}
\end{fmfgraph*}\\
\refstepcounter{diagh}
(\Alph{diagh})
\label{B16f}
\ec
}}
+\;2
\raisebox{-1.5ex}{
\parbox{3cm}{\bc 
\begin{fmfgraph*}(2.5,2.5)
 \stargl
 \starplace{}{}{}{}{}{}{$\nx\Delta$}{55}
 \starplacedummy{}{}{}{}{}{}{}{}
 \startypes{fermion}{}{fermion}{}{boson}{}
 \startypec{boson}{}{fermion}{}{fermion}{}
 \starproj{}{}{}{proj_v}{}{}
 \fmfdot{v1,v2,v3,v4}
\end{fmfgraph*}\\
\refstepcounter{diagh}
(\Alph{diagh})
\label{C16f}
\ec
}}
\label{decompos5}
\ee

    Note that to emphasize the structure of the decomposition (\ref{prop}), 
    Cheng and Tsai have introduced slightly different notations for the two
    $\Delta$-dependent terms, one coming with $\Delta(-k)$ and an overall
    $+$ sign, the other with $\Delta(k)$ and an overall $-$ sign. To ease
    a purely diagrammatic analysis, we depart here from this notational
    convention. we use $\Delta(k) = - \Delta(-k)$ such that all $\Delta$ in
    the diagrams (\ref{decompos}) to (\ref{decompos5}) are understood to 
    represent factors $\Delta(-k)$. As a consequence, if the propagator 
    (\ref{prop}) connects two vertices of the same type, then the orientation 
    of the arrow does not matter, the two $\Delta$-dependent contributions are
    equal and can be summed up. This leads to the factors 2 in front of 
    (\ref{decompos}.\refmy{C12}), (\ref{decompos2}.\refmy{C16}) and 
    (\ref{decompos5}.\refmy{C16f}), while in (\ref{decompos1}) and 
    (\ref{decompos4}) respectively, the arrows point on vertices of different
    types and both $\Delta$-dependent terms have to be dealt with seperatly.\\
    
  \item Using the rules from section \ref{sec2.1.1}, we can decompose each 
    diagram further. Strictly speaking we should use 
    (\ref{maindiagram2}) instead of (\ref{maindiagram}). The additional 
    contributions, however, show an explicit $m$-dependence. They are discussed
    separately in Sec.~\ref{sec5}. We begin with diagram 
    (\ref{decompos}.\refmy{C12}):
\end{enumerate}

\end{fmffile}

%
% Here ends the fourth font ``paperf4''
%
%%%%%%%%%%%%%%%%%%%%%%%%%%%%%%%%%%%%%%%%%%%%%%%%%%%%%%%%%%%%%%%%%%%%%%%
%                                                                     %
% Starting the 5.font-file and adding additional Metafont-Definitions %
%                                                                     %
%%%%%%%%%%%%%%%%%%%%%%%%%%%%%%%%%%%%%%%%%%%%%%%%%%%%%%%%%%%%%%%%%%%%%%%

\begin{fmffile}{paperf5}

%%%%%%%%%%%%%%%%%%%%%%%%%%%%%%%%%%%%%%%%%%%%%%%%%%%%%%%%%%%%%%%%%%%%%%%
%                                                                     %
%       Some Metafont-definitions in addition to feynmf.mf            %
%                                                                     %
%%%%%%%%%%%%%%%%%%%%%%%%%%%%%%%%%%%%%%%%%%%%%%%%%%%%%%%%%%%%%%%%%%%%%%%
%
% Variables:
%

\fmfcurved
\fmfpen{thin} 
\fmfset{curly_len}{3mm}
\fmfset{wiggly_len}{2mm}

%
% Styles:
%

\fmfcmd{ %
 vardef cross_bar (expr p, len, ang) =
  ((-len/2,0)--(len/2,0))
    rotated (ang + angle direction length(p)/2 of p)
    shifted point length(p)/2 of p
 enddef;
 style_def crossed expr p = 
   ccutdraw cross_bar (p, 5mm, 45);
   ccutdraw cross_bar (p, 5mm, -45)
 enddef;}

\fmfcmd{ %
 vardef proj_tarrow (expr p, frac) =
  save a, t, z;
  pair z;
  t1 = frac*length p;
  a = angle direction t1 of p;
  z = point t1 of p;
  (t2,whatever) = p intersectiontimes
    (halfcircle scaled 3arrow_len rotated (a-90) shifted z);
  arrow_head (p, t1, t2, arrow_ang)
enddef;
 style_def projs expr p  =
    cfill (proj_tarrow (reverse p, 0.85));
 enddef;
 style_def projc expr p =
    cfill (proj_tarrow (reverse p, 0.7));
 enddef;}

\fmfcmd{ 
 vardef projl_tarrow (expr p, frac) =
  save a, t, z;
  pair z;
  t1 = frac*length p;
  a = angle direction t1 of p;
  z = point t1 of p;
  (t2,whatever) = p intersectiontimes
    (halfcircle scaled 2arrow_len rotated (a-90) shifted z);
  arrow_head (p, t1, t2, arrow_ang)
 enddef;
 style_def arrowr expr p =
   cfill (projl_tarrow (reverse p, 0.6));
 enddef;
 style_def arrowl expr p  =
   cfill (projl_tarrow (reverse p, 0.6));
 enddef;
}

\fmfcmd{
style_def plains expr p =
 draw p;
 undraw subpath (0,.15) of p;
enddef;}

%%%%%%%%%%%%%%%%%%%%%%%%%%%%%%%%%%%%%%%%%%%%%%%%%%%%%%%%%%%%%%%%%%%%%%%
%                                                                     %
%                  Here the paper continues                           %
%                                                                     %
%%%%%%%%%%%%%%%%%%%%%%%%%%%%%%%%%%%%%%%%%%%%%%%%%%%%%%%%%%%%%%%%%%%%%%%

\be
\raisebox{-1.5ex}{
\parbox{3cm}{\bc 
\begin{fmfgraph*}(2.5,2.5)
 \stargl
 \starplace{}{}{}{}{}{}{$\nx\Delta$}{55}
 \starplacedummy{}{}{}{}{}{}{}{}
 \startypes{boson}{}{boson}{}{boson}{}
 \startypec{boson}{}{boson}{}{boson}{}
 \starproj{}{}{}{proj_v}{}{}
 \fmfdot{v1,v2,v3,v4}
\end{fmfgraph*}\\
\stepcounter{rueck}
\refstepcounter{diagh}
(\Alph{diagh})
\label{A11}
\ec
}}
& \rightarrow & 2 \left[
\raisebox{-1.5ex}{ 
\parbox{3cm}{\bc 
\begin{fmfgraph*}(2.5,2.5)
 \stargl
 \starplace{}{}{}{}{}{}{$\nx\Delta$}{55}
 \starplacedummy{}{}{}{}{}{}{}{}
 \startypes{boson}{}{boson}{arrowr}{boson}{}
 \startypec{boson}{}{boson}{}{boson}{}
 \starproj{}{}{}{}{}{}
 \fmfdot{v2,v3,v4}
\end{fmfgraph*}\\
\refstepcounter{diagh}
(\Alph{diagh})
\label{B11}
\ec
}}
+
\raisebox{-1.5ex}{
\parbox{3cm}{\bc 
\begin{fmfgraph*}(2.5,2.5)
 \stargl
 \starplace{\nx\G}{40}{}{}{}{}{$\nx\Delta$}{55}
 \starplacedummy{}{}{}{}{}{}{}{}
 \startypes{boson}{}{boson}{}{boson}{}
 \startypec{boson}{}{boson}{}{boson}{arrowl}
 \starproj{}{}{}{}{}{}
 \fmfdot{v2,v4}
\end{fmfgraph*}\\
\refstepcounter{diagh}
(\Alph{diagh})
\label{C11}
\ec
}}
-
\raisebox{-1.5ex}{
\parbox{3cm}{\bc 
\begin{fmfgraph*}(2.5,2.5)
 \stargl
 \starplace{\nx\G}{40}{}{}{}{}{$\nx\Delta$}{55}
 \starplacedummy{}{}{}{}{}{}{}{}
 \startypes{boson}{}{boson}{}{boson}{arrowr}
 \startypec{boson}{}{boson}{}{boson}{}
 \starproj{}{}{}{}{}{}
 \fmfdot{v2,v4}
\end{fmfgraph*}\\
\refstepcounter{diagh}
(\Alph{diagh})
\label{D11}
\ec
}}\right. \nonumber\\
& & 
\left. -
\raisebox{-1.5ex}{
\parbox{3cm}{\bc 
\begin{fmfgraph*}(2.5,2.5)
 \stargl
 \starplace{\nx\G}{40}{}{}{\nx\G}{113}{$\nx\Delta$}{55}
 \starplacedummy{}{}{}{}{}{}{}{}
 \startypes{boson}{}{boson}{}{boson}{}
 \startypec{boson}{}{boson}{arrowr}{boson}{}
 \starproj{}{}{}{}{}{}
 \fmfdot{v2}
\end{fmfgraph*}\\
\refstepcounter{diagh}
(\Alph{diagh})
\label{E11}
\ec
}}
-
\raisebox{-1.5ex}{
\parbox{3cm}{\bc 
\begin{fmfgraph*}(2.5,2.5)
 \stargl
 \starplace{\nx\G}{40}{}{}{\nx\G}{113}{$\nx\Delta$}{55}
 \starplacedummy{}{}{}{}{}{}{\nx\G}{-108}
 \startypes{boson}{arrowl}{boson}{}{boson}{}
 \startypec{boson}{}{boson}{}{boson}{}
 \starproj{}{}{}{}{}{}
\end{fmfgraph*}\\
\refstepcounter{diagh}
(\Alph{diagh})
\label{F11}
\ec
}}
+
\raisebox{-1.5ex}{
\parbox{3cm}{\bc 
\begin{fmfgraph*}(2.5,2.5)
 \stargl
 \starplace{\nx\G}{40}{}{}{\nx\G}{113}{$\nx\Delta$}{55}
 \starplacedummy{}{}{}{}{}{}{\nx\G}{-108}
 \startypes{boson}{}{boson}{}{boson}{arrowl}
 \startypec{boson}{}{boson}{}{boson}{}
 \starproj{}{}{}{}{}{}
\end{fmfgraph*}\\
\refstepcounter{diagh}
(\Alph{diagh})
\label{G11}
\ec
}}\right. \nonumber\\
& & 
\left. -
\raisebox{-1.5ex}{
\parbox{3cm}{\bc 
\begin{fmfgraph*}(2.5,2.5)
 \stargl
 \starplace{\nx\G}{40}{\nx\G}{-80}{\nx\G}{113}{$\nx\Delta$}{55}
 \starplacedummy{}{}{}{}{}{}{\nx\G}{-108}
 \startypes{boson}{}{boson}{}{boson}{}
 \startypec{boson}{}{boson}{}{boson}{}
 \starproj{}{proj_v}{}{}{}{}
\end{fmfgraph*}\\
\refstepcounter{diagh}
(\Alph{diagh})
\label{H11}
\ec
}}
-
\raisebox{-1.5ex}{
\parbox{3cm}{\bc 
\begin{fmfgraph*}(2.5,2.5)
 \stargl
 \starplace{\nx\G}{40}{\nx\G}{-80}{{\nx\bf 1}}{112}{$\nx\Delta$}{55}
 \starplacedummy{}{}{}{}{}{}{\nx\G}{-108}
 \startypes{boson}{}{boson}{}{boson}{}
 \startypec{boson}{}{boson}{}{boson}{}
 \starproj{}{}{}{}{}{}
\end{fmfgraph*}\\
\refstepcounter{diagh}
(\Alph{diagh})
\label{I11}
\ec
}}
+\;
\raisebox{-1.5ex}{
\parbox{3cm}{\bc 
\begin{fmfgraph*}(2.5,2.5)
 \stargl
 \starplace{\nx\G}{40}{\nx\G}{155}{\nx\G}{113}{$\nx\Delta$}{55}
 \starplacedummy{}{}{}{}{}{}{\nx\G}{-108}
 \startypes{boson}{}{boson}{}{boson}{}
 \startypec{boson}{}{boson}{}{boson}{}
 \starproj{}{}{}{proj_v}{}{}
\end{fmfgraph*}\\
\refstepcounter{diagh}
(\Alph{diagh})
\label{J11}
\ec
}} \right. \nonumber\\
& & 
\left. +\;
\raisebox{-1.5ex}{
\parbox{3cm}{\bc 
\begin{fmfgraph*}(2.5,2.5)
 \stargl
 \starplace{1}{40}{\nx\G}{155}{\nx\G}{113}{$\nx\Delta$}{55}
 \starplacedummy{}{}{}{}{}{}{\nx\G}{-108}
 \startypes{boson}{}{boson}{}{boson}{}
 \startypec{boson}{}{boson}{}{boson}{}
 \starproj{}{}{}{}{}{}
\end{fmfgraph*}\\
\refstepcounter{diagh}
(\Alph{diagh})
\label{K11}
\ec
}}
-
\raisebox{-1.5ex}{
\parbox{3cm}{\bc 
\begin{fmfgraph*}(2.5,2.5)
 \stargl
 \starplace{\nx\G}{40}{}{}{\nx\G}{-95}{$\nx\Delta$}{55}
 \starplacedummy{}{}{}{}{}{}{}{}
 \startypes{boson}{}{boson}{}{boson}{}
 \startypec{boson}{}{boson}{arrowl}{boson}{}
 \starproj{}{}{}{}{}{}
\end{fmfgraph*}\\
\refstepcounter{diagh}
(\Alph{diagh})
\label{L11}
\ec
}}
+
\raisebox{-1.5ex}{
\parbox{3cm}{\bc 
\begin{fmfgraph*}(2.5,2.5)
 \stargl
 \starplace{\nx\G}{40}{}{}{\nx\G}{-95}{$\nx\Delta$}{55}
 \starplacedummy{}{}{}{}{}{}{}{}
 \startypes{boson}{arrowl}{boson}{}{boson}{}
 \startypec{boson}{}{boson}{}{boson}{}
 \starproj{}{}{}{}{}{}
\end{fmfgraph*}\\
\refstepcounter{diagh}
(\Alph{diagh})
\label{M11}
\ec
}} \right. \nonumber \\
& &
\left. -
\raisebox{-1.5ex}{
\parbox{3cm}{\bc 
\begin{fmfgraph*}(2.5,2.5)
 \stargl
 \starplace{\nx\G}{40}{\nx\G}{10}{\nx\G}{-95}{$\nx\Delta$}{55}
 \starplacedummy{}{}{}{}{}{}{}{}
 \startypes{boson}{}{boson}{}{boson}{}
 \startypec{boson}{}{boson}{}{boson}{arrowr}
 \starproj{}{}{}{}{}{}
\end{fmfgraph*}\\
\refstepcounter{diagh}
(\Alph{diagh})
\label{N11}
\ec
}}
+
\raisebox{-1.5ex}{
\parbox{3cm}{\bc 
\begin{fmfgraph*}(2.5,2.5)
 \stargl
 \starplace{\nx\G}{40}{\nx\G}{10}{{\nx\bf 1}}{-95}{$\nx\Delta$}{55}
 \starplacedummy{}{}{}{}{}{}{\nx\G}{-70}
 \startypes{boson}{}{boson}{}{boson}{}
 \startypec{boson}{}{boson}{}{boson}{}
 \starproj{}{}{}{}{}{}
\end{fmfgraph*}\\
\refstepcounter{diagh}
(\Alph{diagh})
\label{O11}
\ec
}}
+
\raisebox{-1.5ex}{
\parbox{3cm}{\bc 
\begin{fmfgraph*}(2.5,2.5)
 \stargl
 \starplace{\nx\G}{40}{\nx\G}{10}{\nx\G}{-95}{$\nx\Delta$}{55}
 \starplacedummy{}{}{}{}{}{}{\nx\G}{-68}
 \startypes{boson}{}{boson}{}{boson}{}
 \startypec{boson}{}{boson}{}{boson}{}
 \starproj{}{proj_v}{}{}{}{}
\end{fmfgraph*}\\
\refstepcounter{diagh}
(\Alph{diagh})
\label{P11}
\ec
}} \right. \nonumber \\
& & 
\left. -
\raisebox{-1.5ex}{
\parbox{3cm}{\bc 
\begin{fmfgraph*}(2.5,2.5)
 \stargl
 \starplace{\nx\G}{40}{\nx\G}{155}{\nx\G}{-95}{$\nx\Delta$}{55}
 \starplacedummy{}{}{}{}{}{}{}{}
 \startypes{boson}{}{boson}{}{boson}{}
 \startypec{boson}{}{boson}{}{boson}{}
 \starproj{}{}{}{proj_v}{}{}
\end{fmfgraph*}\\
\refstepcounter{diagh}
(\Alph{diagh})
\label{Q11}
\ec
}}
-
\raisebox{-1.5ex}{
\parbox{3cm}{\bc 
\begin{fmfgraph*}(2.5,2.5)
 \stargl
 \starplace{{\nx\bf 1}}{40}{\nx\G}{155}{\nx\G}{-95}{$\nx\Delta$}{55}
 \starplacedummy{}{}{}{}{}{}{}{}
 \startypes{boson}{}{boson}{}{boson}{}
 \startypec{boson}{}{boson}{}{boson}{}
 \starproj{}{}{}{}{}{}
\end{fmfgraph*}\\
\refstepcounter{diagh}
(\Alph{diagh})
\label{R11}
\ec}}\right]
\label{zerstar}
\ee
\begin{enumerate}
    \item[]
    In the same way we treat the star diagrams containing ghosts. Diagram 
    (\ref{decompos1}.\refmy{D15}) can be decomposed as follows:

\be
-
\raisebox{-1.5ex}{
\parbox{3cm}{\bc
\begin{fmfgraph*}(2.5,2.5)
 \stargl
 \starplace{\nx\G}{150}{\nx\G}{-80}{\nx\G}{20}{}{}
 \starplacedummy{$\nx\Delta$}{-55}{}{}{}{}{}{}
 \startypes{boson}{}{boson}{}{boson}{}
 \startypec{boson}{}{boson}{}{boson}{}
 \starproj{}{}{}{proj_h}{}{}
 \fmfdot{v4}
\end{fmfgraph*}\\
\stepcounter{rueck}
\refstepcounter{diagh}
(\Alph{diagh})
\label{A14}
\ec
}}
& \rightarrow &
-
\raisebox{-1.5ex}{
\parbox{3cm}{\bc
\begin{fmfgraph*}(2.5,2.5)
 \stargl
 \starplace{\nx\G}{150}{\nx\G}{-80}{\nx\G}{20}{}{}
 \starplacedummy{$\nx\Delta$}{-55}{}{}{}{}{}{}
 \startypes{boson}{}{boson}{}{boson}{}
 \startypec{boson}{}{boson}{arrowr}{boson}{}
 \starproj{}{}{}{}{}{}
\end{fmfgraph*}\\
\refstepcounter{diagh}
(\Alph{diagh})
\label{B14}
\ec}}
-
\raisebox{-1.5ex}{
\parbox{3cm}{\bc
\begin{fmfgraph*}(2.5,2.5)
 \stargl
 \starplace{\nx\G}{150}{\nx\G}{-80}{\nx\G}{20}{\nx\G}{-107}
 \starplacedummy{$\nx\Delta$}{-55}{}{}{}{}{}{}
 \startypes{boson}{}{boson}{}{boson}{}
 \startypec{boson}{}{boson}{}{boson}{}
 \starproj{}{}{}{}{proj_v}{}
\end{fmfgraph*}\\
\refstepcounter{diagh}
(\Alph{diagh})
\label{C14}
\ec
}}\nonumber\\
& &
+
\raisebox{-1.5ex}{
\parbox{3cm}{\bc
\begin{fmfgraph*}(2.5,2.5)
 \stargl
 \starplace{\nx\G}{150}{\nx\G}{-80}{\nx\G}{20}{}{}
 \starplacedummy{$\nx\Delta$}{-55}{}{}{}{}{}{}
 \startypes{boson}{}{boson}{}{boson}{}
 \startypec{boson}{}{boson}{}{boson}{arrowr}
 \starproj{}{}{}{}{}{}
\end{fmfgraph*}\\
\refstepcounter{diagh}
(\Alph{diagh})
\label{D14}
\ec
}}
+
\raisebox{-1.5ex}{
\parbox{3cm}{\bc
\begin{fmfgraph*}(2.5,2.5)
 \stargl
 \starplace{\nx\G}{150}{\nx\G}{-80}{\nx\G}{20}{\nx\G}{-80}
 \starplacedummy{$\nx\Delta$}{-55}{}{}{}{}{}{}
 \startypes{boson}{}{boson}{}{boson}{}
 \startypec{boson}{}{boson}{}{boson}{}
 \starproj{}{}{}{}{}{proj_v}
\end{fmfgraph*}\\
\refstepcounter{diagh}
(\Alph{diagh})
\label{E14}
\ec
}}
\label{decompos3}
\ee

    Some diagrams already cancel on this level of the calculation. 
    To be specific:\\
    (\ref{zerstar}.\refmy{J11}) cancels (\ref{decompos2}.\refmy{C16}),
    (\ref{zerstar}.\refmy{H11}) cancels (\ref{decompos3}.\refmy{E14}),
    (\ref{zerstar}.\refmy{P11}) cancels (\ref{decompos3}.\refmy{C14}),
    (\ref{zerstar}.\refmy{Q11}) cancels (\ref{decompos1}.\refmy{C12}).\\
    The decomposition and cancellation of the star diagrams containing fermions
    will be discussed below.

  \item[3.] 
    The remaining star diagrams do not cancel each other but are related 
    to diagrams with different topology. To make their cancellation clear we 
    use the identities from section \ref{sec2.1.2}, specially (\ref{rel1}), 
    (\ref{rel2}) and (\ref{rel3}).\\
    We start with relation (\ref{rel1}) and place the $\Delta$ at the upper 
    line near the left vertex. We connect the upper two lines with a 
    three-gluon-vertex, we do the same with the lower ones, and we connect the
    remaining free leg of each vertex. In this way we get: 

\be
+\;2
\raisebox{-1.5ex}{
\parbox{3cm}{\bc 
\begin{fmfgraph*}(2.5,2.5)
 \stargl
 \starplace{}{}{}{}{}{}{$\nx\Delta$}{55}
 \starplacedummy{}{}{}{}{}{}{}{}
 \startypes{boson}{}{boson}{arrowr}{boson}{}
 \startypec{boson}{}{boson}{}{boson}{}
 \starproj{}{}{}{}{}{}
 \fmfdot{v2,v3,v4}
\end{fmfgraph*}\\
\stepcounter{rueck}
\refstepcounter{diagh}
(\Alph{diagh})
\label{A18}
\ec
}}
-
\raisebox{-1.5ex}{
\parbox{3cm}{\bc
 \begin{fmfgraph*}(2.5,2.5)
   \selfgl
   \selfplace{}{}{$\nx\Delta$}{90}{}{}{}{}
   \selfplacedummy{}{}{}{}{}{}{}{}
   \selftypeb{boson}{}{boson}{}{boson}{}{boson}{}
   \selftypes{boson}{arrowr}{boson}{}
   \selfproj{}{}{}{}{}{}
   \fmfdot{v2,v3,v4}
 \end{fmfgraph*}\\
\refstepcounter{diagh}
(\Alph{diagh})
\label{B18}
\ec
}}
= \; -
\raisebox{-1.5ex}{
\parbox{3cm}{\bc
\begin{fmfgraph*}(2.5,2.5)
 \pacgl
 \pacplace{}{}{}{}{$\nx\Delta$}{-105}
 \pacplacedummy{}{}{}{}{}{}
 \pactypes{boson}{}{boson}{}{boson}{}
 \pactypec{boson}{}{boson}{}
 \pacproj{}{}{}{}{proj_v}
 \fmfdot{v1,v2,v3}
\end{fmfgraph*}\\
\refstepcounter{diagh}
(\Alph{diagh})
\label{C18}
\ec
}}.
\label{Label}
\ee

    With this relation diagram (\ref{zerstar}.\refmy{B11}) cancels 
    $\alpha$-dependent parts from the diagrams (\ref{alldia}.\refmy{K9}) 
    and (\ref{alldia}.\refmy{L9}) respectively. Now closing (\ref{rel1}) 
    in different ways and using a ghost-gluon vertex as well as a 
    three-gluon vertex, one can find 12 more relations. With the help of these
    relations the diagrams (\ref{zerstar}.\refmy{C11}), 
    (\ref{zerstar}.\refmy{D11}), (\ref{zerstar}.\refmy{E11}), 
    (\ref{zerstar}.\refmy{L11}) are also cancelled by other diagrams.\\
    For the star diagram, we need no identity derived from 
    (\ref{rel2}). However it is necessary to cancel diagrams originating from 
    (\ref{alldia}.\refmy{K9}) and (\ref{alldia}.\refmy{L9}).\\
    Finally we use  (\ref{rel3}). In an analogous way as above we get 
    e.g.~the following relation:

\be
\raisebox{-1.5ex}{
\parbox{3cm}{\bc 
\begin{fmfgraph*}(2.5,2.5)
 \stargl
 \starplace{1}{40}{\nx\G}{160}{\nx\G}{-95}{$\nx\Delta$}{55}
 \starplacedummy{}{}{}{}{}{}{}{}
 \startypes{boson}{}{boson}{}{boson}{}
 \startypec{boson}{}{boson}{}{boson}{}
 \starproj{}{}{}{}{}{}
 \fmfdot{v4}
\end{fmfgraph*}\\
\stepcounter{rueck}
\refstepcounter{diagh}
(\Alph{diagh})
\label{A19}
\ec
}}
-
\raisebox{-1.5ex}{
\parbox{3cm}{\bc 
\begin{fmfgraph*}(2.5,2.5)
 \stargl
 \starplace{\nx\G}{40}{}{}{\nx\G}{-95}{$\nx\Delta$}{55}
 \starplacedummy{}{}{}{}{}{}{}{}
 \startypes{boson}{arrowl}{boson}{}{boson}{}
 \startypec{boson}{}{boson}{}{boson}{}
 \starproj{}{}{}{}{}{}
 \fmfdot{v4}
\end{fmfgraph*}\\
\refstepcounter{diagh}
(\Alph{diagh})
\label{B19}
\ec
}}
-
\raisebox{-1.5ex}{
\parbox{3cm}{\bc
 \begin{fmfgraph*}(2.5,2.5)
   \selfgl
   \selfplace{\nx\G}{-105}{\nx\G}{75}{}{}{$\nx\Delta$}{-90}
   \selfplacedummy{}{}{}{}{}{}{}{}
   \selftypeb{boson}{}{boson}{}{boson}{}{boson}{}
   \selftypes{boson}{}{boson}{arrowr}
   \selfproj{}{}{}{}{}{}
   \fmfdot{v4}
 \end{fmfgraph*}\\
\refstepcounter{diagh}
(\Alph{diagh})
\label{C19}
\ec
}}= 0.
\ee

    At all one gets five relations out of (\ref{rel3}) which cause the diagrams
    (\ref{zerstar}.\refmy{F11}), (\ref{zerstar}.\refmy{G11}), 
    (\ref{zerstar}.\refmy{I11}), (\ref{zerstar}.\refmy{K11}),
    (\ref{zerstar}.\refmy{M11}), (\ref{zerstar}.\refmy{N11}),
    (\ref{zerstar}.\refmy{O11}), (\ref{zerstar}.\refmy{R11}),
    (\ref{decompos3}.\refmy{B14}), (\ref{decompos3}.\refmy{D14}) to cancel
    each other.\\

  \item[4.] 
    Now we discuss the decomposition and cancellation of the star 
    diagrams with fermions.\\ 
    We only give a brief example of the cancellations. Decomposing 
    e.g.~(\ref{decompos4}.\refmy{C15f}) in the way described above we get:

\be
\raisebox{-1.5ex}{
 \parbox{3cm}{\bc
  \begin{fmfgraph*}(2.5,2.5)
   \stargl
   \starplace{}{}{}{}{}{}{$\nx\Delta$}{55}
   \starplacedummy{}{}{}{}{}{}{}{}
   \startypes{fermion}{}{fermion}{}{fermion}{}
   \startypec{boson}{}{boson}{}{boson}{}
   \starproj{}{}{}{proj_v}{}{}
   \fmfdot{v1,v2,v3,v4}
  \end{fmfgraph*}\\
 \stepcounter{rueck}
 \refstepcounter{diagh}
 (\Alph{diagh})
 \label{A30}
 \ec}}
=
\raisebox{-1.5ex}{
 \parbox{3cm}{\bc
  \begin{fmfgraph*}(2.5,2.5)
   \fmfforce{(0,.5h)}{v1}
   \fmfforce{(.5w,.5h)}{v2}
   \fmfforce{(w,.5h)}{v3}
   \fmf{fermion,left}{v1,v3,v1}
   \fmf{boson,left=.5}{v2,v3,v2}
   \fmf{boson}{v2,v1}
   \fmfv{label=$\nx\Delta$,label.a=80}{v2}
   \fmfdot{v2,v1}
  \end{fmfgraph*}\\
 \refstepcounter{diagh}
 (\Alph{diagh})
 \label{C30}
 \ec}
}
-
\raisebox{-1.5ex}{
 \parbox{3cm}{\bc
  \begin{fmfgraph*}(2.5,2.5)
   \fmfforce{(0,.5h)}{v1}
   \fmfforce{(.5w,.5h)}{v2}
   \fmfforce{(w,.5h)}{v3}
   \fmf{fermion,left}{v1,v3,v1}
   \fmf{boson,left=.5}{v1,v2,v1}
   \fmf{boson}{v2,v3}
   \fmfv{label=$\nx\Delta$,label.a=100}{v2}
   \fmfdot{v2,v3}
  \end{fmfgraph*}\\
 \refstepcounter{diagh}
 (\Alph{diagh})
 \label{B30}
 \ec}
}.
\label{ferap1}
\ee
    On the other hand we close the fermion identity (\ref{rel4}) to find

\be
-
\raisebox{-1.5ex}{
 \parbox{3cm}{\bc
  \begin{fmfgraph*}(2.5,2.5)
   \selfgl
   \selfplace{}{}{}{}{$\nx\Delta$}{80}{}{}
   \selfplacedummy{}{}{}{}{}{}{}{}
   \selftypeb{fermion}{}{fermion}{}{boson}{}{boson}{}
   \selftypes{boson}{arrowl}{boson}{}
   \selfproj{}{}{}{}{}{}
   \fmfdot{v1,v2,v3}
  \end{fmfgraph*}\\
 \stepcounter{rueck}
 \refstepcounter{diagh}
 (\Alph{diagh})
 \label{A32}
 \ec}
}
= \; +
\raisebox{-1.5ex}{
 \parbox{3cm}{\bc
  \begin{fmfgraph*}(2.5,2.5)
   \fmfforce{(0,.5h)}{v1}
   \fmfforce{(.5w,.5h)}{v2}
   \fmfforce{(w,.5h)}{v3}
   \fmf{fermion,left}{v1,v3,v1}
   \fmf{boson,left=.5}{v2,v3,v2}
   \fmf{boson}{v2,v1}
   \fmfv{label=$\nx\Delta$,label.a=82}{v2}
   \fmfdot{v2,v1}
  \end{fmfgraph*}\\
 \refstepcounter{diagh}
 (\Alph{diagh})
 \label{B32}
 \ec}
} -
\raisebox{-1.5ex}{
 \parbox{3cm}{\bc
  \begin{fmfgraph*}(2.5,2.5)
   \fmfforce{(0,.5h)}{v1}
   \fmfforce{(.5w,.5h)}{v2}
   \fmfforce{(w,.5h)}{v3}
   \fmf{fermion,left}{v1,v3,v1}
   \fmf{boson,left=.5}{v1,v2,v1}
   \fmf{boson}{v2,v3}
   \fmfv{label=$\nx\Delta$,label.a=100}{v2}
   \fmfdot{v2,v3}
  \end{fmfgraph*}\\ 
 \refstepcounter{diagh}
 (\Alph{diagh})
 \label{C32}
 \ec}
}. 
\ee

    Hence the diagram (\ref{ferap1}.\refmy{A30}) is cancelled by a diagram 
    originating from (\ref{alldia}.\refmy{L9}). Similarly all 
    $a_{\mu}$-dependent diagrams containing fermions cancel.
\end{enumerate}

In the same way all $a_\mu $-dependent diagrams of (\ref{alldia}) drop out of 
the calculation of the partition function.

\appendix
\setcounter{section}{1}
\section{Feynman Rules}
\label{appb}

We use the following Feynman rules for covariant $\alpha$-gauges in Euklidien
space-time:  

\begin{tabular}{lcc}
{fermion propagator} &
\parbox[b]{7cm}{\bc
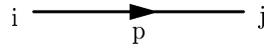
 \begin{fmfgraph*}(3.5,1)
    \fmfincoming{i1,i2}
     \fmfoutgoing{o1,o2}
     \fmfv{label=i}{i1}
     \fmfv{label=j}{o1}
     \fmf{fermion,label=p,label.side=right}{i1,o1}
 \end{fmfgraph*}
\ec}
& $S_{ij} = -\frac{\delta_{ij}}{\slashed{p} - m_f}$\\
{gluon propagator} &
\parbox[b]{7cm}{\bc
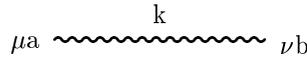
 \begin{fmfgraph*}(3.5,1)
     \fmfincoming{i1,i2}
     \fmfoutgoing{o1,o2}
     \fmfv{label=$\nx\mu$a}{i1}
     \fmfv{label=$\nx\nu$b}{o1}
     \fmf{boson,label=k,label.side=left}{i1,o1}
 \end{fmfgraph*}
\ec}
& $D_{\mu\nu}^{ab} =\frac{\delta^{ab}}{k^2}\left[g_{\mu\nu} - 
                   (1-\alpha)\frac{k_\mu k_\nu}{k^2}\right]$\\
{ghost propagator} &
\parbox[b]{7cm}{\bc
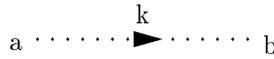
 \begin{fmfgraph*}(3.5,1)
     \fmfincoming{i1,i2}
     \fmfoutgoing{o1,o2}
     \fmfv{label=a}{i1}
     \fmfv{label=b}{o1}
     \fmf{ghost,label=k,label.side=left}{i1,o1}
 \end{fmfgraph*}
\ec}
& $W^{ab} = -\frac{\delta^{ab}}{k^2}$\\
{fermion-gluon vertex} &
\parbox{7cm}{\bc
 \fmfframe(0,1)(0,0)
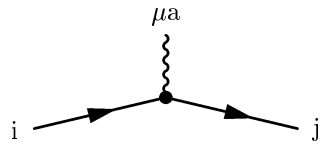
{
 \begin{fmfgraph*}(3.5,2.5)
  \fmfincoming{i1}
  \fmfoutgoing{o1}
   \fmftop{v2} 
   \fmfv{label=$\nx\mu$a}{v2}
   \fmfv{label=i}{i1}
   \fmfv{label=j}{o1}
   \fmf{fermion}{i1,v1}
   \fmf{boson}{v2,v1}
   \fmf{fermion}{v1,o1}
   \fmfdot{v1}
 \end{fmfgraph*}}\ec
}
& $F_{\mu,ij}^a = -g\gamma_\mu t_{ij}^a$\\
{ghost-gluon vertex} &
\parbox{7cm}{\bc
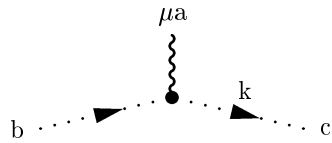
 \begin{fmfgraph*}(3.5,2.5)
  \fmfincoming{i1}
  \fmfoutgoing{o1}
   \fmftop{v2} 
   \fmfv{label=$\nx\mu$a}{v2}
   \fmfv{label=b}{i1}
   \fmfv{label=c}{o1}
   \fmf{ghost}{i1,v1}
   \fmf{boson}{v2,v1}
   \fmf{ghost,label=k,label.side=left}{v1,o1}
   \fmfdot{v1}
 \end{fmfgraph*}\ec
}
& $\Gamma_{\mu}^{abc} = -igf^{abc}k_\mu$
\end{tabular}

\begin{tabular}{lcc}
{three-gluon vertex} &
\parbox{7cm}{\bc
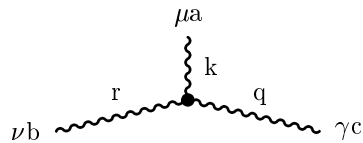
 \begin{fmfgraph*}(3.5,2.5)
  \fmfincoming{i1}
  \fmfoutgoing{o1}
   \fmftop{v2} 
   \fmfv{label=$\nx\mu$a}{v2}
   \fmfv{label=$\nx\nu$b}{i1}
   \fmfv{label=$\nx\gamma$c}{o1}
   \fmf{boson,label=r,l.s=left}{i1,v1}
   \fmf{boson,label=k,l.s=left}{v2,v1}
   \fmf{boson,label=q,l.s=left}{v1,o1}
   \fmfdot{v1}
 \end{fmfgraph*}\ec
}
& \parbox{5cm}{
\be
\Gamma^{abc}_{\mu\nu\gamma}(k,r,q) & = &  
    -igf^{abc}\left[g_{\nu\gamma}(r-q)_{\mu}\right.\nonumber\\
& & \quad \left. +g_{\mu\nu}(k-r)_{\gamma} + g_{\gamma\mu}(q-k)_{\nu}\right]
\nonumber\ee}\\
{four-gluon vertex} &
\parbox{7cm}{\bc
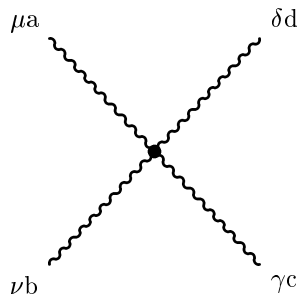
 \begin{fmfgraph*}(3.5,3)
   \fmfincoming{i1,i2}
   \fmfoutgoing{o1,o2}
   \fmfv{label=$\nx\mu$a}{i2}
   \fmfv{label=$\nx\delta$d}{o2}
   \fmfv{label=$\nx\gamma$c}{o1}
   \fmfv{label=$\nx\nu$b}{i1}
   \fmf{boson}{i1,v1} 
   \fmf{boson}{i2,v1}
   \fmf{boson}{v1,o1}
   \fmf{boson}{v1,o2}
   \fmfdot{v1}
 \end{fmfgraph*}\ec
}
& \parbox{5cm}{\be
Q_{\mu\nu\gamma\delta}^{abcd} & = & - g^2
       \left[f^{ade}f^{ebc}(g_{\mu\nu}g_{\delta\gamma}
                         - g_{\mu\gamma}g_{\delta\nu})\right.\nonumber\\
   & &    \left. +f^{abe}f^{edc}(g_{\mu\delta}g_{\nu\gamma}
          -g_{\mu\gamma}g_{\delta\nu})\right.\nonumber\\
   & &    \left. + f^{ace}f^{edb}(g_{\mu\delta}g_{\gamma\nu}
                          - g_{\mu\nu}g_{\delta\gamma})\right]\nonumber
\ee}
\end{tabular}
\vspace{2cm}

In static resummed theory discussed in this paper, the gluon propagator 
changes to 

\be
  D_{\mu\nu}^{ab} =\frac{\delta^{ab}}{k^2}\left[g_{\mu\nu} 
  -\delta_{\mu 0}\delta_{\nu 0}\delta_{k_0 0}\frac{m^2}{m^2 + k^2}
- (1-\alpha)\frac{k_\mu k_\nu}{k^2}\right]
\ee

and one gets an additional mass-counterterm vertex:
\vspace{1cm}

\begin{tabular}{lcc}
 mass-counterterm vertex &
\parbox[b]{7cm}{\bc
 \begin{fmfgraph*}(3.5,.1)
  \fmfincoming{i1,i2}
  \fmfoutgoing{o1,o2}
  \fmfv{label=a}{i1}
  \fmfv{label=b}{o1}
  \fmf{boson,label=$\nx\mu\nx\qquad\nx\nu$,l.s=right}{i1,o1}
  \fmf{crossed}{i1,o1}
 \end{fmfgraph*}
\ec}
& 
$\delta^{ab}\delta_{\nu 0}\delta_{\mu 0}\delta_{k_0 0}m^2$\\
\end{tabular}
\vspace{1cm}

From these the vacuum rules are easily recovered: 
\begin{enumerate}
 \item Replace the Matsubara sum by an integral as described in 
       Sec.~\ref{sec3.1}. 
 \item Multiply every vertex with a factor $i$ and every propagator with
       $(-i)$. 
 \item Correct the 
$\frac{1}{k^2}$-factor in the ghost and gluon propagator to
$\frac{1}{k^2+i\varepsilon}$, in the fermion-propagator analogously to 
$\frac{1}{\slashed{p}-m+i\varepsilon}$.
\end{enumerate}

\section{Additional Fermion Vertex}
\label{appc}

Here we define the four point gluon-fermion vertex used in (\ref{rel4}). 
We consider the following diagram, using the fermion rule (\ref{femrul}):

\be
\parbox{3.5cm}{\bc
\begin{fmfgraph*}(3,5)
 \fmfforce{(0,h)}{v1}
 \fmfforce{(0,.5h)}{v2}
 \fmfforce{(0,0)}{v3}
 \fmfforce{(.5w,h)}{v4}
 \fmfforce{(.5w,h)}{v9}
 \fmfforce{(.5w,.5h)}{v5}
 \fmfforce{(.5w,.5h)}{v10}
 \fmfforce{(.5w,0)}{v6}
 \fmfforce{(.5w,0)}{v11}
 \fmfforce{(w,h)}{v7}
 \fmfforce{(w,0)}{v8}
 \fmf{boson}{v1,v4}
 \fmf{boson,label=$p$}{v2,v5}
 \fmf{projc}{v5,v2}
 \fmf{boson}{v3,v6}
 \fmf{fermion,label=$q$}{v4,v5}
 \fmf{fermion,label=$p+q$}{v5,v6}
 \fmf{fermion}{v6,v8}
 \fmf{fermion}{v7,v4}
 \fmfv{label=$\nx\gamma$c}{v1}
 \fmfv{label=$\nx\mu$a}{v2}
 \fmfv{label=$\nx\nu$b}{v3}
 \fmfv{label=i,label.a=40}{v4}
 \fmfv{label=j,label.a=-60}{v9}
 \fmfv{label=m,label.a=60}{v5}
 \fmfv{label=n,label.a=-60}{v10}
 \fmfv{label=k,label.a=60}{v6}
 \fmfv{label=l,label.a=-30}{v11}
\end{fmfgraph*}
\ec}
&=&
\parbox{3.5cm}{\bc
\begin{fmfgraph*}(3,5)
 \fmfforce{(0,h)}{v1}
 \fmfforce{(0,.5h)}{v2}
 \fmfforce{(0,0)}{v3}
 \fmfforce{(.5w,h)}{v4}
 \fmfforce{(.5w,h)}{v9}
 \fmfforce{(.5w,.5h)}{v5}
 \fmfforce{(.5w,.5h)}{v10}
 \fmfforce{(.5w,0)}{v6}
 \fmfforce{(.5w,0)}{v11}
 \fmfforce{(w,h)}{v7}
 \fmfforce{(w,0)}{v8}
 \fmf{boson}{v1,v4}
 \fmf{boson,label=$p$}{v2,v6}
 \fmf{boson}{v3,v6}
 \fmf{fermion,label=$q$}{v4,v6}
 \fmf{fermion}{v6,v8}
 \fmf{fermion}{v7,v4}
 \fmfv{label=$\nx\gamma$c}{v1}
 \fmfv{label=a}{v2}
 \fmfv{label=$\nx\nu$b}{v3}
 \fmfv{label=i,label.a=40}{v4}
 \fmfv{label=j,label.a=-60}{v9}
 \fmfv{label=k,label.a=60}{v6}
 \fmfv{label=l,label.a=-30}{v11}
\end{fmfgraph*}\ec}
- 
\parbox{3.5cm}{\bc
\begin{fmfgraph*}(3,5)
 \fmfforce{(0,h)}{v1}
 \fmfforce{(0,.5h)}{v2}
 \fmfforce{(0,0)}{v3}
 \fmfforce{(.5w,h)}{v4}
 \fmfforce{(.5w,h)}{v9}
 \fmfforce{(.5w,.5h)}{v5}
 \fmfforce{(.5w,.5h)}{v10}
 \fmfforce{(.5w,0)}{v6}
 \fmfforce{(.5w,0)}{v11}
 \fmfforce{(w,h)}{v7}
 \fmfforce{(w,0)}{v8}
 \fmf{boson}{v1,v4}
 \fmf{boson,label=$p$}{v2,v4}
 \fmf{boson}{v3,v6}
 \fmf{fermion,label=$p+q$}{v4,v6}
 \fmf{fermion}{v6,v8}
 \fmf{fermion}{v7,v4}
 \fmfv{label=$\nx\gamma$c}{v1}
 \fmfv{label=a}{v2}
 \fmfv{label=$\nx\nu$b}{v3}
 \fmfv{label=i,label.a=40}{v4}
 \fmfv{label=j,label.a=-60}{v9}
 \fmfv{label=k,label.a=60}{v6}
 \fmfv{label=l,label.a=-30}{v11}
\end{fmfgraph*}\ec}.
\ee

\be 
  F^{c}_{\gamma ,ij}S^{jm}(q)p^{\mu}F^{a}_{\mu ,mn}S^{nk}(p+q)F^{b}_{\nu ,kl}
& = & 
  F^{c}_{\gamma ,ij}\left[i\mbox{\em g}\left(t^{a}\right)^{jk}
      \left(\frac{1}{\slashed{q}-m_f+i\varepsilon} -
            \frac{1}{\slashed{p}+\slashed{q}-m_f+i\varepsilon}\right)\right]
  F^{b}_{\nu ,kl}\nonumber\\
& = &
  F^{c}_{\gamma ,ij}S^{jk}(q)\underbrace{
    \left[-i\mbox{\em g}^{2}\gamma_{\nu}\left(t^{a}t^{b}\right)^{kl}\right]
                               }_{\Upsilon^{ab}_{\nu ,kl}} - \underbrace{
    \left[-i\mbox{\em g}^{2}\gamma_{\gamma}\left(t^{c}t^{a}\right)^{ij}\right]
                               }_{\Upsilon^{ca}_{\gamma ,ij}}
  S_{jk}(p+q)F^{b}_{\nu ,kl}.
\label{lastrule}
\ee

From eq.~(\ref{lastrule}) we can read off the four point fermion-gluon vertex:

\be
\Upsilon^{ab}_{\mu ,ij} = -i\mbox{\em g}^{2}\gamma_{\mu}\left(t^{a}t^{b}\right)^{ij}
\ee

Note that $\Upsilon^{ab}_{\mu ,ij} \not= \Upsilon^{ba}_{\mu ,ij}$.

%%%%%%%%%%%%%%%%%%%%%%%%%%%%%%%%%%%%%%%%%%%%%%%%%%%%%%%%%%%%%%%%%%%%%%

\end{fmffile}

\end{document}